\documentclass{aa}%
\usepackage{graphicx}
\usepackage{xcolor}
\usepackage{subfigure}
\usepackage{subcaption}

\usepackage[switch]{lineno}

\usepackage{txfonts}
\definecolor{myblue}{rgb}{0.00, 0.0, 0.9}
\definecolor{myred}{rgb}{0.90, 0.0, 0.0}
\definecolor{mygreen}{rgb}{0.0, 0.7, 0.0}
\usepackage[breaklinks,  citecolor=myblue, linkcolor=myred, urlcolor=purple, colorlinks=true, debug, baseurl=' ']{hyperref}

\usepackage{orcidlink}
\usepackage{ulem}

\begin{document} 
   \title{TRIShUL: Technique for Reconstructing magnetic Interstellar Structure Using starLight polarization\thanks{The GitHub link for the code will be made available upon the acceptance of the paper.}}

\author{Namita Uppal\inst{1}
          \and
          Konstantinos Tassis\inst{1,2}
          \and
          Vasiliki Pavlidou\inst{1,2}
          \and
          Vincent Pelgrims\inst{3}
          \and
          Myrto Falalaki\inst{2}
          }

   \institute{Institute of Astrophysics, Foundation for Research and Technology-Hellas, Vasilika Vouton, GR-71110 Heraklion, Greece \\
   \email{namita@ia.forth.gr}
    \and
    Department of Physics and Institute of Theoretical and Computational Physics, University of Crete, Voutes Campus, GR-70013, Heraklion, Greece
     \and
    Universit{\'e} Libre de Bruxelles, Science Faculty CP230, B-1050 Brussels, Belgium\\
            }

 
  \abstract
{We present a novel technique for decomposition of line-of-sight (LOS) stellar polarization as a function of distance, aimed at reconstructing three dimensional (3D) plane-of-sky (POS) magnetic structures in the interstellar medium (ISM). The method is based on the assumption that the observed polarization arises from discrete, thin dust layers located at varying distances along the LOS. Using a simple and intuitive frequentist framework, our method identifies structural changes in the distance-sorted cumulative Mahalanobis distance between Stokes parameters ($q$ and $u$) to detect the locations of dust layers and estimates their associated physical properties (parallax and Stokes parameters) necessary to construct 3D maps.  
We benchmark the method using mock datasets representative of high-Galactic-latitude regions, incorporating realistic parallax uncertainties from Gaia and expected polarization measurements from the upcoming \textsc{Pasiphae} survey. Our tests show that the method reliably recovers the distances and polarization properties of dust clouds when the polarization signal exceeds 0.1\%, and the effective fraction of background stars is greater than 10\% in our tested samples having $\sim 345$ stars. The effect of background star fraction on the performance becomes less critical with increasing amplitude of the polarization source field from the dust cloud.
We apply our method to existing polarization data from two illustrative sightlines, one at intermediate-high Galactic latitude and one near the Galactic plane, with known tomographic solutions, finding excellent agreement with the literature and demonstrating its accuracy across both regions.
We compare the performances of our method with those of the Bayesian method  BISP-1. While both methods effectively recover dust cloud properties, our approach is prior-free and computationally more efficient in determining the optimal number of clouds along the LOS. These advantages make our method more flexible and broadly applicable for multi-layer dust cloud reconstruction for the upcoming era of large-scale stellar polarization surveys.
}

   \maketitle
%
\section{Introduction}\label{sec:1}
Our vantage point within the Galactic disk complicates the study of the global structure of the Galaxy, as multiple features are superimposed along each line-of-sight (LOS).  Studies of the Milky Way interstellar medium (ISM) have traditionally relied on two-dimensional (2D) sky projections \citep[e.g.,][]{Schlegel1998, planck2016, Lenz2017, Chiang2023}. However, recent advancements have enabled the exploration of the third spatial dimension: distance.
Gaia mission, with its precise parallax measurements of over billions of stars \citep[e.g.,][]{GaiaDR2, GaiaDR3, bailer2021}, has been instrumental in this progress, allowing for detailed 3D mapping of the dust density in the ISM \citep[e.g.,][]{Lallement2019, Green2019, Lallement2022, vergely2022, edenhofer2023}.

The enhanced resolution of 3D studies has provided new insights into local ISM structures, such as the Orion complex and other isolated molecular clouds \citep[e.g.,][]{Doi2021, RezaeiOrion, Tahani2022Orion}. These studies, as well as other methods independent of stellar distances \citep[e.g.,][]{TritsisMusca}, highlight the critical role of magnetic fields in shaping the morphology and formation of these regions. Magnetic fields significantly influence the dynamics of the ISM at various scales \citep[e.g.,][and reference therein]{Pattle2023}, and shape the large-scale structures of the disk and halo of the Galaxy \citep[e.g.,][]{Beck2015, Clark2019}.  However, despite their importance, our three-dimensional understanding of Galactic magnetic fields remains limited due to the lack of direct, spatially resolved measurements.

Various observational tracers, such as Zeeman splitting, Faraday rotation, synchrotron emission, and polarized thermal dust emission, have been used to study magnetic fields in the Galaxy. Each, however, has its own limitations: Zeeman splitting is very weak, and it is difficult to detect \citep{Beck2013}. Faraday rotation studies only probe the LOS information \citep[e.g.,][]{Tahani2018, Dickey2022} and is sensitive to nearby small-scale structures \citep[][]{Wolleben2010}. Our understanding of the magnetic field both in external galaxies and in the Milky Way has been greatly enhanced by observations of synchrotron and thermal dust emission. However, these tracers only provide plane-of-sky (POS) integrated information along the LOS, making it challenging to resolve 3D structures of the magnetic field where multiple dust clouds overlap. In addition, various magnetic-field models \citep[][]{Sun2008, Jansson2012, Jaffe2019, Unger2024, Korochkin2025}, each with differing morphology, fit equally well the polarized synchrotron and Faraday rotation or dust emission \citep{planck2016}. Therefore, integrated polarization data do not suffice to lift the existing degeneracies \citep{Unger2024}. For example, \cite{pelgrims2025} showed that all fits of large-scale GMF could be biased by the presence of the Local Bubble.

On the other hand, starlight polarization, first observed by \citet{Hall1949} and \citet{Hiltner1949}, offers a complementary probe.  Aspherical dust grains align their short axes parallel to the ambient magnetic field. The background starlight, when interacts with such grains, becomes partially linearly polarized due to dichroic extinction \citep{Andersson2015}. When combined with precise stellar distances from Gaia, starlight polarization enables tomographic reconstruction of the POS component of the magnetic field, offering a unique opportunity to disentangle overlapping magnetic structures along the LOS \citep{Panopoulou2019, Doi2021, Pelgrims2024, Yenifer2025}. The approach is uniquely feasible with only starlight polarization, as other magnetic field tracers lack the precise distance estimations necessary to separate the magnetic fields of different clouds along the LOS.

Recently, 3D tomographic maps of different regions of the  Galaxy have been constructed using starlight polarization \citep[e.g.,][]{Panopoulou2019, Doi2021, Pelgrims2023, Pelgrims2024, Doi2024, Uppal2024, Yenifer2025}. Several regions of the sky have been mapped in linear polarization observations within the disk and in more diffuse regions \citep[][]{GPIPS, Versteeg2023, Uppal2024, Pelgrims2024, Panopoulou2025}. The upcoming \textsc{Pasiphae} survey \citep{Tassis2018} and the SOUTH POL survey \citep{Magalhaes2005} promise significant advancements by providing stellar polarization data for millions of stars covering a large area of the sky. In the upcoming era of large-scale polarization surveys, it is essential to develop algorithms that solve the inversion problem to unravel the intricate 3D structure of the magnetized ISM from starlight polarization data and stellar distances. 

The first such automated framework is \texttt{BISP-1} \citep{Pelgrims2023}, developed for the upcoming high-latitude polarization survey, \textsc{Pasiphae}. It employs a Bayesian formalism for decomposing the starlight polarization with distance. Although originally designed for high-latitude regions, it has also been successfully applied to lines of sight in the Galactic plane \citep{Uppal2024, Bijas2024, Yenifer2025}. 
The current implementation of \texttt{BISP-1} can model up to five thin dust layers along the LOS, which is a valid assumption for high-latitude fields, however, in the Galactic plane, characterized by higher dust densities and more complex structures, this assumption may be limiting. Moreover, evaluating multiple models (e.g., 1- to 5-layer configurations) requires exploring a high-dimensional parameter space, making the method computationally expensive (e.g., requires $\sim 1$ hour to process one LOS on Apple M1 chip having 16 GB RAM).
 
In this paper, we present a novel and independent non-Bayesian method, named \texttt{TRIShUL}, developed for starlight polarization tomography. This method works on a per-line-of-sight basis. In Sect.~\ref{sec:2}, we describe our methodology and its key components designed to identify the optimal number of dust layers along the LOS. We also explain how the method estimates the average properties of each dust layer, such as parallax and Stokes parameters, along with their associated uncertainties. A rigorous testing for benchmarking the performance and identifying the limitations of our method in low polarization ($\le 1\%$) and low signal-to-noise (S/N) is presented in Sect.~\ref{sec:3}. In Sect.~\ref{sec:4}, we present a detailed performance comparison between our method and \texttt{BISP-1} by applying both to mock dataset representing 1- and 2-layer sightlines. In Sect.~\ref{sec:5} we applied our method to published stellar polarization observations, demonstrating its applicability to both high-latitude regions and fields near the Galactic plane. We finally summarize and conclude our work in Sect.~\ref{sec:6}.

\section{Methodology: \texttt{TRIShUL} framework}\label{sec:2}
Our method is based on the fact that starlight becomes partially polarized when it passes through the dust clouds present along the LOS. 
As a result, when an ensemble of stars at different distances along a LOS is analyzed, the presence of multiple dust layers can be inferred from sudden ``jumps'' in the polarization signal as a function of distance. To capture these features, our approach relies on the following assumptions. 

Firstly, we use stellar polarization measurements within a finite circular area centered on each LOS. The area is chosen to be small enough to avoid significant large-scale plane-of-sky polarization variations caused by spatially varying dust structures and large-scale ISM turbulence, so that a meaningful mean polarization can be defined. The scatter in the relative Stokes parameters $(q,u)$ reflects the contributions from small-scale inhomogeneities and observational noise.

Secondly, we assume that polarization is induced by a set of discrete, thin dust layers along the LOS. This assumption is justified in the low polarization regime, where each layer contributes independently to the total observed polarization and can be modeled as the vector sum of the Stokes parameters from all foreground layers \citep[Appendix B of][]{patat2010}.

Thirdly, we assume that the observed polarization is entirely due to the intervening magnetized ISM, with no contribution from stellar intrinsic components such as circumstellar disks or asymmetric stellar envelopes \citep[e.g.,][]{cotton2016, Bailey2024}.

Our method, named \texttt{TRIShUL}, draws inspiration from Indian mythology, where it means ``trident.'' The three prongs of the \texttt{TRIShUL} represent the three key steps in our approach, the details of which are represented in the following subsections along with the post-processing steps for calculating the physical parameters associated with the identified layers.

\subsection{Calculating cumulative Mahalanobis distance}\label{sec:2.1}
\begin{figure} 
   \resizebox{\hsize}{!}{\includegraphics{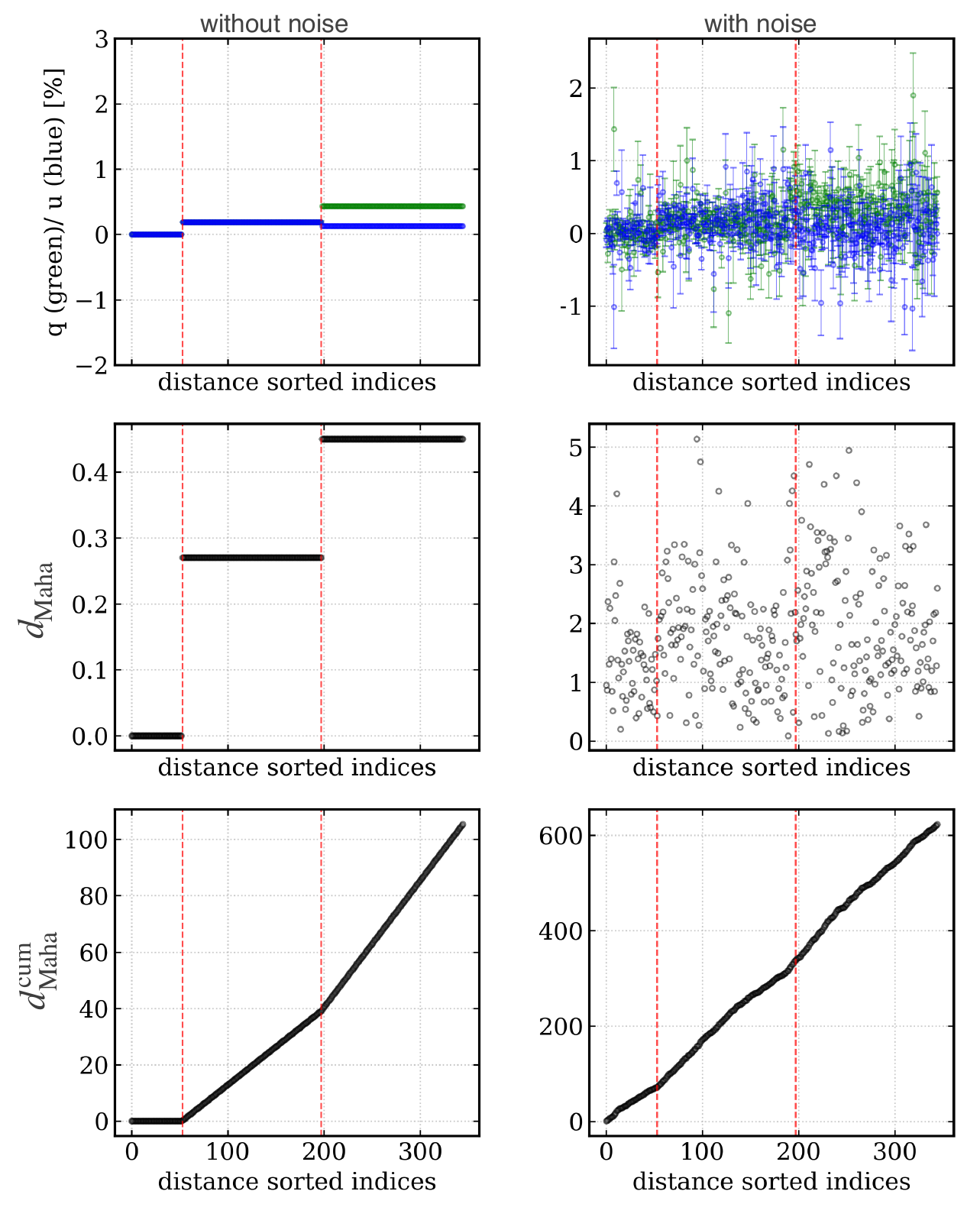}} 
     \caption{Simulated LOS with top-left panel showing $q$ (in green) and $u$ (in blue) as a function of distance sorted stellar indices without incorporating turbulence (internal scattering) and measurement uncertainties. The corresponding Mahalanobis distance ($d_{\rm{Maha}}$) and its cumulative effect ($d_{\rm{Maha}}^{\rm{cum}}$) are presented in the middle and bottom-left panels. The respective panel on the right side corresponds to the data, including turbulence-induced intrinsic scatter and measurement uncertainties. The red dashed lines in each panel mark the positions of the clouds.}
     \label{fig:Fig1}
\end{figure}

Early efforts to decompose starlight polarization as a function of distance were initiated by \citet{Andersson2005}, and \citet{Pavel2014}. 
Traditional approaches to tomographic decomposition of the magnetized ISM often rely on detecting sudden changes or jumps in polarization as a function of stellar distance \citep[e.g.,][]{Panopoulou2019, Doi2021b, Doi2024, Uppal2024}. An example of step changes in the polarization (Stokes parameters), when dust clouds uniformly polarize an ensemble of background stars, is shown in the top-left panel of Fig~\ref{fig:Fig1}.
However, in reality, various factors, including turbulence in the ISM, polarization S/N, fluctuations in dust density, and variations in polarization efficiency, collectively obscure the detection of these discrete jumps, making their identification challenging (similar to the top-right panel of  Fig~\ref{fig:Fig1}). 

\citet{Doi2024} employed the \textit{strucchange} package in R \citep{Zeileis2003} to detect structural changes in the Stokes parameters as a function of the stellar distance, enabling the estimation of dust cloud distances along the LOS. This approach is effective when observed polarization jumps are distinctly seen above the noise level. However, in low-polarization regimes, commonly observed at high Galactic latitudes, or in low polarization S/N, the accurate identification of such discrete jumps becomes significantly more difficult (for example, in the cases similar to Fig.~\ref{fig:Fig1}, top-right panel). To address this, our method introduces a novel approach: we track changes in the slope of the cumulative Mahalanobis distance computed from the Stokes parameters of stars sorted by distance along LOS. The Mahalanobis distance ($d_{\rm{Maha}}$) of Stokes parameters,  $S^{T} = [q, u]$ from point (0,0) on the polarization plane is given by
\begin{equation}
    d_{\rm{Maha}}(q,u) = \sqrt{S^{T}\Sigma^{-1}S} \, ,
\end{equation}
where, $\Sigma$ denotes the covariance matrix of $q$ and $u$ parameters of a star. For a set of $N$ stars with individual Stokes parameters $(q_i, u_i)$ along the LOS, the cumulative Mahalanobis distance up to the $j$-th star is computed as: 
\begin{equation}
    d_{\rm{Maha}}^{\rm{cum}} (j) = \sum_{i = 1}^{j} d_{\rm{Maha}}(q_i, u_i) \, .
\end{equation}

We make the following key choices in our method as compared to others found in the literature.
We use Mahalanobis distance to represent the polarization signal for two primary reasons. 
First, it combines the information from both Stokes parameters ($q$ and $u$) into a single statistically robust scalar, preserving sensitivity to variations in either component while avoiding the computational complexity of analyzing them separately. 
 Second, and more importantly, unlike the degree of polarization, which is positively biased in the low S/N regime \citep{DClarke}, the Mahalanobis distance provides an unbiased statistical measure by incorporating individual uncertainties and their covariance.
 This makes it a robust and reliable indicator of polarization changes along the LOS, especially when identifying subtle jumps caused by dust layers in noisy observational data.

Instead of using physical distances or parallaxes directly on the $X$-axis, we sort stars by distance and use their indices as a proxy for the position of stars along the LOS. This choice avoids artifacts arising from the non-uniform spacing of stars in distance (or parallax space), which can influence the estimation of slope or trend changes in Mahalanobis distance profiles. In addition, distance itself is not a directly measured quantity (derived from parallax or photometry), hence involving model-dependent assumptions and uncertainties. Alternative representations, such as using distance modulus or parallax on the $X$-axis, may complicate interpretation due to their non-linear relationship with physical distance. As part of our method development, we explored all these alternatives and found that using distance-sorted indices best serves our goal. This approach not only regularizes sampling along the LOS but also enables a more intuitive interpretation of cloud locations in terms of the relative number of stars located before and after a dust layer. The effect of the distance or parallax uncertainties on the distance sorted indices is discussed in Sect.~\ref{sec:2.3}.

To enhance the detectability of weak polarization jumps, particularly in noisy and low-polarization environments, we use the cumulative Mahalanobis distances rather than the direct values. The summation process acts as a low-pass filter, suppresses the high-frequency fluctuations arising from measurement errors and ISM turbulence, while amplifying the global trends associated with actual dust layers. This technique is widely used in signal processing of noisy data \citep[e.g.,][]{cumsum1993, cumsum2010, cumsum2022}. In the cumulative framework of \texttt{TRIShUL}, the presence of a dust layer is reflected not as an abrupt jump (which may be difficult to identify in noisy data), but as a distinct change in the slope of the profile.

The effect of cumulative Mahalanobis distance with indices over a low polarization signal is illustrated in Fig.~\ref{fig:Fig1}. We use a simulated dataset of starlight polarization with distance from a toy model introduced in \cite{Pelgrims2023}. In the model, each layer along LOS acts as a polarizing screen, inducing a mean polarization and scatter about it, which reflects the effects from turbulence-induced magnetic fluctuations in the beam. The simulated data contains two dust layers at distances $d$ = 350~pc and 900~pc along the LOS. A jump in the $q$, $u$, and Mahalanobis distance ($d_{\rm{Maha}}$) is seen at the location of the dust clouds in the top and middle-left panel of Fig.\ref{fig:Fig1}. The cumulative profile of this data shows a changing slope near the location of the dust clouds, as shown in the bottom-left panel of Fig.~\ref {fig:Fig1}). Incorporating Gaussian noise and scatter in the data around the mean polarization (as seen in the right panels of Fig.~\ref {fig:Fig1}), can make it challenging to directly identify the cloud from individual polarization measurements (top-right panel).  However, the cumulative Mahalanobis profile (bottom-right panel) has a visible changing slope near the actual dust layers. 

\subsection{Detecting breakpoints}\label{sec:2.2}
The second key step of \texttt{TRIShUL} involves the automated detection of breakpoints in the distance-sorted cumulative Mahalanobis distance profile, i.e., positions along the $X$-axis where the slope changes significantly. For this, we apply statistical change-point detection techniques using the \textit{strucchange} package in R \citep{Zeileis2003}. 
The cumulative Mahalanobis profile is modeled as piecewise linear functions of distance-sorted stellar indices, and \textit{strucchange} identifies locations where regression parameters of these models change significantly.
 We can control the analysis by specifying the minimum number of points per segment and the maximum number of breaks to test. The package implements the \cite{Bai2002} methodology, which uses dynamic programming and performs several statistical tests to efficiently detect multiple structural changes. 
To prevent over-fitting, the optimal number of breaks (best-fit solution) is selected using the Bayesian Information Criterion (BIC), ensuring a balance between model complexity and goodness-of-fit \citep{schwarz1978estimating}. The package also provides confidence intervals for each breakpoint, quantifying the associated statistical uncertainty. More details about the package can be found in \cite{Zeileis2002, Zeileis2003}.
 
\begin{figure}
   \centering
   \subfigure{\includegraphics[width=0.4\textwidth]{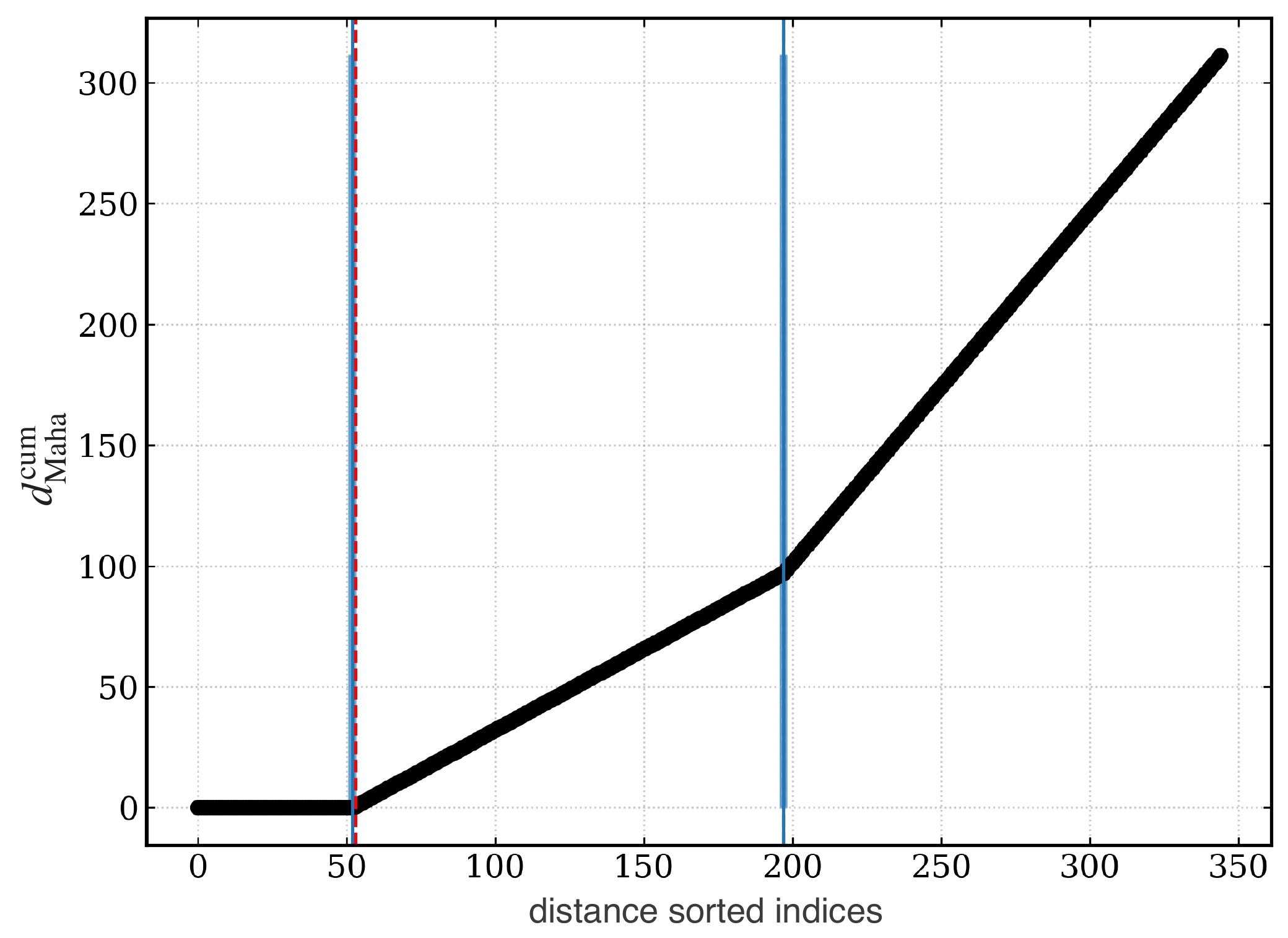}}
   
   \subfigure{\includegraphics[width=0.4\textwidth]{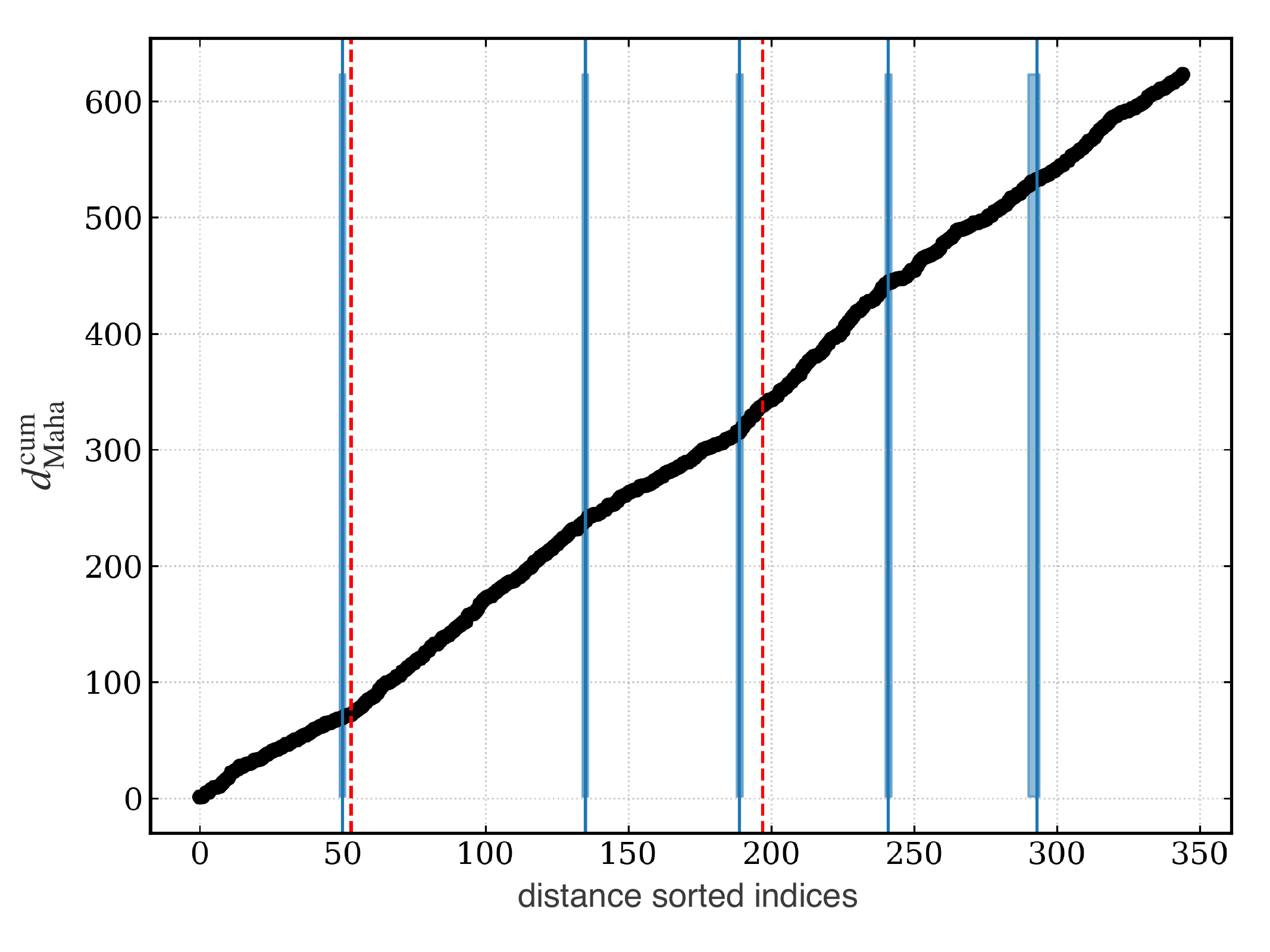}} 
      \caption{Breakpoints (blue lines) detected in the distance-sorted cumulative Mahalanobis distance profiles using the R \textit{strucchange} package. The top panel corresponds to the mock data shown in the bottom-left panel of Fig.~\ref{fig:Fig1}, while the bottom panel corresponds to the bottom-right panel of Fig.~\ref{fig:Fig1}. The vertical dashed red lines indicate the input distances of the dust layers used in the simulations.} \label{fig: Fig2}
   \end{figure}

An example of the application of the algorithm to the simulated data presented in Fig.~\ref{fig:Fig1} is shown in Fig.~\ref{fig: Fig2}, both with (bottom) and without (top) measurement uncertainties, respectively. 
The bottom panel of Fig.~\ref{fig: Fig2} shows that the breakpoint detection algorithm (marked by blue lines) effectively recovers the locations of the true dust layers (shown in red dashed lines) from the cumulative Mahalanobis distance profile but also identifies spurious breaks.  The false identifications likely arise from the accumulation of fluctuations (caused by both observational noise and ISM turbulence), which, when summed cumulatively, can mimic a genuine change of slope. To recover the correct number of dust clouds along the LOS, it is necessary to distinguish the true breaks from the spurious ones. This is done through a filtering procedure explained next, which forms the third and most critical aspect of our method.

\subsection{Rejecting spurious breaks}\label{sec:2.3}
Physically, a true dust layer should produce a significant change in the average Stokes parameters ($q$ and $u$) across its location, whereas spurious breaks are expected to show statistically similar distributions of ($q$, $u$) on either side. To implement this, we split the data at each detected breakpoint into segments and apply a Hotelling’s T-squared test \citep{Hotelling1992} to compare the ($q$, $u$) distributions on either side, using the \textit{multivariate\_ttest} function from the Python \textit{Pingouin} package \citep{pingouin2018}. This test requires a minimum of five stars in each segment.

The standard implementation of the test does not incorporate measurement uncertainties in $q$ and $u$ parameters, which are critical in our case.  We therefore reformulate the null hypothesis (i.e., the absence of a cloud) by modeling $(q,u)$ pairs from both segments as a bivariate normal distribution, with the mean given by the inverse-covariance weighted average and the covariance by the unbiased weighted sample covariance of the combined data. Measurement noise is incorporated by performing Monte Carlo simulations, i.e., synthetic datasets for each segment are drawn from the estimated bivariate distribution, with both intrinsic dispersion and individual measurement uncertainties of each star added to the covariance of the distribution.  The contribution of the intrinsic dispersion is estimated by subtracting the average contribution of measurement noise from the weighted covariance of the combined data. 
We reapply Hotelling’s $T^2$ test on each simulated segment under the null hypothesis to determine how frequently the test statistic exceeds the observed value. A break is considered statistically significant if fewer than a specific percentage (e.g., threshold = 1\%) of the simulations produced a test statistic more than the observed one.
This threshold is left as a user-defined parameter, as it depends on the level of uncertainties in the polarization measurements.

Furthermore, in our approach, we only use the absolute distance information in defining the indices, without considering their uncertainties. Consequently, stars near the breakpoints may be misassigned between the segments.
 These stars, which we refer to as ``jumpers'', can potentially change the weighted mean $q$ and $u$ of the segments, leading to spurious detections of the cloud layers. To mitigate this effect, we removed stars within $3\sigma$ of the breakpoint in parallax space, provided they constituted no more than 10\% of the stars in the corresponding segment.  This criterion strikes a balance between removing potential jumpers and retaining a sufficient number of stars to assess the statistical significance of each layer.
Hotelling’s T2 test is applied to this restricted sample to determine whether the break remained statistically significant under the null hypothesis. 

We applied these tests iteratively to each identified break along the LOS, sequentially redefining the segments based on the acceptance and/or rejection of the previous break. In this dynamic procedure, the segments may vary depending on the direction of analysis, i.e., forward: proceeding from minimum to maximum distance or backward: from maximum to minimum.  A true physical break is expected to show statistical significance regardless of the direction of analysis.  Breakpoints that are statistically significant in only one direction are flagged as uncertain (flagged as `1'), indicating potential inconsistency and the need for further scrutiny of the polarization data or adjustment of the hyperparameters of the \textit{strucchange} algorithm (see Appendix~\ref{A1} for more details on the nature of flagged cases). Valid reconstruction, consistent in both directions, are flagged as `0' (valid). 

\subsection{Post-processing}\label{sec:2.4}
The three-core steps of \texttt{TRIShUL} described above yield breakpoints in index space, along with associated $\sim 1\sigma$ confidence intervals. The next step involves translating these break positions (stellar indices) into physically meaningful parameters. This includes estimating the average parallax (proxy for distance) and mean Stokes parameters (proxy of degree of polarization and polarization angle, which indicate the POS magnetic field orientation) of the identified clouds along with their associated credible intervals.  
\subsubsection{Parallax estimation}\label{sec:2.4.1}
We map the confidence interval bounds from index space to parallax space by selecting the parallaxes of the stars corresponding to the lower and upper index bounds, rounded to the nearest integers using the floor and ceiling functions. The parallax of the star at the breakpoint index is taken as the representative of the distance of the dust layer, with its statistical uncertainty approximated as half the difference in parallaxes corresponding to the upper and lower bounds. Since the breakpoint method does not account for observational uncertainties in either polarization ($q$, $u$) or parallax, the uncertainties estimated above would be underestimated. To account for parallax measurement errors, we add in quadrature the median parallax uncertainty of stars within the confidence interval to the breakpoint-based parallax uncertainty.

In addition to the statistical uncertainties, we also account for potential systematic effects in the estimated breakpoints. For this, we segment the data around each statistically significant breakpoint. The left-side uncertainty is calculated using the weighted mean Stokes parameters ($q$, $u$) of the right-hand segment as a reference and incrementally includes stars from the left segment away from the break, recalculating the mean at each step. The process is stopped when the difference from the reference exceeds $2\sigma$ in either $q$ or $u$. The parallax of the last included star defined the left-side systematic boundary. We tested for $1\sigma$, $2\sigma$, and $3\sigma$ on the test samples discussed in Sect.~\ref{sec:3} and found that $2\sigma$ reasonably aligns with the true location of the layers.
The same procedure is applied in reverse to estimate the right-side boundary. These bounds provide a measure of systematic uncertainty in the location of the inferred dust layer.

To ensure that the identified dust layers are physically distinct, we retain only those layers that are separated by at least $2\sigma$ in parallax, where $\sigma$ is the quadrature sum of the maximum (statistical or systematic) uncertainties on the right side of the foreground break and the left side of the subsequent one. Adjacent layers with separations below this threshold are merged together. The location of the merged layer is calculated as a weighted mean parallax, using inverse maximum uncertainties as weights. Left and right uncertainties are propagated separately to preserve asymmetry. We also update the flagged status of the merged layer such that if both original layers had a flag = `0' or `1', the merged layer is assigned a flag = `0'; if one was flagged as `0' and the other as `1', the flag of the merged layer is designated as `01'.

\subsubsection{Estimating polarization properties}\label{sec:2.4.2}
To estimate the average polarization properties of the identified clouds, we consider that the Stokes parameters of the stars in the background of multiple clouds result from the cumulative contributions from individual clouds. This approximation is valid for LOSs in diffuse ISM where the dust clouds act as weak polarizers \citep{patat2010}.   
The Stokes parameters $q$ and $u$ for each dust layer are computed by subtracting the weighted average Stokes parameters of the stars in the foreground segment  (preceding the cloud) from those in the background segment (following the cloud) around the dust cloud location. In defining these segments, stars whose parallax (distances) fall within the uncertainty range (systematic or statistical) of the position of the cloud layer are excluded to avoid contamination from ``jumper'' stars that may not be part of the defined segment. The weighted means of the Stokes parameters are computed using the full covariance matrix, ensuring proper error propagation. To report the uncertainties in the estimated $q$ and $u$ parameters of the layers, we use the error on the weighted mean.

\subsection{Special Treatment of the Foreground Dust Layer}\label{sec:2.5}
Observational data sometimes exhibit non-zero Stokes parameters ($q$, $u$) of the very first stars, indicating the presence of a dust layer foreground to the observed sample. Additionally, the breakpoint detection algorithm may fail to identify a foreground break before which the mean $q$ and $u$ are expected to be statistically consistent with zero, primarily due to insufficient number of nearby stars to assess the statistical significance of the first layer. To ensure such early-LOS polarization is not overlooked, we must treat the foreground layer specially. 

In our method, we include an optional correction step that assigns an upper limit to the location of a potential foreground dust layer. 
To estimate this upper limit, we search for a foreground region in which the weighted average Stokes parameters $q$ and $u$ are consistent with zero within a $2\sigma$ confidence interval. This search is conducted starting from the location of the first cloud detected by the breakpoint algorithm and moving toward stars at lower distances, adding one star at a time. The index of the last star at which the accumulated group Stokes parameters become consistent with zero is designated as the boundary of the first dust layer. If no such boundary is found, we assign the position of the first star as the upper limit on the cloud distance. The statistical upper limit on index-based uncertainty associated with this boundary is approximated using the next star beyond the identified index for the layer. These index-based estimates are then mapped to parallax as discussed in Sect.~\ref{sec:2.4.1}, but only positive bounds (upper limits) should be reported in such cases. In this special case, the positive bounds indicate the positive uncertainties in the upper limit on the distance of the first layer.

\section{Performance testing}\label{sec:3}
The performance of the method is expected to be influenced by the amplitude of the polarization signal induced by the LOS cloud, the number of stars effectively sampling a given cloud, the S/N ratio of the polarization measurements, the degree of intrinsic scatter, and the number of clouds along a LOS. 
To test these effects, we use mock data representing typical one- and two-cloud LOS scenarios at intermediate and high Galactic latitudes that will be targeted in \textsc{Pasiphae}. The polarization induced by the clouds to the background stars depends on polarization efficiency (maximum polarization induced to the stars per unit reddening), and apparent 3D orientation of the magnetic field ($B$) in the cloud, which is described by the inclination angle of magnetic field line with respect to POS ($\gamma_{B_{\rm{reg}}}$) and position angle of POS component of magnetic field ($\psi_{B_{\rm{reg}}}$) \citep{Andersson2015, Henseley2021}.  The mock data for benchmarking the method is generated using a toy model described in \cite{Pelgrims2023}, where the total 3D magnetic field is modelled as a sum of regular ($B_\text{reg}$) and stochastic component ($B_\text{sto}$) as $B_\text{tot} = B_\text{reg} + A_\text{turb}\; B_\text{sto}$. The parameter $A_\text{turb}$ measures the amplitude of the stochastic component with respect to the regular one and takes care of the expected inhomogeneities in the polarization source field. Further details about the toy model and mock samples can be found in Sect.~4.1 and Appendices~A3 and~A4 of \cite{Pelgrims2023}.

Our simulations assume a $1^\circ$ beam containing $\sim345$ stars along the LOS. A fixed intrinsic scatter of $A_{\text{turb}} = 0.2$ is adopted, and we vary the distance(s) of the cloud layer(s) to vary the fraction of stars lying in the background of the cloud(s). For each configuration, we generate multiple realizations by varying the POS orientation of the regular magnetic field and the random seed of the stochastic magnetic field. These realizations form the basis for testing our method to identify the number of clouds along the LOS and to recover their parallaxes and polarization properties.
To interpret the results, we categorize the outcomes of our method into four categories:
(i) False negative (FN), where no clouds are detected or a cloud is missed in two-cloud cases, indicating method failure; 
(ii) False positives (FP), where more clouds are identified than the number of clouds input to the mock data;
(iii) Flagged or uncertain cases as discusses in Sect.~\ref{sec:2.3}, and Sect.~\ref{sec:2.4.1};
(iv) Valid reconstructions, where the number of identified clouds matches the input layers in mock data.

\subsection{One cloud cases}\label{sec:3.1}
In the single cloud setup, only one cloud is placed along the LOS.  The distance of the layer is varied such that fraction of stars located behind the cloud ($f_{\rm{bg}}$) is $90\%$, $70\%$, $50\%$, $30\%$, and $10\%$ corresponding to the distances of approximately 270, 565, 790, 1050, and 1712~pc, respectively. The inclination of the regular component of magnetic field with respect to the POS is fixed at $\gamma_{B_{\rm{reg}}} = 0^\circ$. To test the performance of the method in the low polarization regime, we varied $p_{\rm{max}}$ from $0.05\%$ to $0.3\%$ with a step size of $0.05\%$ and also considered higher values of $0.5\%$, $0.75\%$, and $1.0\%$. We further consider 10 realizations of each pair of $p_{\rm{max}}$ and $f_{\rm{bg}}$, varying the POS position angle of $B_{\rm{reg}}$. All these combinations led to 450 simulated samples of starlight polarization data, where we apply our method to identify the number of clouds, estimate their parallax, and polarization properties. 

\begin{figure}{\includegraphics[width=0.45\textwidth]{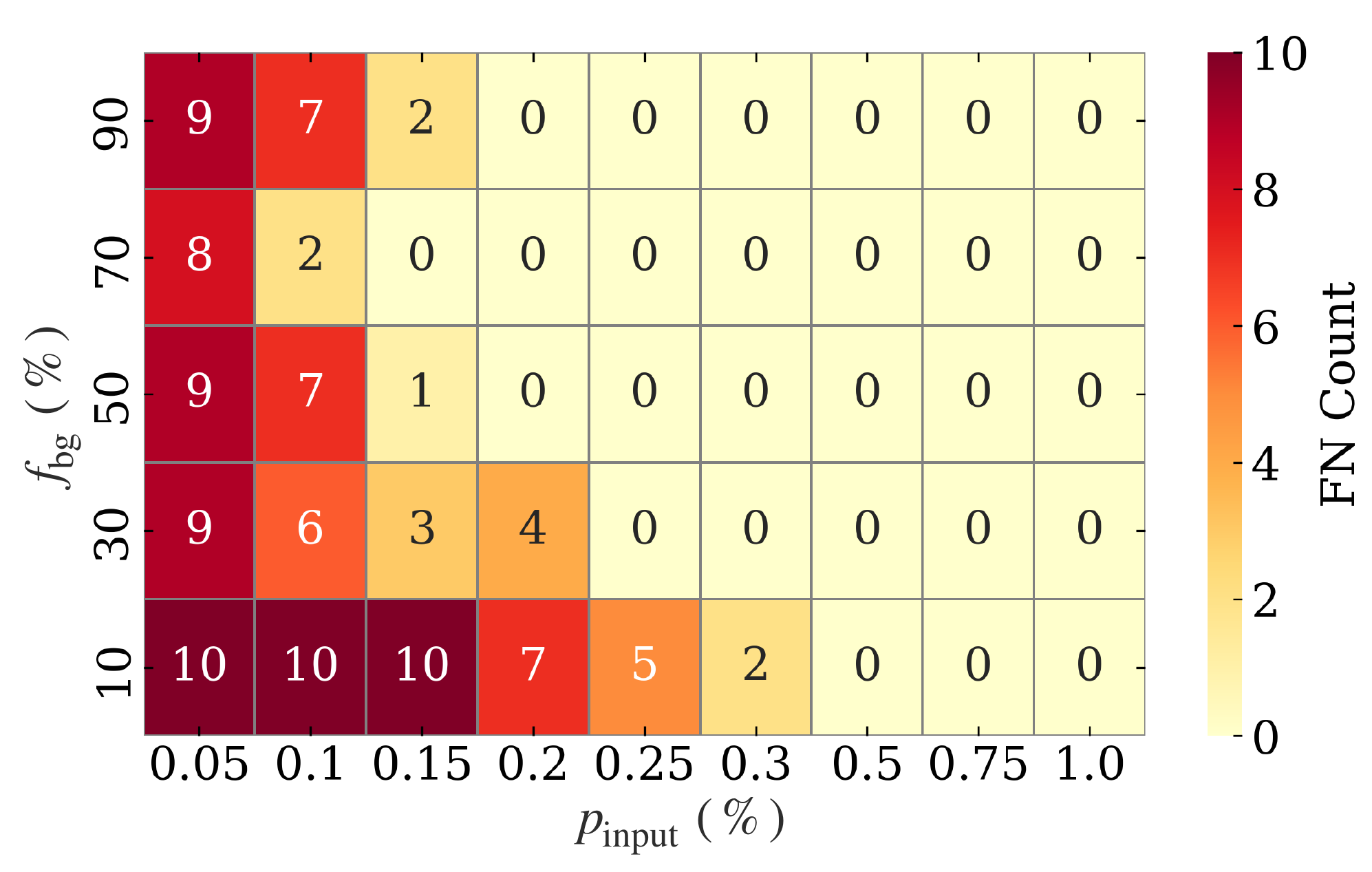}}
      \caption{Heat map illustrating the number of false negative cases out of 450 single-cloud mock samples, each with a combination of $p_{\rm{input}}$ and $f_{\rm{bg}}$, evaluated across 10 samples of $\psi$ ranging from $0^\circ$ to $180^\circ$. The color intensity and the number within each cell represent the resulting count of FNs.}\label{fig:Fig3}
   \end{figure}

We impose a fixed minimum cut on the number of stars per segment to be five (set by Hotelling's $T^2$ statistical test). Each sample is tested for one to five layers and uses the minimum BIC to find the optimal solution. The null hypothesis for statistically significant layers (see Sect. \ref{sec:2.3}) is rejected at a 99\% confidence level. Out of 450 simulated cases, our method successfully recovers the cloud layers in the majority of the cases. However, a subset of $\sim 25$\% of cases results in FNs where no cloud layer is identified (case of failure). This does not imply a general failure of the method in 25\% of cases; rather, these outcomes are closely linked to the effect of the degree of polarization induced by the cloud and the fraction of stars located behind it. Figure~\ref{fig:Fig3} shows the number of FN cases as a function of the input degree of polarization ($p_{\rm{max}}$) and the background star fraction ($f_{\rm{bg}}$). For $p_{\rm{max}} \le 0.1\%$, the polarization signal becomes too weak to be detected, leading to a higher FN rate. False negatives also extend towards the higher degree of polarization, but only when the background star fraction becomes as low as 10\% ($\sim 34$ stars). This is due to an insufficient number of background stars to confidently detect the slope change in the Mahalanobis distance.

Only $\sim 3\%$ of the test samples resulted in FPs, where our method identifies two layers instead of one.  In most of these cases, one layer corresponds to the true physical cloud, while the other shows polarization consistent with zero within $2\sigma$. This suggests that the latter does not represent a physically distinct dust component and should not be treated as a separate cloud. These spurious layers can be removed, and the polarization properties of the remaining clouds should then be recalculated. In addition to FN and FP, only $\sim 2\%$ of the cases are classified as uncertain or flagged. These correspond to sightlines where at least one detected cloud is flagged due to ambiguity in the decomposition.  Uncertain cases do not indicate a complete failure as the method still identifies dust layers, but these cases indicate that the LOS dust structure is complex or unclear, making the interpretation of the decomposition less straightforward. Therefore, the results from flagged sightlines should be used with caution (see Appendix~\ref{A1} for more details).  
Heat maps illustrating FP and flagged cases, similar to the false negatives, are shown in Appendix~\ref{A3}. Unlike FN cases, these do not show any significant dependence on $p_{\text{max}}$ or $f_{\text{bg}}$. The remaining majority of the cases yield a single cloud with a flag 0, indicating a reliable and statistically significant result. These cases are summarized in the bottom panel of Fig.~\ref{fig:A3}.

To assess the accuracy of the calculated parallaxes and Stokes parameters of the valid identified dust layers, we avoid using the input values employed in generating the synthetic catalogs. Instead, we segment the dataset at the input cloud distances and estimate the mean and covariance of the polarization signal attributed to the cloud by analyzing the Stokes parameters of stars located behind it. These values, referred to as the true polarization properties ($\hat{\bar{q}}_\text{true}, \hat{\bar{u}}_\text{true}$), serve as an empirical benchmark. For parallax, we adopt the parallax of the star closest to the input cloud distance, denoted as $\varpi_{\rm{true}}$. 

\begin{figure}
   \centering
   \subfigure{\includegraphics[width=0.4\textwidth]{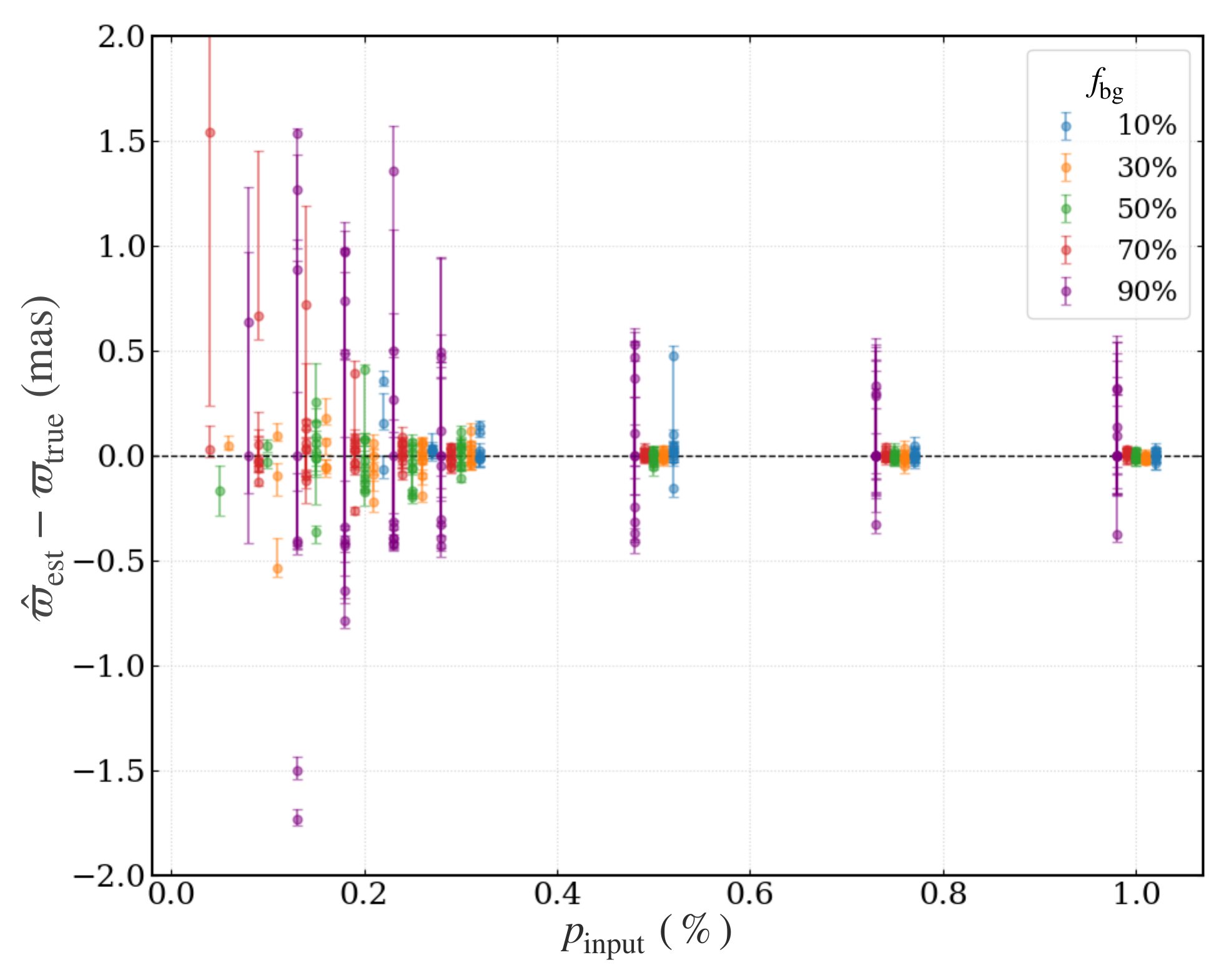}} 
   \subfigure{\includegraphics[width=0.4\textwidth]{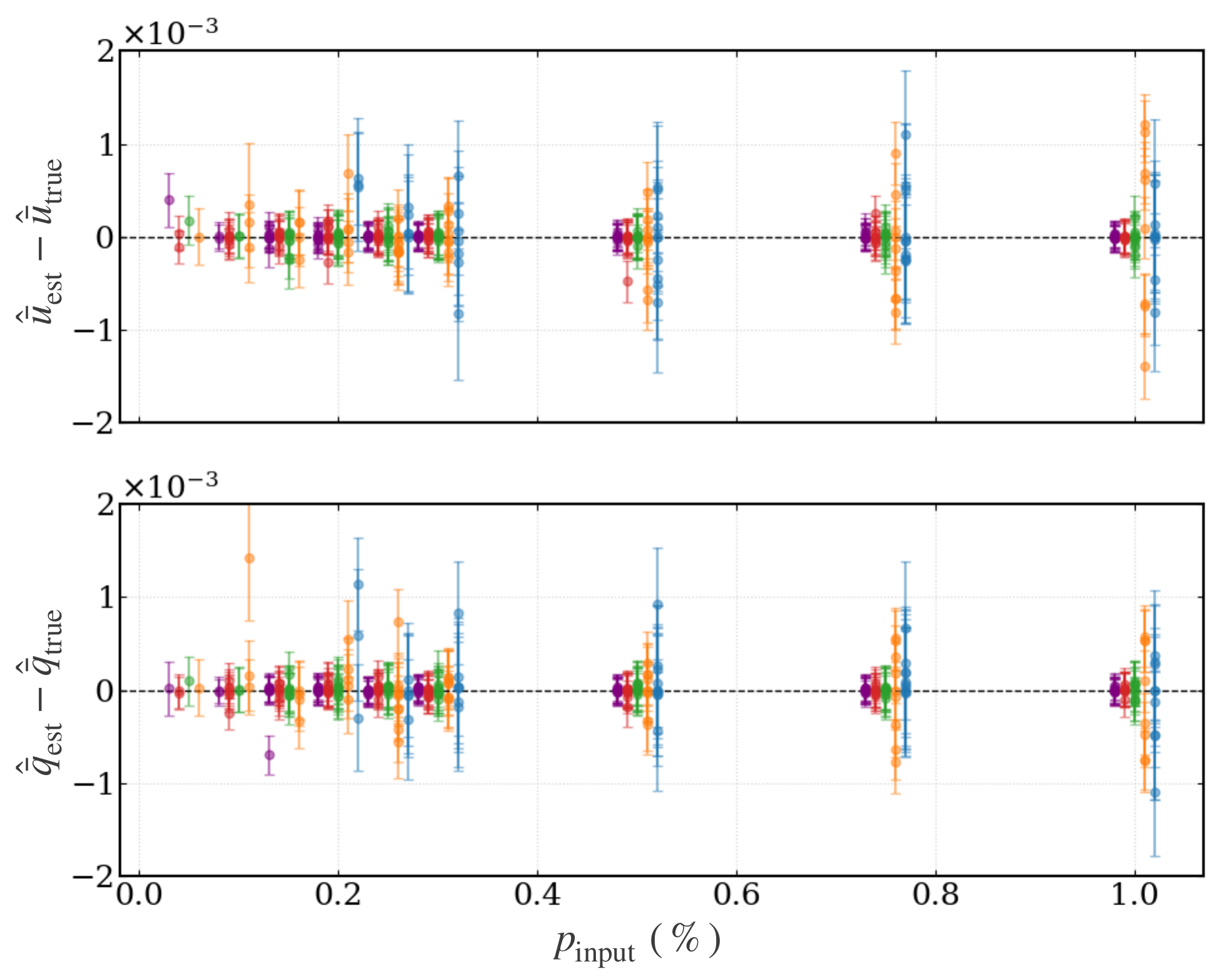}}
   \caption{Performance of \texttt{TRIShUL} as a function of input polarization signal for different cloud distances probed by $f_{\rm{bg}}$ (in different colors). The top panel shows the difference between the estimated parallax ($\varpi_{\rm{est}}$) and the parallax of the star closest to the input layer ($\varpi_{\rm{true}}$). The bottom panel corresponds to the difference in estimated Stokes parameters ($\hat{\bar{q}}_{\rm{est}}$, $\hat{\bar{u}}_{\rm{est}}$) from the corresponding true values ($\hat{\bar{q}}_{\rm{true}}$, $\hat{\bar{u}}_{\rm{true}}$). The results corresponding to different $f_{bg}$ are shifted by 0.02\% on the $X$-axis for visual clarity. Dashed horizontal lines are the reference to a perfect match between estimated and true parameters.} \label{fig:Fig4}
   \end{figure}

We plot the differences between the estimated (est) and true values (true) of both parallax (top panel) and Stokes parameters (lower panels) for the identified cloud layers in Fig.~\ref{fig:Fig4}.
This figure illustrates the performance of the valid reconstructions across varying input degree of polarization ($p_{\rm{input}}$ on the $X$-axis), cloud distances ($f_{\rm{bg}}$, indicated by different colors), and polarization angles (number of points per $p_{\rm{input}}$ and $f_{\rm{bg}}$). In the top panel, the error bars on parallax represent the larger of the statistical or systematic uncertainties. The error bars in the bottom panel reflect the uncertainties in the estimated Stokes parameters, as described in Sect.~\ref{sec:2.4.2}. A horizontal offset of 0.02\% is applied to each $f_{\rm{bg}}$ group to prevent overlap of data points clustered around the same $p_{\rm{input}}$ values. The black horizontal line in all the panels represents perfect agreement between estimated and true values. In both panels, all data points seem to cluster around zero on the $Y$-axis, indicating strong agreement and high accuracy in the recovered values except at low polarizations, where deviations become more pronounced. The precision of the estimated parallax (top panel) deteriorates at $f_{\rm{bg}} = 90\%$ (purple points), as indicated by the larger uncertainties. This is due to the cumulative nature of the Mahalanobis distance we used in our method. With only $\sim 34$ stars (10\% of the stars) located in the foreground, the cumulative profile builds up slowly from near-zero values, causing a shift in the location of the detected slope change and leading to less precise parallax estimates. Nevertheless, in most cases, the discrepancies remain close to zero within the error bars. Overall, in 90\% of the reliable cases, the estimated parallaxes agree with the true layer parallaxes within $2\sigma$ uncertainties. An opposite trend is observed in the lower panels for the $q$ and $u$ estimates, where low precision occurs at $f_{\rm{bg}} = 10\%$ (i.e., when only 10\% of stars are present behind the cloud layer). This is also caused by low number statistics as the average $q$ and $u$ are derived from the background stars. A smaller sample size, especially with higher measurement uncertainties at the fainter end, compromises the precision of the weighted mean calculation. Nevertheless, the estimated $q$ and $u$ values agree with the true values within combined $2\sigma$ intervals in $99.5\%$ and $90.5\%$ of the cases, respectively.

We also investigate the effect of the inclination angle of the magnetic field with the POS on the performance of our method and find no observable trend. The detailed analysis is provided in Appendix~\ref{A2}. 

\subsection{Two cloud cases}\label{sec:3.2}
To test our method for cases with two clouds along the LOS, we use the samples described in Sect.~4.4 of \cite{Pelgrims2023} to test our method. 
The first cloud is fixed at 350 pc, representative of the Local Bubble boundary at high Galactic latitude \citep[e.g.,][]{Skalidis2019, Pelgrims2020}. This distance corresponds to $\sim55$ stars located foreground the cloud for $1^\circ$ beam-size (345 stars). 
The inclination angle of the magnetic field for the nearby cloud is fixed at $30^\circ$, and $p_{\rm{max}}$ is fixed, so that $p_C = 0.2\%$. The polarization angle for the nearby cloud is set at $22.5^\circ$. According to the discussion in Sect.~\ref{sec:3.1}, any cloud with such properties is expected to be well recovered by our method.

The performance of our method on 2-cloud cases is expected to depend on the distance between the two layers, as they alter the number of foreground and background stars to the clouds, relative polarization amplitude, and possibly on the difference in the orientation of the magnetic field.
Hence, the second cloud layer is added along the LOS with varying distance such that $f_{\rm{bg2}} = 90\%$, $70\%$, $50\%$, $30\%$, and $10\%$ of the stars in the background of the nearby cloud are also background of the more distant cloud. The corresponding distances of the second cloud would approximately be 520, 685, 900, 1150, 1800~pc, respectively. The polarization of the second cloud ($p_{\rm{input}}^{(2)}$) takes values of 0.05\%, 0.10\%, 0.15\%, 0.20\%, 0.25\%, 0.50\%, 0.75\%, and 1.00\% and relative position angle of the magnetic field between the two clouds ($\Delta \psi$) in range $[0^\circ, 180^\circ]$ with steps of $30^\circ$. The inclination angle of the magnetic field of the second cloud is fixed to zero, and $A_{\rm{turb}}$ to $0.2$. Five random realizations are used for each set, thus constructing 1200 simulated samples for the magnetized ISM structures along the LOS to test our method. 

Similar to the one-cloud case, we applied our method to test one to five layers, keeping the minimum number of stars per segment fixed to five in the two-cloud scenario as well. Out of 1200 cases, only 8 cases show complete failure in detecting both the clouds, while in 38\% of the cases, only one of the two layers are identified. The combined statistics of FN cases as a function of input parameters, $p_{\rm{input}}^{(2)}$, $\Delta \psi$, and $f_{\rm{bg2}}$ are shown in Fig.~\ref{fig:FigA4} of Appendix~\ref{A4}. 
Consistent with findings from the previous section, the results from 2-cloud samples also demonstrate that clouds with low polarization signals ($\le 0.1\%$) or located at large distances ($f_{\rm{bg2}} = 10\%$) are more challenging to recover. The rate of failure is notably higher in the two-cloud scenario as compared to the single-cloud case, especially for $f_{\rm{bg2}} = 90\%$, i.e, when the two layers are close to each other. Accurate decomposition in the two-cloud configuration in these cases becomes particularly challenging because there are too few stars between the layers or an insufficient number of background stars behind the more distant cloud. 
In such cases, the algorithm often detects only one cloud, failing to resolve the second layer. For example, the cases when $p_{C1} > p_{C2} $, the identified layer corresponds to the nearby one. However, as $p_{C2}$  exceeds $p_{C1}$, the decomposition results in the identification of the distant layer only. In 57\% of the cases, we are able to identify both clouds, but $\sim 2\%$ of these cases have one of the layers flagged as uncertain.  In addition to FN and flagged cases, $\sim 4\%$ of the reconstructions exhibit FP results.  As discussed in the single-cloud case, the number of FPs and uncertain cases in the 2-cloud case can also be decreased when layers with polarization consistent with zero are excluded. 
\begin{figure}
   \centering
   \subfigure{\includegraphics[width=0.4\textwidth]{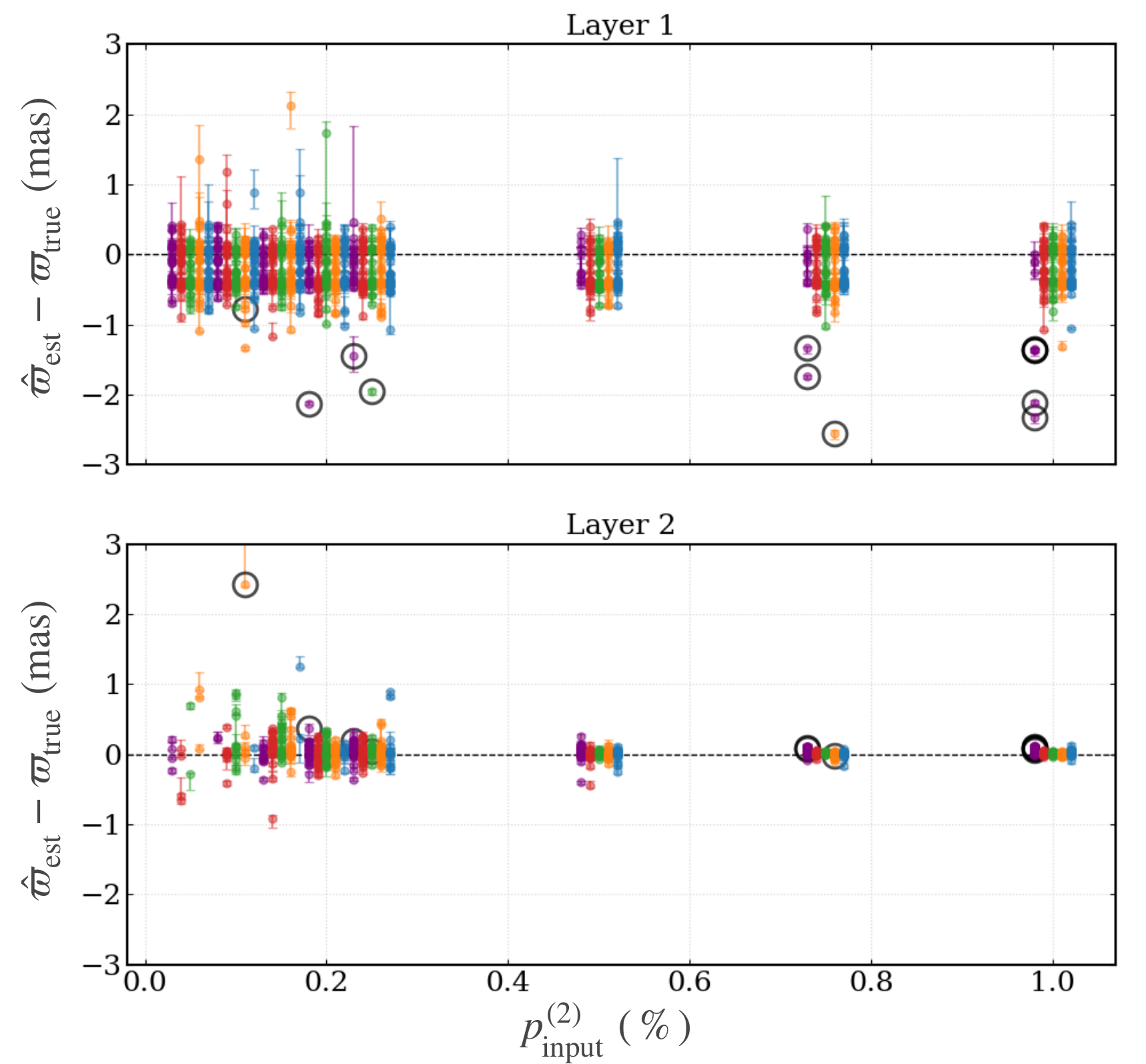}}
   
   \subfigure{\includegraphics[width=0.45\textwidth]{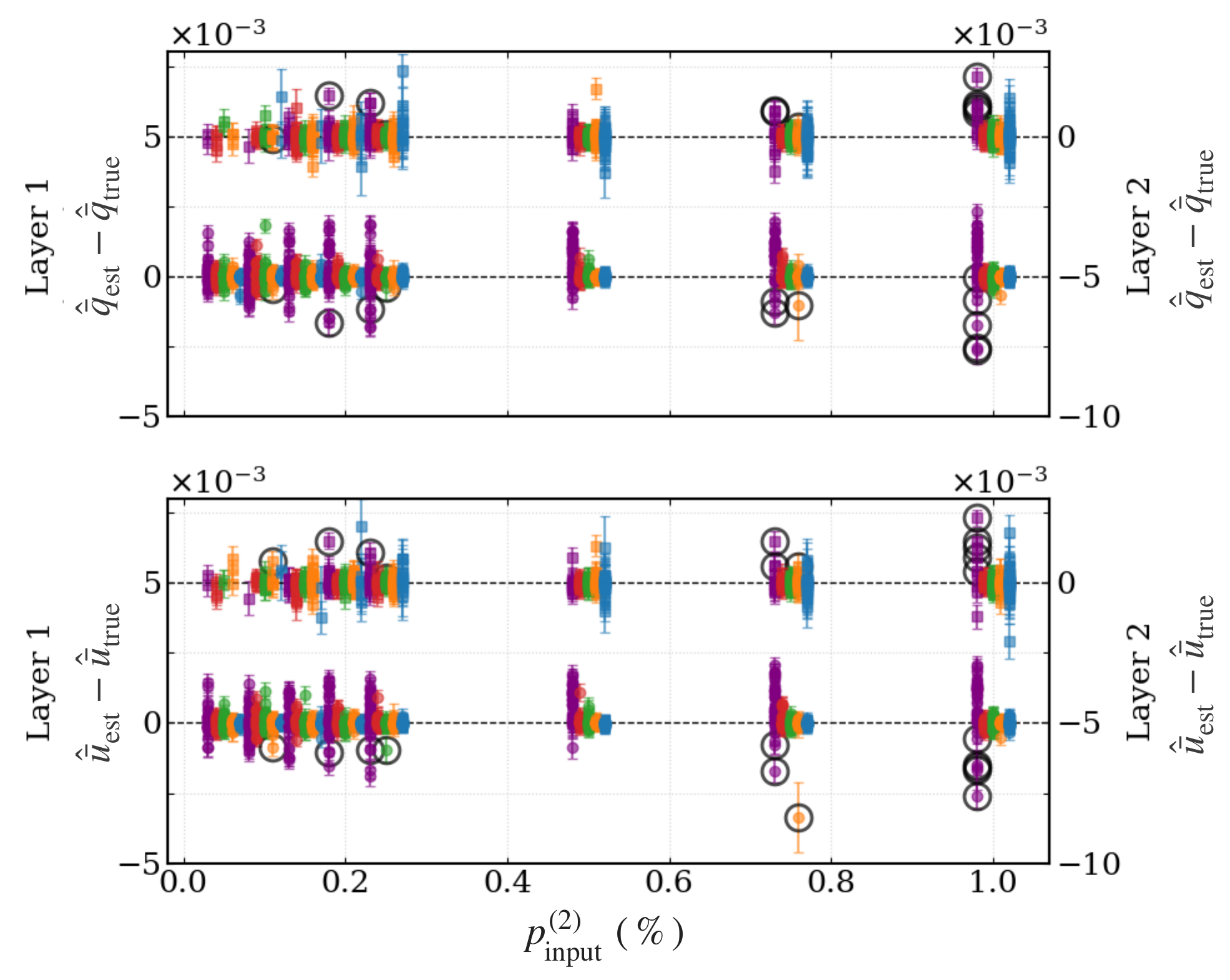}}
   \caption{Performance testing of our method in recovering parallax (top panel) and Stokes parameters (bottom panel) of both layers in 2-cloud sample tests. The color scheme is defined for different set of $f_{\rm{bg2}}$ and is similar to $f_{\rm{bg}}$ in Fig.~\ref{fig:Fig4}. The horizontal dashed lines correspond to a perfect match between estimated and true values. Open black circles indicate cases where an additional, spurious layer beyond the second layer is identified, while the first layer is missed. More details about the figure are deferred to the text in Sect.~\ref{sec:3.2}  } \label{fig:Fig5}
   \end{figure}

For the valid reconstructions, excluding the flagged and FP cases, the accuracy of our method in estimating both the parallax and polarization properties of the two clouds is illustrated in Fig.~\ref{fig:Fig5}. The two-panels at the top show the parallax accuracy for each layer, represented as the difference between estimated and true values. Different colors represent varying values of $f_{\rm{bg2}}$, following the same color scheme and offset as of $f_{\rm{bg}}$ in Fig.~\ref{fig:Fig4}. The data points in both the sub-panels cluster closely around zero, indicating strong agreement between the estimated and true parallaxes. 
The upper sub-panel (Layer 1) shows a larger number of points at low $p_{\rm{input}}^{(2)}$ compared to Layer~2 (bottom sub-panel). This is primarily coming from FN cases, where only the nearby cloud is detected for $p_{C2} < p_{C1}$ as explained earlier. Even when both the layers are detected, the estimated parallax of layer~2 at low input polarization ($p_{\rm{input}}^{(2)}< 0.1\%$) deviates more than $3\sigma$ from the true value. This is because weak input signals make the second slope change harder to identify compared to Layer~1, where input-polarization is fixed at a relatively higher value (= 0.2\%). Moreover, for low polarization, restricting the maximum number of breaks to five in identifying structural changes in Mahalanobis distance may be limiting, especially when multiple small variations arising from physical structure and the measurement noise have comparable amplitudes. 
In addition, the parallax estimates for Layer~1 show larger uncertainties and scatter compared to Layer~2. As explained in Sect.~\ref{sec:3.1}, the cumulative profile builds up gradually from near-zero values, making the first slope change harder to identify. Once the profile is established, subsequent layers exhibit a clearer slope change, leading to more precise and accurate identification of Layer~2. The open black circles in Fig.~\ref{fig:Fig5} represent cases in which the algorithm correctly identifies the second cloud but detects a spurious, distant layer in place of the nearby one. These points are not associated with the nearby cloud but are shown to highlight false detections. Among the valid reconstructions, the estimated parallaxes match the true values within $2\sigma$ in 78\% of the cases for the layer~1 and 75\% for the layer~2.

The accuracy of the Stokes $q$ and $u$ parameters for the test samples is shown in the lower two panels of Fig.~\ref{fig:Fig5}. Circular markers represent Layer~1 and square markers are used for Layer~2. To separate the visual representation of both layers, the left $Y$-axis corresponds to Layer~1 while the right $Y$-axis to Layer~2.
The tight clustering of points around zero in both sub-panels indicates good agreement between estimated and true values. The purple points (corresponding to $f_{\rm{bg2}} = 90\%$), however, show noticeably larger scatter and uncertainties, especially for Layer~1. This excess scatter is driven by FN cases where one of the two layers is missed, largely because only $\sim 30$ stars lie between the layers. Since the Stokes parameters $q$ and $u$ are derived after segmenting the data by identified cloud layers (as described in Sect.~\ref{sec:2.4.1}), missing a layer leads to an incomplete decomposition and consequently biased polarization estimates.
 The blue points, corresponding to $f_{\rm{bg2}} = 10\%$, also exhibit larger uncertainties in the $q$ and $u$ differences, particularly for Layer~2. This is primarily due to the limited number of background stars behind the more distant cloud, which reduces the precision of the Stokes parameter estimates, similar to one cloud cases in Sect.~\ref{sec:3.1}. The plotted uncertainties represent the quadrature sum of the uncertainties in both the estimated and true Stokes parameters. In around 89\% of the cases, the estimated $q$ and $u$ agree well with the true values within 2$\sigma$ uncertainties. 

It is worth noting that many of the cases where the estimated parallax and polarization properties deviate from the true values, stem from inaccuracies in the locations of the breakpoints identified by the \textit{strucchange} algorithm (see Appendix~\ref{A1} for more details). When a true breakpoint corresponding to a cloud location is missed in the initial detection step, the error propagates through the subsequent analysis, ultimately causing departures from the true cloud properties.
 We are using a maximum of five breakpoints in our current implementation for performance testing, and the final output is selected by optimizing solutions over one to five breaks. However, increasing the maximum number of breaks in \textit{strucchange} could potentially improve the recovery of slope changes closer to the true locations of the clouds and thereby reduce the number of flagged cases, and also enhance the precision and accuracy of the method. However, it may increase FP rates and also increase computational time, as more spurious breaks would need to be filtered out. Nevertheless, the current results are satisfactory, with most estimates lying within $3\sigma$ of the true parallax. 

\section{Comparison with \texttt{BISP-1}}\label{sec:4}
We compare the performance of \texttt{TRIShUL} with the LOS inversion method developed with a Bayesian framework (\texttt{BISP-1}) by \cite{Pelgrims2023}. This method models between one and five thin dust layers along the LOS, each described by six parameters: $(q, u, \varpi, C_q, C_u, C_{qu})$. The algorithm employs the nested sampling implemented via \texttt{dynesty} code \citep{Speagle2020}, where it uses the sampling points (called live points) to explore the parameter space and estimate the posterior distribution on model parameters. The results presented in Sect.~4.3 and Sect.~4.4 of their paper show the outcome of the \texttt{BISP-1} algorithm using a one-layer model for single cloud samples and a two-layer model for two cloud samples. They also used prior information on cloud parallaxes to analyze the two cloud samples. For a fair comparison with \texttt{TRIShUL}, we reapply the \texttt{BISP-1} algorithm on mock samples, testing one to five-layer models, without imposing any external priors on cloud parallax. In our re-analyses, we use flat priors on Stokes parameters on $q$ and $u$ within $\pm 2\%$, and explore the parameter space by considering 1000 live points and sampling until we reach a total tolerance in the estimated log-evidence \citep[check Eq.~15 of][]{Pelgrims2023} below 0.1. All the pathological cases, where the parallax posterior peaks at the prior boundaries (indicating prior dominating solutions), are excluded, and the optimal model is selected using the minimum Akaike Information Criterion (AIC) criterion \citep{AIC1974}. The AIC criteria measure of the amount of information that is lost by representing the data with a given model.

For consistency, we allow upto five layers in \texttt{TRIShUL} with a minimum of five stars per segment, and used 99\% confidence level to reject the null hypothesis for valid breakpoints. We limit our comparison to one set of single- and two-layer cases in the low-polarization regime from Sect.~4.5 of the \texttt{BISP-1} paper.

\subsection{Comparison of one cloud mock sample}\label{sec:4.1}
We consider the mock LOSs containing a single dust cloud located at 350~pc, having polarization $p_C^{\rm{true}} = 0.2\%$, magnetic field inclination angle $\gamma_{B_{\text{reg}}} = 30^\circ$, and intrinsic scatter $A_{\text{turb}} = 0.2$. The POS position angle of the magnetic field is varied from $0^\circ$ to $162^\circ$ in increments of $18^\circ$, and for each configuration, we generate 10 random realizations, resulting in a total of 100 samples.   Among the 100 samples, \texttt{TRIShUL} consistently identifies a single layer, with two FPs but no FNs, and one reconstruction is flagged as uncertain.

For \texttt{BISP-1}, we adopted the default prior on parallax set by the five-layer model and inspected the resulting posteriors across all models. 
In nearly all cases, \texttt{BISP-1} favored a one-layer model, with only two exceptions where a 2-layer model is preferred based on AIC. However, we note that for these two cases, the probability that the 1-layer model minimizes the loss of information based on the models' AIC is above 30\%. Therefore, the 1-layer model cannot be rejected, and we treat these cases as equivalent to uncertain or flagged cases of \texttt{TRIShUL}. 

\begin{figure}
   \centering
   \subfigure{\includegraphics[width=0.4\textwidth]{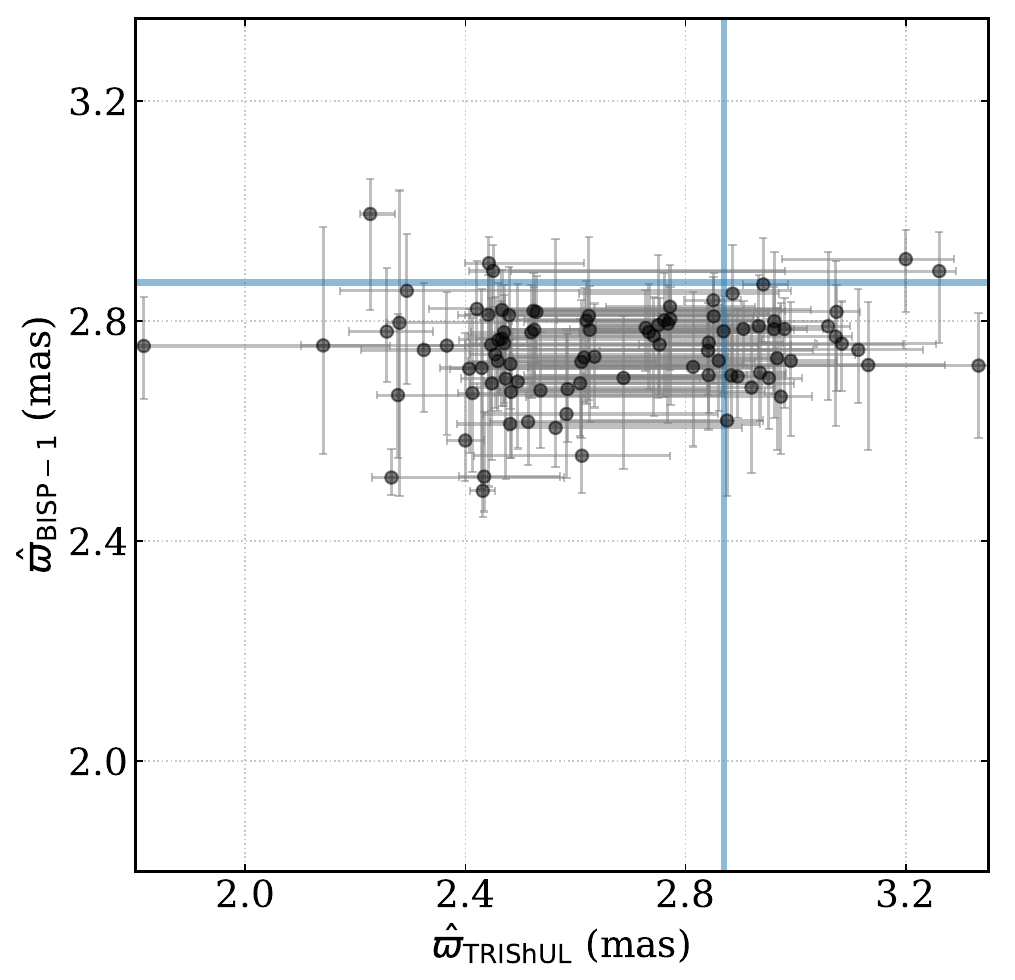}}
   
   \subfigure{\includegraphics[width=0.4\textwidth]{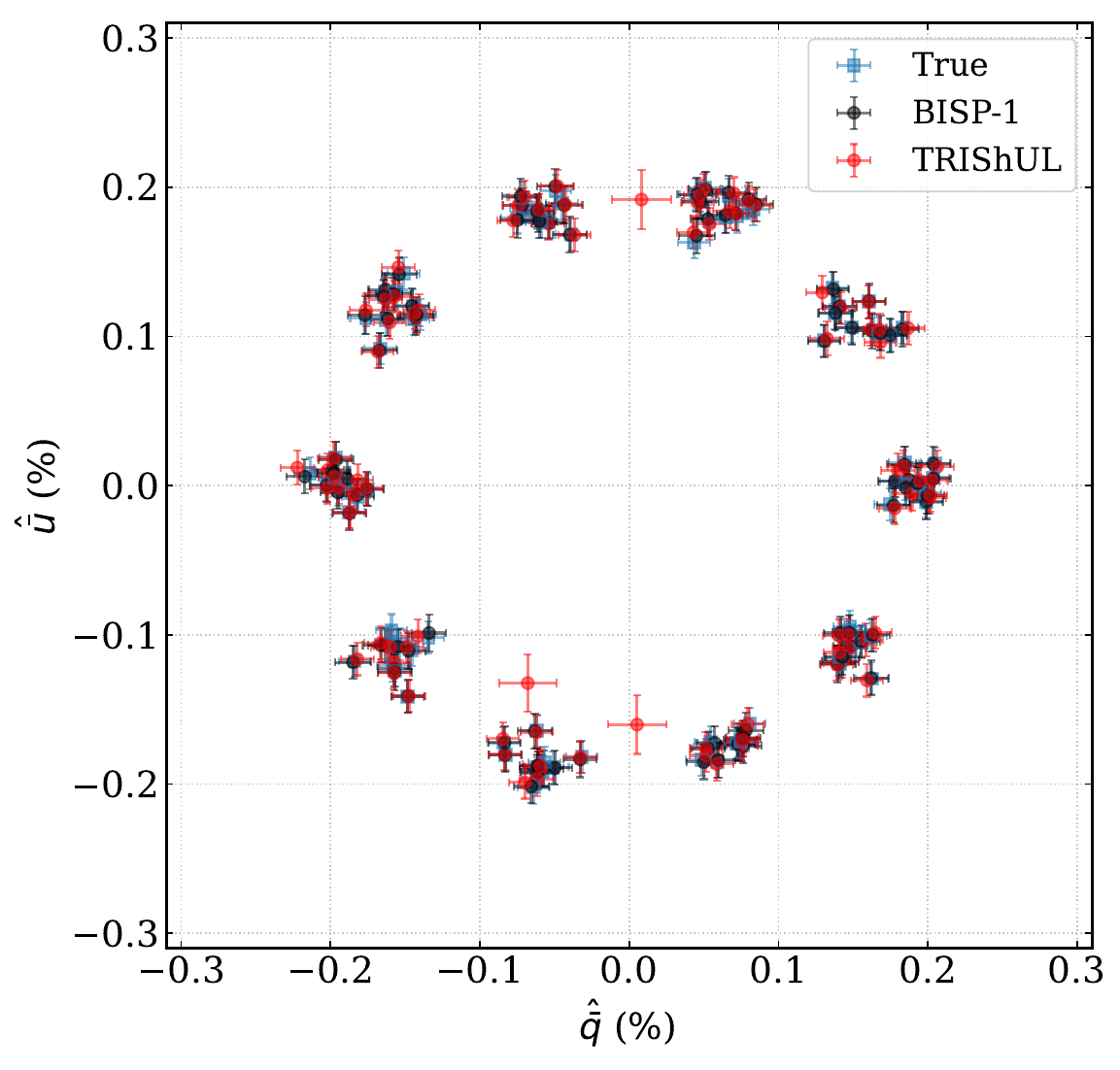}}
      \caption{Comparison of estimated parameters from \texttt{BISP-1} (black) and \texttt{TRIShUL} (red) with the true values (blue) for parallax (top panel) and Stokes parameters (bottom panel) in one-cloud mock catalogs.} \label{fig:Fig6}
   \end{figure}
To compare the accuracy of the two methods, we exclude reconstructions with FPs and flagged cases. Figure~\ref{fig:Fig6} presents a comparison of the retrieved parallaxes of the LOS cloud from both methods. We plot the mode of the posterior parallax for \texttt{BISP-1} with its $16\rm{th}$ and $84\rm{th}$ percentile as asymmetric uncertainties on the $Y$-axis, against the estimate of \texttt{TRIShUL} on the $X$-axis. Uncertainties in estimates from \texttt{TRIShUL} are defined by the larger of the statistical and systematic errors on each side. 
To visualize the deviation of the estimated values from the true parallax, blue horizontal and vertical lines are plotted in the figure. 
In most cases, both methods yield parallax estimates that are systematically lower than the true values. This is due to the low number of stars in the foreground of the layer, as explained in Sect.~\ref{sec:3.1} and Sect.~\ref{sec:3.2}. It is evident from the figure that the uncertainty estimates in  \texttt{TRIShUL} are larger than those of \texttt{BISP-1}. Due to larger uncertainties, \texttt{TRIShUL} captures the true value within $1\sigma$ in $73\%$ of the cases, while \texttt{BISP-1} achieves  $\sim 31\%$ coverage. At the 2$\sigma$ level, the coverage improves to 84\% for \texttt{TRIShUL} and 97\% for \texttt{BISP-1}. This indicates that \texttt{TRIShUL} provides results with lower precision than \texttt{BISP-1}.

The polarization properties of the clouds were computed based on both the true cloud layers and those identified by \texttt{TRIShUL} as described in Sect.~\ref{sec:2.4.2}. These are shown in blue and red points, respectively, in the bottom panel of Fig.~\ref{fig:Fig6}. For comparison, the \texttt{BISP-1} results are overplotted as black points, representing the mean Stokes parameters and their standard deviations from the posterior distributions.
The estimates from \texttt{TRIShUL}, and \texttt{BISP-1} both closely align with the true values in almost all reconstructions, with only a few exceptions. These findings demonstrate that both methods yield accurate estimates of polarization properties for the single-layer mock datasets.

\subsection{Comparison of 2-clouds mock samples}\label{sec:4.2}
In this section, we consider the cases when two clouds are introduced along the LOS from Sect.~4.5.2 of \cite{Pelgrims2023}.
The nearby cloud is fixed at 350 pc with polarization degree of $0.2\%$, magnetic-field position angle $\psi_{B_{\rm{reg}}} = 22.5^\circ$, inclination $\gamma_{B_{\rm{reg}}} = 30^\circ$, and $A_{\rm{turb}} = 0.2$. A second cloud is added at 685~pc ($f_{\rm{bg2}} = 70\%$), with a degree of polarization set to $0.2\%$, and an inclination angle of the magnetic field to $0^\circ$. The relative position angle of the magnetic field between the two clouds ($\Delta\psi$) is varied in the range $0^\circ$ to $150^\circ$ in steps of $30^\circ$. For each configuration, 10 random realizations are generated, yielding 60 mock samples in total.

As reported in the Sect.~4.5.2 of \texttt{BISP-1} the analysis yielded 44 non-pathological parallax posterior distributions out of 60 mock samples, with the AIC favoring a two-layer model in 43 of these 44 cases. However, this result was achieved by imposing informative priors on cloud distances, i.e,  assuming the presence of a nearby cloud within 100 to 600 pc and a second, more distant cloud within 300 to 3500 pc, and comparing the best AIC between 1- and 2-layer models only.

We reanalyzed 2-cloud mock samples with \texttt{BISP-1} without considering any external prior on parallax and updated parallax priors for all the models to priors informed by the five-layer configuration, instead of relying on the differing default priors typically associated with each model for a consistent comparison across models\footnote{The choice of cloud parallax priors must be made with caution. We tested comparing models with different prior settings for the layers in each model, independent of each other. This resulted in 56 out of 60 cases incorrectly favoring a single-layer model (FNs), despite the presence of 2-layers.}.  
Out of 60 samples, 13 favor the one-layer model more than the two-layer model, thus constituting FNs. The remaining 47 cases led to a two-layer model as the best fit to the sample data.

In contrast, \texttt{TRIShUL} requires no prior information about cloud distances and performs robustly on the same two-cloud mock samples, yielding only two FN cases and no FPs. Two additional cases are flagged as uncertain.
Overall, the comparative statistics between the two methods highlights the effectiveness of \texttt{TRIShUL} in identifying number of cloud along the LOS, without relying on the prior knowledge of cloud distance.

\begin{figure}
   \centering
   \subfigure{\includegraphics[width=0.4\textwidth]{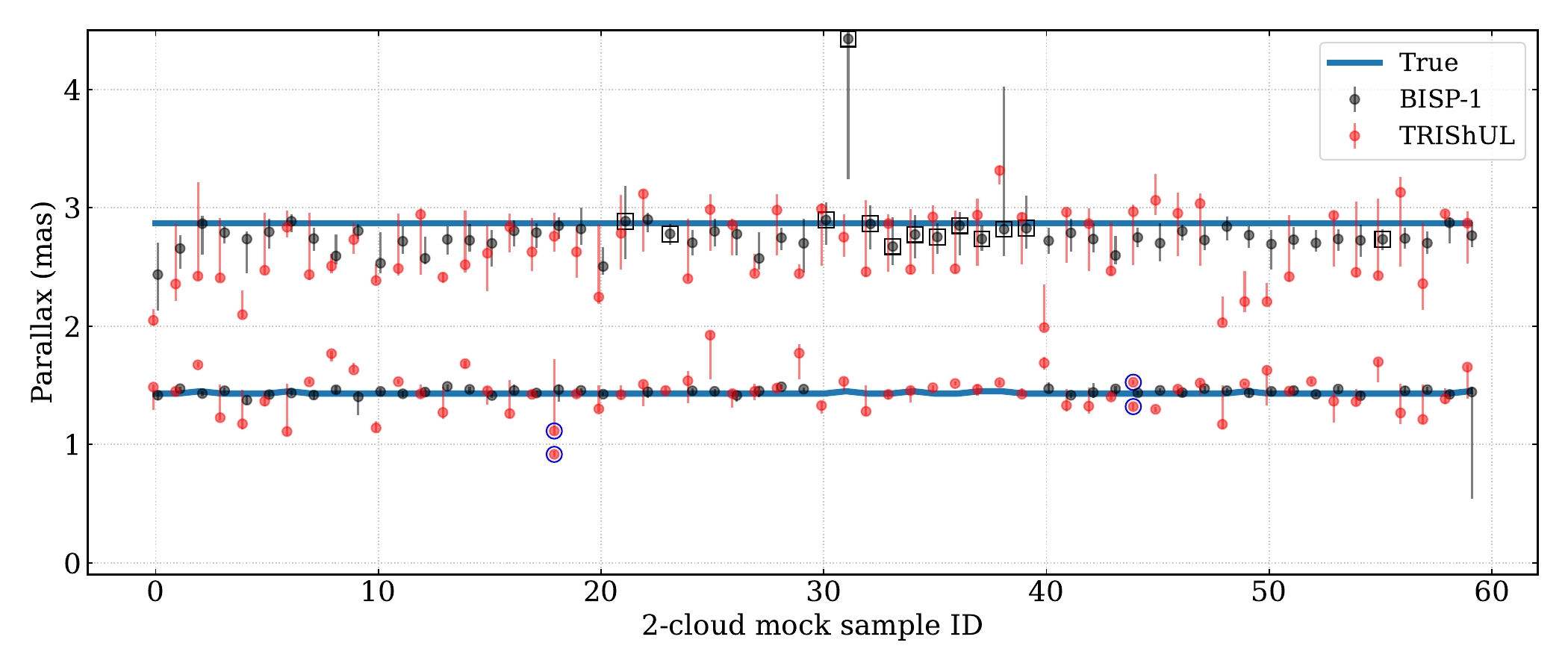}}
   
   \subfigure{\includegraphics[width=0.4\textwidth]{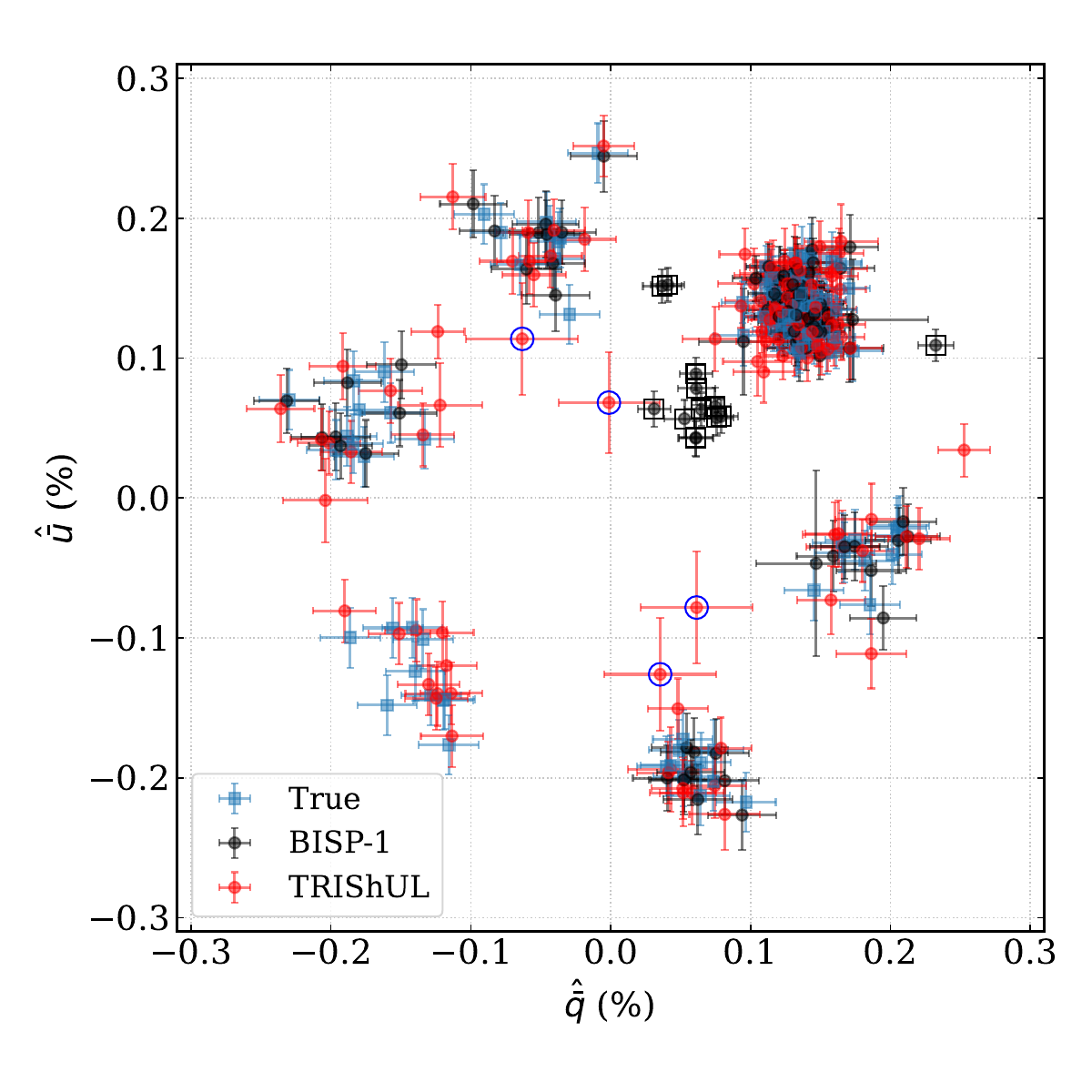}}
      \caption{Comparison of the estimated parameters from \texttt{BISP-1} (black points) and \texttt{TRIShUL} (red points) with the true value (blue points) for parallax (top panel) and the Stokes parameters (bottom panel) of mock catalogs with two clouds. Dark-blue open circles indicates the flagged cases in \texttt{TRIShUL} and black open squares corresponds to FN cases in \texttt{BISP-1}.} \label{fig:Fig7}
   \end{figure}

The accuracy of the two methods in recovering the parallax and polarization properties of the identified clouds is illustrated in the top and bottom panels of Fig.~\ref {fig:Fig7}, respectively. 
 In both panels, the results from \texttt{TRIShUL} are shown in red, while \texttt{BISP-1} outputs are represented by black markers. For the comparison of estimated parallax with the true values, the parallax based on input cloud distances is highlighted as a blue line in the top panel. Similarly, in the bottom, the true Stokes parameters, extracted by segmenting the data from the input distances are shown in blue points along with the weighted error on the means. Flagged cases from \texttt{TRIShUL} are indicated by blue open circles in both panels. While the FNs of \texttt{BISP-1} are marked by black open squares. 
 
 The top panel displays the estimated parallaxes from both methods for each LOS, sorted by 10 random samples per difference in position angle of the two clouds as sample ID. In most cases, the black points lie closer to the blue line with smaller error bars, indicating that \texttt{BISP-1} yields higher accuracy and precision compared to \texttt{TRIShUL}, which shows greater scatter and larger uncertainties. Despite this, most of our results remain within $3\sigma$ of the true values, indicating good accuracy but a lower precision. The precision of \texttt{TRIShUL} could be improved by relaxing the cap of five maximum breaks in breakpoint detection. 
 Moreover, it is worth noting that \texttt{TRIShUL} successfully and accurately resolves sightlines that \texttt{BISP-1} fails to identify, particularly for sample IDs between 30 to 40, which correspond to a specific configuration where $\Delta\psi = 90^\circ$.

 Similar to parallax, the Stokes parameters estimated from both methods, represented in the bottom-panel, generally agree with the true values within uncertainties. Excluding the flagged cases (indicated by open blue circles), \texttt{TRIShUL} closely recovers the true polarization in most cases, with only a few exceptions.  \texttt{BISP-1} also recovers Stokes parameters consistently across most cases. False negative cases show significant deviation from the true estimates (marked by black open squares in Fig.~\ref{fig:Fig7}), which arise because the estimates are based on a single layer along the LOS rather than two. 
 
In terms of computational efficiency,  \texttt{TRIShUL} completes the tomographic decomposition of a typical LOS within only $\sim 2$ minutes when testing up to five layers, whereas \texttt{BISP-1} requires approximately 70~minutes to test all five models for the same LOS on same machine (Apple MacBook Air with M1 chip, 8-core CPU, 16 GB RAM).
Employing parallel processing for the null-hypothesis testing by distributing random draws for Monte Carlo simulations across multiple cores further reduces the runtime of \texttt{TRIShUL} by nearly half, with the same set of criteria as mentioned above. These advantages make \texttt{TRIShUL} computationally efficient, and broadly applicable method for multi-layer dust cloud reconstruction.

\section{Application to observational data}\label{sec:5}
In this section, we demonstrate the applicability of our method to real stellar polarization data. To assess its reliability across diverse ISM environments, we apply our method to two contrasting Galactic regions: (i) towards the Galactic plane, where the ISM is dense; and (ii) towards high Galactic latitudes, where the ISM is more diffuse and exhibits lower polarization. 

Intrinsically polarized stars are expected to show a higher degree of polarization than neighboring stars and unrelated polarization angle. Such stars should be removed from the data before applying our method for LOS tomography as mentioned in Sect.~\ref{sec:2}. 

\subsection{Towards high Galactic latitudes}\label{sec:5.1}
We applied \texttt{TRIShUL} to R-band stellar polarization data from the high Galactic latitude regions centered around $(l,b) =  (104^\circ.08, 22^\circ.31)$ and $(103^\circ.90, 21^\circ.97)$. The polarization data set is obtained from \cite{Panopoulou2019} and consists of $\sim 100$ stars towards both regions. The former sightline is reported to indicate the presence of at least two clouds (2-cloud LOS, hereafter), and the latter has only a single cloud (1-cloud LOS, hereafter) based on HI spectra. 

\cite{Pelgrims2023} applied \texttt{BISP-1} to these LOSs and successfully identified a single-layer structure along the 1-cloud LOS. However, towards the 2-cloud LOS, their analysis favored a 1-layer model as the most probable rather than a 2-layer model. They noted that while the 1-layer model was preferred, their results did not entirely rule out the possibility of a second layer along that LOS. They found that, among the tested models, the 2-layer model shows a probability of about $42\%$ to be the model that actually minimizes the loss of information. 

\begin{figure}
   \centering
\subfigure{\includegraphics[width=0.4\textwidth]{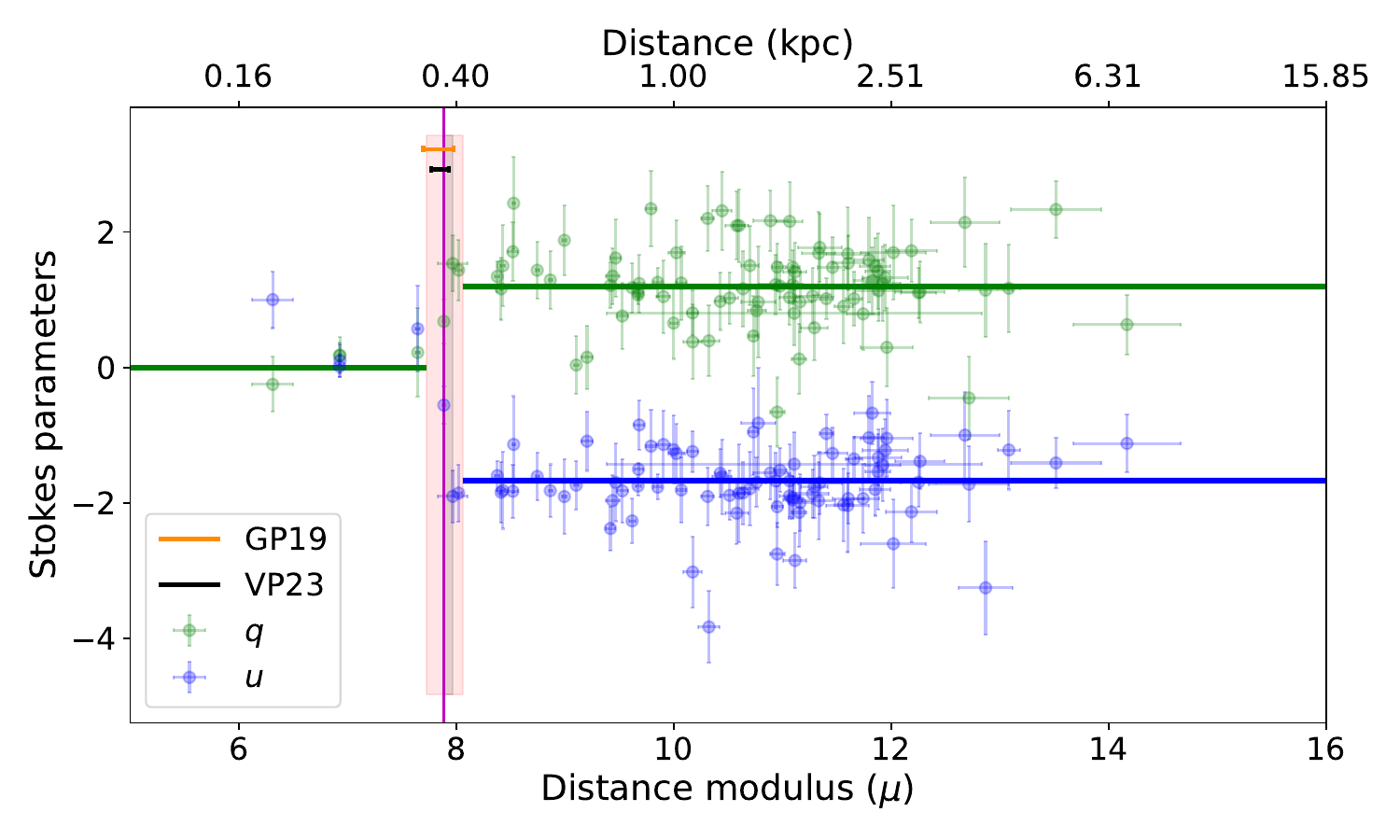}}
\subfigure{\includegraphics[width=0.4\textwidth]{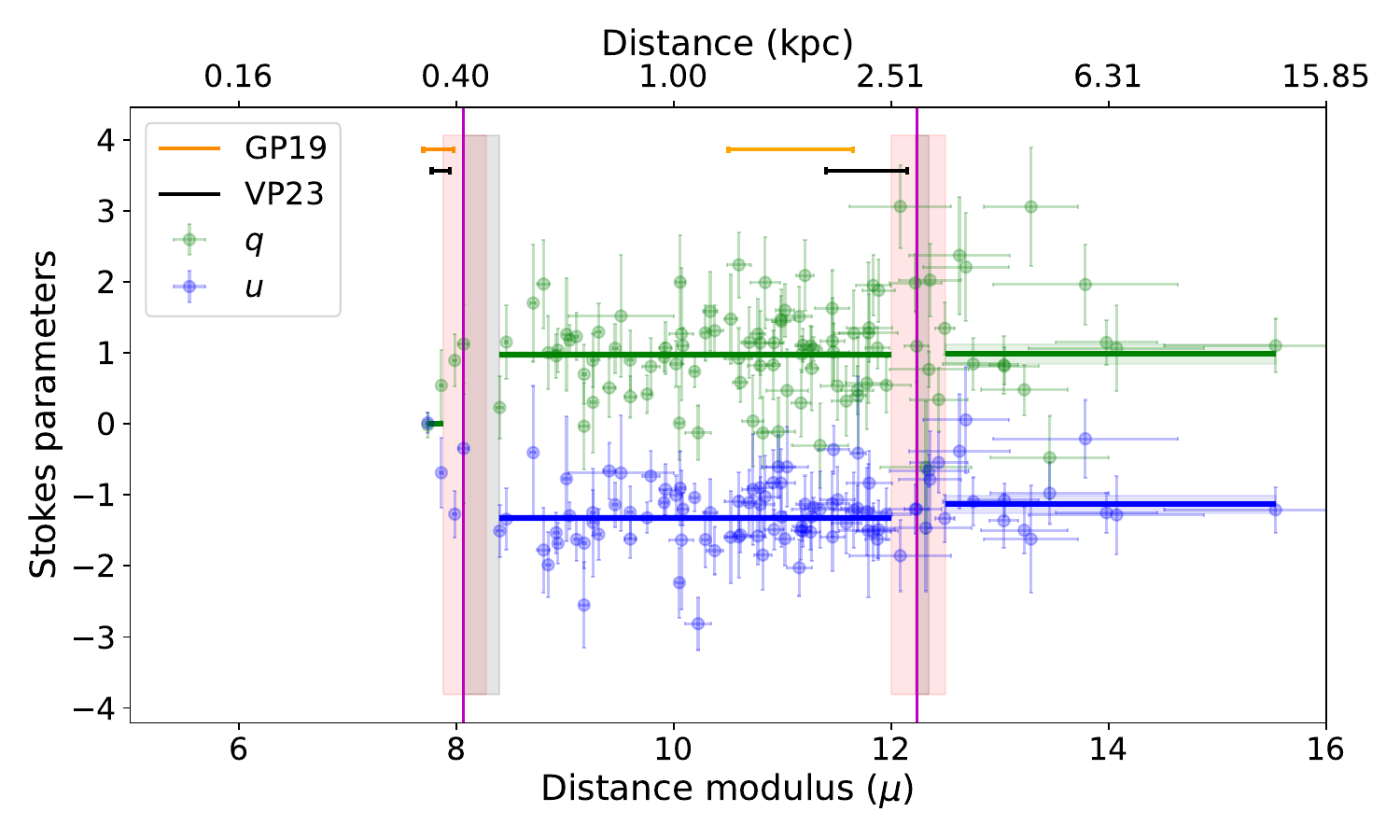}}
      \caption{LOS decomposition of starlight polarization by \texttt{TRIShUL} displayed in $(q,u) - \mu$ ($q$ in green, $u$ in blue) plane for the polarization data from \cite{Panopoulou2019}. Vertical lines indicate the positions of cloud layers estimated by our method, with shaded regions representing statistical (pink) and systematic (gray) uncertainties. The top panel corresponds to the 1-cloud region, and the bottom panel to the 2-cloud region. The estimated distance range from \cite{Panopoulou2019} and \cite{Pelgrims2023} are marked in orange and black color as GP19 and VP23, respectively.} \label{fig:Fig8}
   \end{figure}

We applied \texttt{TRIShUL} to both datasets by setting the minimum number of stars per segment fixed to five, and no constraints on the maximum number of breakpoints.
The Spurious breaks are subsequently rejected to isolate statistically significant cloud layers with null-hypothesis testing at 99\% confidence level. Our approach robustly identified a single polarization layer in the 1-cloud LOS (see top-panel of Fig.~\ref{fig:Fig8}). In contrast to \texttt{BISP-1} results, our method successfully detected two distinct layers towards the 2-cloud LOS shown in the bottom panel of Fig.~\ref{fig:Fig8}. 
The estimated distances of the identified layers in both regions agree with those reported by \cite[GP19,][]{Panopoulou2019}  and \cite[VP23,][]{Pelgrims2023} within $2\sigma$ uncertainties, as shown by the orange and black lines in the plot marking the distance ranges labeled GP19 and VP23, respectively.
We computed the polarization properties of the identified layers in both regions and compared them with the literature values in Table \ref{table:1}. The values derived from our method closely match the literature within the $1\sigma$ uncertainty level.
\begin{table}
\caption{Comparison of polarization properties of the clouds estimated from our method with the results from the literature.}              
\label{table:1}      
\centering                                      
\begin{tabular}{c c c c}          
\hline\hline
parameter & GP19 & VP23 & \texttt{TRIShUL} \\[.5ex]   
\hline %
\hline
\\[-1.5ex]
1-cloud & & & \\[.5ex]   
\hline                                   
    $p_{C} (\%)$ & $2.04 \pm 0.04$ & $1.62 \pm 0.04$ & $ 2.05 \pm 0.04$ \\[.5ex]         
    $\psi_{B} (^\circ)$ &$-27.5 \pm 0.6$ & $-26.0 \pm 0.9$ & $-27.3 \pm 0.6$ \\[.5ex]   
\hline
\\[-1.5ex]
2-cloud & & &\\[.5ex]   
\hline
$p_{C1} (\%)$ & $1.65 \pm 0.04$ & $1.64 \pm 0.05$  & $ 1.63 \pm 0.04$\\[.5ex]   
$\psi_{B}^{C1} (^\circ)$  & $-27.3 \pm 0.8$ & $-27.1 \pm 1.0 $ &$ -26.8 \pm 0.8$\\[.5ex]   
$p_{C2} (\%)$  & $0.28 \pm 0.07$ & $0.32^{0.15}_{-0.12}$ &$ 0.26 \pm 0.09$ \\[.5ex]   
$\psi_{B}^{C2} (^\circ) $ & $36.0 \pm 8.0$ & $20.6 ^{+18.3}_{-12.6}$ &$ 38.4 \pm 11.8$ \\[.5ex]   
\hline                                             
\end{tabular}
\end{table}
\subsection{Towards the Galactic plane}\label{sec:5.2}
Regions near the Galactic plane contain denser structures and typically exhibit higher polarization in contrast to high-latitude regions. However, the turbulence in the ISM is expected to be higher in these areas, leading to greater scatter in starlight polarization, hence making it more challenging to identify and characterize LOS dust clouds and their polarization properties. 

 We applied \texttt{TRIShUL} to the polarization data of the distant open cluster Berkeley 19 ($l = 176^\circ.919$, $b = - 3^\circ.612$), selected from the cluster sample observed in \citet{Uppal2024}. 
As one of the most distant clusters in their study, Berkeley 19 provides a long path length, allowing us to probe multiple dust layers. Moreover, it shows the highest scatter in polarization degree among all clusters in the sample, making it particularly challenging to visually infer cloud locations from polarization alone (see Fig. 5e in their paper). 

\cite{Uppal2024} used \texttt{BISP-1}  to quantify the distance of the LOS dust cloud by incorporating external priors, i.e., they adopted a flat prior for the parallax of the nearby layer (with a minimum distance of 50 pc) and a Gaussian prior for the second layer with a mean of 1.58 mas (corresponding to a distance modulus of $\mu = 9$) with a standard deviation of 0.5 mas. We reanalyzed the decomposition with \texttt{BISP-1} without incorporating any prior knowledge on the distance of the layers, and found that the distance of the identified layers remains consistent within the reported uncertainty levels. 

\begin{figure} 
   \resizebox{\hsize}{!}{\includegraphics{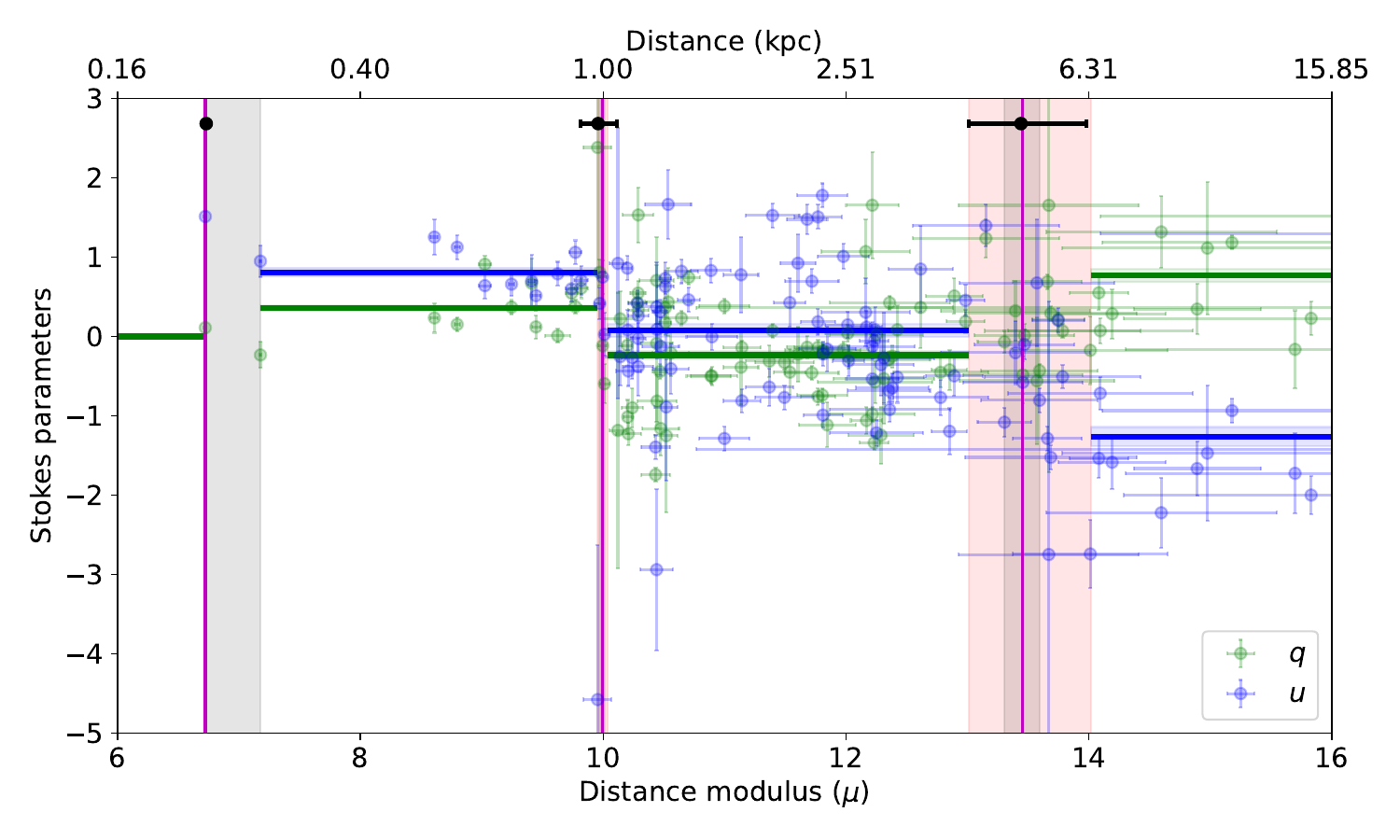}}
     \caption{LOS decomposition of starlight polarization by \texttt{TRIShUL} towards Galactic open cluster Berkeley 19 in $(q,u)-\mu$ plane. The color scheme follows that of Fig.~\ref{fig:Fig8}. The black point denotes the mean parallax of the cloud as inferred from the posterior distribution of \texttt{BISP-1}, with the error bar representing the standard deviation.}
     \label{fig:Fig9}
\end{figure}

We applied our algorithm, \texttt{TRIShUL}, to the same dataset by cleaning the data and removing the outliers following the procedure provided in Sect.~4.2.2 of \cite{Uppal2024}. We again fixed a minimum of five stars per segment and identified three distinct layers along the LOS. The first layer represents an upper limit on the distance, as the Stokes parameters of the nearest foreground stars deviate significantly from zero, indicating the presence of a polarizing dust layer in the foreground (see Sect.~\ref{sec:2.5}) of all the stars. The identified layers are marked in the $(q,u)-\mu$ plane with vertical lines shown in Fig.~\ref{fig:Fig9}. The estimated distances closely match the mean and standard deviation of the posterior distribution from \texttt{BISP-1}, obtained without applying any prior to cloud distance, as indicated by the black points and error-bars.

\section{Summary and conclusion}\label{sec:6}
In this work, we introduced and demonstrated \texttt{TRIShUL}, a novel, data-driven method for tomographic decomposition of the magnetized ISM using starlight polarization. Developed using a frequentist framework, the method is both conceptually straightforward and computationally efficient. \texttt{TRIShUL} employs a unique combination of cumulative Mahalanobis distance as a function of distance-sorted indices to detect structural breaks that indicate the presence of dust layers along the LOS. This cumulative approach allows reliable detection of dust layers, even those inducing weak polarization and in low S/N conditions. The effectiveness of the cumulative approach has also been recently demonstrated by \cite{Zhang2025} in identifying extinction jumps along the LOS for constructing 3D extinction maps. However, their strategy is not directly applicable to polarization: unlike extinction, which increases monotonically and whose cumulative differences from the mean can enhance discontinuities, polarization can exhibit depolarization effects, making the averaging approach unreliable. Tests with our simulated data confirm that the same algorithm is unable to identify the correct break in polarization.

While the cumulative process enhances sensitivity to subtle structural changes, it may also introduce spurious breaks caused by local fluctuations in Mahalanobis distance, attributable to measurement uncertainties or scatter in the data. To address this, we designed a robust strategy to eliminate such false detections. 
We estimate average parallaxes and Stokes parameters of the identified layers, incorporating a full error budget to quantify both statistical and systematic uncertainties. This enables reliable inference of physical properties such as cloud distance, degree of polarization, and position angle, or the orientation of the POS magnetic field. 
The entire framework, including Mahalanobis distance calculation, break detection (testing one to five breaks), spurious break rejection, and parameter estimation along the uncertainty intervals, completes in under two minutes on a laptop, highlighting its suitability for large-scale survey applications.
One of the caveats of our method is that while we used uncertainties in the parallax in the rejection process, they have not been considered in the identification of the breaks themselves.

We benchmarked and validated the performance of \texttt{TRIShUL} using synthetic polarization data constructed by \citep{Pelgrims2023}. We showed that \texttt{TRIShUL} successfully recovers LOS dust clouds and their polarization properties as soon as the induced degree of polarization exceeds 0.1\% which has more than 10\% of the stars lying in the background of the cloud or in between two clouds of LOS having 350 stars. 
It is important to note that our algorithm is able to detect structural breaks in distance-sorted Mahalanobis distance even for the extreme cases (low polarization $(< 0.1\%)$ and low number statistics background to the cloud). However, such detections are typically excluded in the spurious layer rejection step because of their low statistical significance. This limits the sensitivity of the method to identify extremely weak features. Enhancements to address this regime will be explored in future work. Nevertheless, the current sensitivity limits of \texttt{TRIShUL} are well matched to the typical detection limits $(\sim 0.1\%)$ of most of the existing polarization instruments \citep[e.g.,][]{Maharana2024}.

We further demonstrated the applicability of \texttt{TRIShUL} to real observational data across diverse ISM environments - in a high Galactic latitude region and a region near the Galactic plane. 
In both cases, the estimated cloud distances and polarization properties were consistent with values reported in the literature, within uncertainties.

We conducted a detailed comparison between our method, \texttt{TRIShUL}, and the recently developed Bayesian inference-based algorithm, \texttt{BISP-1}, which performs fully automated tomographic decomposition of starlight polarization.  We observed that \texttt{BISP-1} generally yields more precise estimates due to its probabilistic framework, but they are sensitive to the choice of the priors on cloud parallaxes. In contrast, \texttt{TRIShUL} does not require any external prior information and provides results that are statistically robust, albeit with somewhat larger uncertainties. Another key distinction between the two methods lies in the computational efficiency. \texttt{TRIShUL} employs a direct, data-driven frequentist approach to identify the structural breaks and reject the spurious ones. These processes are more intuitive and significantly faster for testing $1-$ to $N-$breaks, compared to \texttt{BISP-1}.  \texttt{BISP-1} requires extensive posterior sampling over a high-dimensional parameter space, and its computation time increases substantially as the number of model layers grows. However, unlike \texttt{TRIShUL}, \texttt{BISP-1} has the advantage of also accounting for and modeling out the intrinsic scatter arising due to turbulence in the ISM, which is not currently estimated in our approach.

Overall, \texttt{TRIShUL} presents a computationally efficient, statistically robust, and prior-free multi-layer decomposition approach for polarization tomography, well-suited for the upcoming era of large-scale polarization surveys. 
The complementary strengths of \texttt{TRIShUL} and \texttt{BISP-1} can be effectively harnessed by combining the two methods: initial results from \texttt{TRIShUL} can be used to inform the priors for \texttt{BISP-1}. This hybrid strategy enables fast, reliable decomposition with \texttt{TRIShUL}, followed by high-precision, statistically rigorous refinement through Bayesian inference.
\begin{acknowledgements}
This research is funded by the European Union. Views and opinions expressed are, however, those of the author(s) only and do not necessarily reflect those of the European Union or the European Research Council Executive Agency. Neither the European Union nor the granting authority can be held responsible for them. This work is supported by an ERC grant, MW-ATLAS project no. 101166905. We thank Dr. G. V. Panopoulou for their constructive feedback.  
N.U. acknowledges support from the European Research Council (ERC) under the Horizon ERC Grants 2021 programme under grant agreement No. 101040021. K.T. acknowledges the support by the TITAN ERA Chair project (contract no. 101086741) within the Horizon Europe Framework Program of the European Commission. V. Pelgrims acknowledges funding from a Marie Curie Action of the European Union (grant agreement No. 101107047). N.U. would like to thank D. Blinov for discussions and insights from the coding perspective. 
\end{acknowledgements}
\bibliographystyle{aa} 
\bibliography{ref}

\begin{thebibliography}{67}
\expandafter\ifx\csname natexlab\endcsname\relax\def\natexlab#1{#1}\fi

\bibitem[{{Akaike}(1974)}]{AIC1974}
{Akaike}, H. 1974, IEEE Transactions on Automatic Control, 19, 716

\bibitem[{{Andersson} {et~al.}(2015){Andersson}, {Lazarian}, \&
  {Vaillancourt}}]{Andersson2015}
{Andersson}, B.~G., {Lazarian}, A., \& {Vaillancourt}, J.~E. 2015, \araa, 53,
  501

\bibitem[{{Andersson} \& {Potter}(2005)}]{Andersson2005}
{Andersson}, B.~G. \& {Potter}, S.~B. 2005, \mnras, 356, 1088

\bibitem[{{Angarita} {et~al.}(2025){Angarita}, {Versteeg}, {Haverkorn},
  {Pelgrims}, {Rodrigues}, {Magalh{\~a}es}, {Santos-Lima}, \&
  {Kawabata}}]{Yenifer2025}
{Angarita}, Y., {Versteeg}, M.~J.~F., {Haverkorn}, M., {et~al.} 2025, arXiv
  e-prints, arXiv:2506.01564

\bibitem[{Bai \& Perron(2002)}]{Bai2002}
Bai, J. \& Perron, P. 2002, Journal of Applied Econometrics, 18, 1–22

\bibitem[{{Bailer-Jones} {et~al.}(2021){Bailer-Jones}, {Rybizki}, {Fouesneau},
  {Demleitner}, \& {Andrae}}]{bailer2021}
{Bailer-Jones}, C.~A.~L., {Rybizki}, J., {Fouesneau}, M., {Demleitner}, M., \&
  {Andrae}, R. 2021, \aj, 161, 147

\bibitem[{{Bailey} {et~al.}(2024){Bailey}, {Lewis}, {Howarth}, {Cotton},
  {Marshall}, \& {Kedziora-Chudczer}}]{Bailey2024}
{Bailey}, J., {Lewis}, F., {Howarth}, I.~D., {et~al.} 2024, \apj, 972, 103

\bibitem[{{Beck}(2015)}]{Beck2015}
{Beck}, R. 2015, \aapr, 24, 4

\bibitem[{{Beck} \& {Wielebinski}(2013)}]{Beck2013}
{Beck}, R. \& {Wielebinski}, R. 2013, in Planets, Stars and Stellar Systems.
  Volume 5: Galactic Structure and Stellar Populations, ed. T.~D. {Oswalt} \&
  G.~{Gilmore}, Vol.~5, 641

\bibitem[{{Bijas} {et~al.}(2024){Bijas}, {Eswaraiah}, {Sandhyarani}, {Jose}, \&
  {Gopinathan}}]{Bijas2024}
{Bijas}, N., {Eswaraiah}, C., {Sandhyarani}, P., {Jose}, J., \& {Gopinathan},
  M. 2024, \mnras, 529, 4234

\bibitem[{{Chiang}(2023)}]{Chiang2023}
{Chiang}, Y.-K. 2023, \apj, 958, 118

\bibitem[{{Clark} \& {Hensley}(2019)}]{Clark2019}
{Clark}, S.~E. \& {Hensley}, B.~S. 2019, \apj, 887, 136

\bibitem[{{Clarke}(2010)}]{DClarke}
{Clarke}, D. 2010, {Stellar Polarimetry}

\bibitem[{{Clemens} {et~al.}(2012){Clemens}, {Pinnick}, {Pavel}, \&
  {Taylor}}]{GPIPS}
{Clemens}, D.~P., {Pinnick}, A.~F., {Pavel}, M.~D., \& {Taylor}, B.~W. 2012,
  \apjs, 200, 19

\bibitem[{{Cotton} {et~al.}(2016){Cotton}, {Bailey}, {Kedziora-Chudczer},
  {Bott}, {Lucas}, {Hough}, \& {Marshall}}]{cotton2016}
{Cotton}, D.~V., {Bailey}, J., {Kedziora-Chudczer}, L., {et~al.} 2016, \mnras,
  455, 1607

\bibitem[{{Dickey} {et~al.}(2022){Dickey}, {West}, {Thomson}, {Landecker},
  {Bracco}, {Carretti}, {Han}, {Hill}, {Ma}, {Mao}, {Ordog}, {Brown},
  {Douglas}, {Erceg}, {Jeli{\'c}}, {Kothes}, \& {Wolleben}}]{Dickey2022}
{Dickey}, J.~M., {West}, J., {Thomson}, A. J.~M., {et~al.} 2022, \apj, 940, 75

\bibitem[{{Doi} {et~al.}(2021{\natexlab{a}}){Doi}, {Hasegawa}, {Bastien},
  {Tahani}, {Arzoumanian}, {Coud{\'e}}, {Matsumura}, {Sadavoy}, {Hull},
  {Shimajiri}, {Furuya}, {Johnstone}, {Plume}, {Inutsuka}, {Kwon}, \&
  {Tamura}}]{Doi2021}
{Doi}, Y., {Hasegawa}, T., {Bastien}, P., {et~al.} 2021{\natexlab{a}}, \apj,
  914, 122

\bibitem[{{Doi} {et~al.}(2021{\natexlab{b}}){Doi}, {Hasegawa}, {Bastien},
  {Tahani}, {Arzoumanian}, {Coud{\'e}}, {Matsumura}, {Sadavoy}, {Hull},
  {Shimajiri}, {Furuya}, {Johnstone}, {Plume}, {Inutsuka}, {Kwon}, \&
  {Tamura}}]{Doi2021b}
{Doi}, Y., {Hasegawa}, T., {Bastien}, P., {et~al.} 2021{\natexlab{b}}, \apj,
  914, 122

\bibitem[{{Doi} {et~al.}(2024){Doi}, {Nakamura}, {Kawabata}, {Matsumura},
  {Akitaya}, {Coud{\'e}}, {Rodrigues}, {Kwon}, {Tamura}, {Tahani},
  {Magalh{\~a}es}, {Santos-Lima}, {Angarita}, {Versteeg}, {Haverkorn},
  {Hasegawa}, {Sadavoy}, {Arzoumanian}, \& {Bastien}}]{Doi2024}
{Doi}, Y., {Nakamura}, K., {Kawabata}, K.~S., {et~al.} 2024, \apj, 961, 13

\bibitem[{{Edenhofer} {et~al.}(2024){Edenhofer}, {Zucker}, {Frank}, {Saydjari},
  {Speagle}, {Finkbeiner}, \& {En{\ss}lin}}]{edenhofer2023}
{Edenhofer}, G., {Zucker}, C., {Frank}, P., {et~al.} 2024, \aap, 685, A82

\bibitem[{{Fridman}(2010)}]{cumsum2010}
{Fridman}, P.~A. 2010, \mnras, 409, 808

\bibitem[{{Gaia Collaboration} {et~al.}(2018){Gaia Collaboration}, {Brown},
  {Vallenari}, {Prusti}, {de Bruijne}, {Babusiaux}, {Bailer-Jones}, {Biermann},
  {Evans}, {Eyer}, {Jansen}, {Jordi}, {Klioner}, {Lammers}, {Lindegren},
  {Luri}, {Mignard}, {Panem}, {Pourbaix}, {Randich}, {Sartoretti}, {Siddiqui},
  {Soubiran}, {van Leeuwen}, {Walton}, {Arenou}, {Bastian}, {Cropper},
  {Drimmel}, {Katz}, {Lattanzi}, {Bakker}, {Cacciari}, {Casta{\~n}eda},
  {Chaoul}, {Cheek}, {De Angeli}, {Fabricius}, {Guerra}, {Holl}, {Masana},
  {Messineo}, {Mowlavi}, {Nienartowicz}, {Panuzzo}, {Portell}, {Riello},
  {Seabroke}, {Tanga}, {Th{\'e}venin}, {Gracia-Abril}, {Comoretto},
  {Garcia-Reinaldos}, {Teyssier}, {Altmann}, {Andrae}, {Audard},
  {Bellas-Velidis}, {Benson}, {Berthier}, {Blomme}, {Burgess}, {Busso},
  {Carry}, {Cellino}, {Clementini}, {Clotet}, {Creevey}, {Davidson}, {De
  Ridder}, {Delchambre}, {Dell'Oro}, {Ducourant},
  {Fern{\'a}ndez-Hern{\'a}ndez}, {Fouesneau}, {Fr{\'e}mat}, {Galluccio},
  {Garc{\'\i}a-Torres}, {Gonz{\'a}lez-N{\'u}{\~n}ez}, {Gonz{\'a}lez-Vidal},
  {Gosset}, {Guy}, {Halbwachs}, {Hambly}, {Harrison}, {Hern{\'a}ndez},
  {Hestroffer}, {Hodgkin}, {Hutton}, {Jasniewicz}, {Jean-Antoine-Piccolo},
  {Jordan}, {Korn}, {Krone-Martins}, {Lanzafame}, {Lebzelter}, {L{\"o}ffler},
  {Manteiga}, {Marrese}, {Mart{\'\i}n-Fleitas}, {Moitinho}, {Mora}, {Muinonen},
  {Osinde}, {Pancino}, {Pauwels}, {Petit}, {Recio-Blanco}, {Richards},
  {Rimoldini}, {Robin}, {Sarro}, {Siopis}, {Smith}, {Sozzetti}, {S{\"u}veges},
  {Torra}, {van Reeven}, {Abbas}, {Abreu Aramburu}, {Accart}, {Aerts},
  {Altavilla}, {{\'A}lvarez}, {Alvarez}, {Alves}, {Anderson}, {Andrei},
  {Anglada Varela}, {Antiche}, {Antoja}, {Arcay}, {Astraatmadja}, {Bach},
  {Baker}, {Balaguer-N{\'u}{\~n}ez}, {Balm}, {Barache}, {Barata}, {Barbato},
  {Barblan}, {Barklem}, {Barrado}, {Barros}, {Barstow}, {Bartholom{\'e}
  Mu{\~n}oz}, {Bassilana}, {Becciani}, {Bellazzini}, {Berihuete}, {Bertone},
  {Bianchi}, {Bienaym{\'e}}, {Blanco-Cuaresma}, {Boch}, {Boeche}, {Bombrun},
  {Borrachero}, {Bossini}, {Bouquillon}, {Bourda}, {Bragaglia}, {Bramante},
  {Breddels}, {Bressan}, {Brouillet}, {Br{\"u}semeister}, {Brugaletta},
  {Bucciarelli}, {Burlacu}, {Busonero}, {Butkevich}, {Buzzi}, {Caffau},
  {Cancelliere}, {Cannizzaro}, {Cantat-Gaudin}, {Carballo}, {Carlucci},
  {Carrasco}, {Casamiquela}, {Castellani}, {Castro-Ginard}, {Charlot},
  {Chemin}, {Chiavassa}, {Cocozza}, {Costigan}, {Cowell}, {Crifo}, {Crosta},
  {Crowley}, {Cuypers}, {Dafonte}, {Damerdji}, {Dapergolas}, {David}, {David},
  {de Laverny}, {De Luise}, {De March}, {de Martino}, {de Souza}, {de Torres},
  {Debosscher}, {del Pozo}, {Delbo}, {Delgado}, {Delgado}, {Di Matteo},
  {Diakite}, {Diener}, {Distefano}, {Dolding}, {Drazinos}, {Dur{\'a}n},
  {Edvardsson}, {Enke}, {Eriksson}, {Esquej}, {Eynard Bontemps}, {Fabre},
  {Fabrizio}, {Faigler}, {Falc{\~a}o}, {Farr{\`a}s Casas}, {Federici},
  {Fedorets}, {Fernique}, {Figueras}, {Filippi}, {Findeisen}, {Fonti},
  {Fraile}, {Fraser}, {Fr{\'e}zouls}, {Gai}, {Galleti}, {Garabato},
  {Garc{\'\i}a-Sedano}, {Garofalo}, {Garralda}, {Gavel}, {Gavras}, {Gerssen},
  {Geyer}, {Giacobbe}, {Gilmore}, {Girona}, {Giuffrida}, {Glass}, {Gomes},
  {Granvik}, {Gueguen}, {Guerrier}, {Guiraud}, {Guti{\'e}rrez-S{\'a}nchez},
  {Haigron}, {Hatzidimitriou}, {Hauser}, {Haywood}, {Heiter}, {Helmi}, {Heu},
  {Hilger}, {Hobbs}, {Hofmann}, {Holland}, {Huckle}, {Hypki}, {Icardi},
  {Jan{\ss}en}, {Jevardat de Fombelle}, {Jonker}, {Juh{\'a}sz}, {Julbe},
  {Karampelas}, {Kewley}, {Klar}, {Kochoska}, {Kohley}, {Kolenberg},
  {Kontizas}, {Kontizas}, {Koposov}, {Kordopatis}, {Kostrzewa-Rutkowska},
  {Koubsky}, {Lambert}, {Lanza}, {Lasne}, {Lavigne}, {Le Fustec}, {Le
  Poncin-Lafitte}, {Lebreton}, {Leccia}, {Leclerc}, {Lecoeur-Taibi},
  {Lenhardt}, {Leroux}, {Liao}, {Licata}, {Lindstr{\o}m}, {Lister}, {Livanou},
  {Lobel}, {L{\'o}pez}, {Managau}, {Mann}, {Mantelet}, {Marchal}, {Marchant},
  {Marconi}, {Marinoni}, {Marschalk{\'o}}, {Marshall}, {Martino}, {Marton},
  {Mary}, {Massari}, {Matijevi{\v{c}}}, {Mazeh}, {McMillan}, {Messina},
  {Michalik}, {Millar}, {Molina}, {Molinaro}, {Moln{\'a}r}, {Montegriffo},
  {Mor}, {Morbidelli}, {Morel}, {Morris}, {Mulone}, {Muraveva}, {Musella},
  {Nelemans}, {Nicastro}, {Noval}, {O'Mullane}, {Ord{\'e}novic},
  {Ord{\'o}{\~n}ez-Blanco}, {Osborne}, {Pagani}, {Pagano}, {Pailler},
  {Palacin}, {Palaversa}, {Panahi}, {Pawlak}, {Piersimoni}, {Pineau}, {Plachy},
  {Plum}, {Poggio}, {Poujoulet}, {Pr{\v{s}}a}, {Pulone}, {Racero}, {Ragaini},
  {Rambaux}, {Ramos-Lerate}, {Regibo}, {Reyl{\'e}}, {Riclet}, {Ripepi}, {Riva},
  {Rivard}, {Rixon}, {Roegiers}, {Roelens}, {Romero-G{\'o}mez}, {Rowell},
  {Royer}, {Ruiz-Dern}, {Sadowski}, {Sagrist{\`a} Sell{\'e}s}, {Sahlmann},
  {Salgado}, {Salguero}, {Sanna}, {Santana-Ros}, {Sarasso}, {Savietto},
  {Schultheis}, {Sciacca}, {Segol}, {Segovia}, {S{\'e}gransan}, {Shih},
  {Siltala}, {Silva}, {Smart}, {Smith}, {Solano}, {Solitro}, {Sordo}, {Soria
  Nieto}, {Souchay}, {Spagna}, {Spoto}, {Stampa}, {Steele},
  {Steidelm{\"u}ller}, {Stephenson}, {Stoev}, {Suess}, {Surdej}, {Szabados},
  {Szegedi-Elek}, {Tapiador}, {Taris}, {Tauran}, {Taylor}, {Teixeira},
  {Terrett}, {Teyssand ier}, {Thuillot}, {Titarenko}, {Torra Clotet}, {Turon},
  {Ulla}, {Utrilla}, {Uzzi}, {Vaillant}, {Valentini}, {Valette}, {van Elteren},
  {Van Hemelryck}, {van Leeuwen}, {Vaschetto}, {Vecchiato}, {Veljanoski},
  {Viala}, {Vicente}, {Vogt}, {von Essen}, {Voss}, {Votruba}, {Voutsinas},
  {Walmsley}, {Weiler}, {Wertz}, {Wevers}, {Wyrzykowski}, {Yoldas},
  {{\v{Z}}erjal}, {Ziaeepour}, {Zorec}, {Zschocke}, {Zucker}, {Zurbach}, \&
  {Zwitter}}]{GaiaDR2}
{Gaia Collaboration}, {Brown}, A.~G.~A., {Vallenari}, A., {et~al.} 2018, \aap,
  616, A1

\bibitem[{{Gaia Collaboration} {et~al.}(2023){Gaia Collaboration}, {Vallenari},
  {Brown}, {Prusti}, {de Bruijne}, {Arenou}, {Babusiaux}, {Biermann},
  {Creevey}, {Ducourant}, {Evans}, {Eyer}, {Guerra}, {Hutton}, {Jordi},
  {Klioner}, {Lammers}, {Lindegren}, {Luri}, {Mignard}, {Panem}, {Pourbaix},
  {Randich}, {Sartoretti}, {Soubiran}, {Tanga}, {Walton}, {Bailer-Jones},
  {Bastian}, {Drimmel}, {Jansen}, {Katz}, {Lattanzi}, {van Leeuwen}, {Bakker},
  {Cacciari}, {Casta{\~n}eda}, {De Angeli}, {Fabricius}, {Fouesneau},
  {Fr{\'e}mat}, {Galluccio}, {Guerrier}, {Heiter}, {Masana}, {Messineo},
  {Mowlavi}, {Nicolas}, {Nienartowicz}, {Pailler}, {Panuzzo}, {Riclet}, {Roux},
  {Seabroke}, {Sordo}, {Th{\'e}venin}, {Gracia-Abril}, {Portell}, {Teyssier},
  {Altmann}, {Andrae}, {Audard}, {Bellas-Velidis}, {Benson}, {Berthier},
  {Blomme}, {Burgess}, {Busonero}, {Busso}, {C{\'a}novas}, {Carry}, {Cellino},
  {Cheek}, {Clementini}, {Damerdji}, {Davidson}, {de Teodoro}, {Nu{\~n}ez
  Campos}, {Delchambre}, {Dell'Oro}, {Esquej}, {Fern{\'a}ndez-Hern{\'a}ndez},
  {Fraile}, {Garabato}, {Garc{\'\i}a-Lario}, {Gosset}, {Haigron}, {Halbwachs},
  {Hambly}, {Harrison}, {Hern{\'a}ndez}, {Hestroffer}, {Hodgkin}, {Holl},
  {Jan{\ss}en}, {Jevardat de Fombelle}, {Jordan}, {Krone-Martins}, {Lanzafame},
  {L{\"o}ffler}, {Marchal}, {Marrese}, {Moitinho}, {Muinonen}, {Osborne},
  {Pancino}, {Pauwels}, {Recio-Blanco}, {Reyl{\'e}}, {Riello}, {Rimoldini},
  {Roegiers}, {Rybizki}, {Sarro}, {Siopis}, {Smith}, {Sozzetti}, {Utrilla},
  {van Leeuwen}, {Abbas}, {{\'A}brah{\'a}m}, {Abreu Aramburu}, {Aerts},
  {Aguado}, {Ajaj}, {Aldea-Montero}, {Altavilla}, {{\'A}lvarez}, {Alves},
  {Anders}, {Anderson}, {Anglada Varela}, {Antoja}, {Baines}, {Baker},
  {Balaguer-N{\'u}{\~n}ez}, {Balbinot}, {Balog}, {Barache}, {Barbato},
  {Barros}, {Barstow}, {Bartolom{\'e}}, {Bassilana}, {Bauchet}, {Becciani},
  {Bellazzini}, {Berihuete}, {Bernet}, {Bertone}, {Bianchi}, {Binnenfeld},
  {Blanco-Cuaresma}, {Blazere}, {Boch}, {Bombrun}, {Bossini}, {Bouquillon},
  {Bragaglia}, {Bramante}, {Breedt}, {Bressan}, {Brouillet}, {Brugaletta},
  {Bucciarelli}, {Burlacu}, {Butkevich}, {Buzzi}, {Caffau}, {Cancelliere},
  {Cantat-Gaudin}, {Carballo}, {Carlucci}, {Carnerero}, {Carrasco},
  {Casamiquela}, {Castellani}, {Castro-Ginard}, {Chaoul}, {Charlot}, {Chemin},
  {Chiaramida}, {Chiavassa}, {Chornay}, {Comoretto}, {Contursi}, {Cooper},
  {Cornez}, {Cowell}, {Crifo}, {Cropper}, {Crosta}, {Crowley}, {Dafonte},
  {Dapergolas}, {David}, {David}, {de Laverny}, {De Luise}, {De March}, {De
  Ridder}, {de Souza}, {de Torres}, {del Peloso}, {del Pozo}, {Delbo},
  {Delgado}, {Delisle}, {Demouchy}, {Dharmawardena}, {Di Matteo}, {Diakite},
  {Diener}, {Distefano}, {Dolding}, {Edvardsson}, {Enke}, {Fabre}, {Fabrizio},
  {Faigler}, {Fedorets}, {Fernique}, {Fienga}, {Figueras}, {Fournier},
  {Fouron}, {Fragkoudi}, {Gai}, {Garcia-Gutierrez}, {Garcia-Reinaldos},
  {Garc{\'\i}a-Torres}, {Garofalo}, {Gavel}, {Gavras}, {Gerlach}, {Geyer},
  {Giacobbe}, {Gilmore}, {Girona}, {Giuffrida}, {Gomel}, {Gomez},
  {Gonz{\'a}lez-N{\'u}{\~n}ez}, {Gonz{\'a}lez-Santamar{\'\i}a},
  {Gonz{\'a}lez-Vidal}, {Granvik}, {Guillout}, {Guiraud},
  {Guti{\'e}rrez-S{\'a}nchez}, {Guy}, {Hatzidimitriou}, {Hauser}, {Haywood},
  {Helmer}, {Helmi}, {Sarmiento}, {Hidalgo}, {Hilger}, {H{\l}adczuk}, {Hobbs},
  {Holland}, {Huckle}, {Jardine}, {Jasniewicz}, {Jean-Antoine Piccolo},
  {Jim{\'e}nez-Arranz}, {Jorissen}, {Juaristi Campillo}, {Julbe}, {Karbevska},
  {Kervella}, {Khanna}, {Kontizas}, {Kordopatis}, {Korn}, {K{\'o}sp{\'a}l},
  {Kostrzewa-Rutkowska}, {Kruszy{\'n}ska}, {Kun}, {Laizeau}, {Lambert},
  {Lanza}, {Lasne}, {Le Campion}, {Lebreton}, {Lebzelter}, {Leccia}, {Leclerc},
  {Lecoeur-Taibi}, {Liao}, {Licata}, {Lindstr{\o}m}, {Lister}, {Livanou},
  {Lobel}, {Lorca}, {Loup}, {Madrero Pardo}, {Magdaleno Romeo}, {Managau},
  {Mann}, {Manteiga}, {Marchant}, {Marconi}, {Marcos}, {Marcos Santos},
  {Mar{\'\i}n Pina}, {Marinoni}, {Marocco}, {Marshall}, {Martin Polo},
  {Mart{\'\i}n-Fleitas}, {Marton}, {Mary}, {Masip}, {Massari},
  {Mastrobuono-Battisti}, {Mazeh}, {McMillan}, {Messina}, {Michalik}, {Millar},
  {Mints}, {Molina}, {Molinaro}, {Moln{\'a}r}, {Monari}, {Mongui{\'o}},
  {Montegriffo}, {Montero}, {Mor}, {Mora}, {Morbidelli}, {Morel}, {Morris},
  {Muraveva}, {Murphy}, {Musella}, {Nagy}, {Noval}, {Oca{\~n}a}, {Ogden},
  {Ordenovic}, {Osinde}, {Pagani}, {Pagano}, {Palaversa}, {Palicio},
  {Pallas-Quintela}, {Panahi}, {Payne-Wardenaar}, {Pe{\~n}alosa Esteller},
  {Penttil{\"a}}, {Pichon}, {Piersimoni}, {Pineau}, {Plachy}, {Plum}, {Poggio},
  {Pr{\v{s}}a}, {Pulone}, {Racero}, {Ragaini}, {Rainer}, {Raiteri}, {Rambaux},
  {Ramos}, {Ramos-Lerate}, {Re Fiorentin}, {Regibo}, {Richards}, {Rios Diaz},
  {Ripepi}, {Riva}, {Rix}, {Rixon}, {Robichon}, {Robin}, {Robin}, {Roelens},
  {Rogues}, {Rohrbasser}, {Romero-G{\'o}mez}, {Rowell}, {Royer}, {Ruz Mieres},
  {Rybicki}, {Sadowski}, {S{\'a}ez N{\'u}{\~n}ez}, {Sagrist{\`a} Sell{\'e}s},
  {Sahlmann}, {Salguero}, {Samaras}, {Sanchez Gimenez}, {Sanna},
  {Santove{\~n}a}, {Sarasso}, {Schultheis}, {Sciacca}, {Segol}, {Segovia},
  {S{\'e}gransan}, {Semeux}, {Shahaf}, {Siddiqui}, {Siebert}, {Siltala},
  {Silvelo}, {Slezak}, {Slezak}, {Smart}, {Snaith}, {Solano}, {Solitro},
  {Souami}, {Souchay}, {Spagna}, {Spina}, {Spoto}, {Steele},
  {Steidelm{\"u}ller}, {Stephenson}, {S{\"u}veges}, {Surdej}, {Szabados},
  {Szegedi-Elek}, {Taris}, {Taylor}, {Teixeira}, {Tolomei}, {Tonello}, {Torra},
  {Torra}, {Torralba Elipe}, {Trabucchi}, {Tsounis}, {Turon}, {Ulla}, {Unger},
  {Vaillant}, {van Dillen}, {van Reeven}, {Vanel}, {Vecchiato}, {Viala},
  {Vicente}, {Voutsinas}, {Weiler}, {Wevers}, {Wyrzykowski}, {Yoldas}, {Yvard},
  {Zhao}, {Zorec}, {Zucker}, \& {Zwitter}}]{GaiaDR3}
{Gaia Collaboration}, {Vallenari}, A., {Brown}, A.~G.~A., {et~al.} 2023, \aap,
  674, A1

\bibitem[{{Green} {et~al.}(2019){Green}, {Schlafly}, {Zucker}, {Speagle}, \&
  {Finkbeiner}}]{Green2019}
{Green}, G.~M., {Schlafly}, E., {Zucker}, C., {Speagle}, J.~S., \&
  {Finkbeiner}, D. 2019, \apj, 887, 93

\bibitem[{{Hall}(1949)}]{Hall1949}
{Hall}, J.~S. 1949, Science, 109, 166

\bibitem[{{Hensley} \& {Draine}(2021)}]{Henseley2021}
{Hensley}, B.~S. \& {Draine}, B.~T. 2021, \apj, 906, 73

\bibitem[{{Hiltner}(1949)}]{Hiltner1949}
{Hiltner}, W.~A. 1949, Science, 109, 165

\bibitem[{Hotelling(1992)}]{Hotelling1992}
Hotelling, H. 1992, The Generalization of Student’s Ratio (Springer New
  York), 54–65

\bibitem[{{Jaffe}(2019)}]{Jaffe2019}
{Jaffe}, T.~R. 2019, Galaxies, 7, 52

\bibitem[{{Jansson} \& {Farrar}(2012)}]{Jansson2012}
{Jansson}, R. \& {Farrar}, G.~R. 2012, \apj, 757, 14

\bibitem[{{Korochkin} {et~al.}(2025){Korochkin}, {Semikoz}, \&
  {Tinyakov}}]{Korochkin2025}
{Korochkin}, A., {Semikoz}, D., \& {Tinyakov}, P. 2025, \aap, 693, A284

\bibitem[{{Lallement} {et~al.}(2019){Lallement}, {Babusiaux}, {Vergely},
  {Katz}, {Arenou}, {Valette}, {Hottier}, \& {Capitanio}}]{Lallement2019}
{Lallement}, R., {Babusiaux}, C., {Vergely}, J.~L., {et~al.} 2019, \aap, 625,
  A135

\bibitem[{{Lallement} {et~al.}(2022){Lallement}, {Vergely}, {Babusiaux}, \&
  {Cox}}]{Lallement2022}
{Lallement}, R., {Vergely}, J.~L., {Babusiaux}, C., \& {Cox}, N.~L.~J. 2022,
  \aap, 661, A147

\bibitem[{{Lenz} {et~al.}(2017){Lenz}, {Hensley}, \& {Dor{\'e}}}]{Lenz2017}
{Lenz}, D., {Hensley}, B.~S., \& {Dor{\'e}}, O. 2017, \apj, 846, 38

\bibitem[{{Lombard} \& {Koen}(1993)}]{cumsum1993}
{Lombard}, F. \& {Koen}, C. 1993, \mnras, 263, 309

\bibitem[{{Magalh{\~a}es} {et~al.}(2005){Magalh{\~a}es}, {Pereyra},
  {Melgarejo}, {de Matos}, {Carciofi}, {Benedito}, {Valentim}, {Vidotto}, {da
  Silva}, {de Souza}, {Faria}, \& {Gabriel}}]{Magalhaes2005}
{Magalh{\~a}es}, A.~M., {Pereyra}, A., {Melgarejo}, R., {et~al.} 2005, in
  Astronomical Society of the Pacific Conference Series, Vol. 343, Astronomical
  Polarimetry: Current Status and Future Directions, ed. A.~{Adamson},
  C.~{Aspin}, C.~{Davis}, \& T.~{Fujiyoshi}, 305

\bibitem[{{Maharana} {et~al.}(2024){Maharana}, {Blinov}, {Ramaprakash},
  {Pavlidou}, \& {Tassis}}]{Maharana2024}
{Maharana}, S., {Blinov}, D., {Ramaprakash}, A.~N., {Pavlidou}, V., \&
  {Tassis}, K. 2024, arXiv e-prints, arXiv:2407.13470

\bibitem[{{Panopoulou} {et~al.}(2025){Panopoulou}, {Markopoulioti}, {Bouzelou},
  {Millar-Blanchaer}, {Tinyanont}, {Blinov}, {Pelgrims}, {Johnson}, {Skalidis},
  \& {Soam}}]{Panopoulou2025}
{Panopoulou}, G.~V., {Markopoulioti}, L., {Bouzelou}, F., {et~al.} 2025, \apjs,
  276, 15

\bibitem[{{Panopoulou} {et~al.}(2019){Panopoulou}, {Tassis}, {Skalidis},
  {Blinov}, {Liodakis}, {Pavlidou}, {Potter}, {Ramaprakash}, {Readhead}, \&
  {Wehus}}]{Panopoulou2019}
{Panopoulou}, G.~V., {Tassis}, K., {Skalidis}, R., {et~al.} 2019, \apj, 872, 56

\bibitem[{{Patat} {et~al.}(2010){Patat}, {Maund}, {Benetti}, {Botticella},
  {Cappellaro}, {Harutyunyan}, \& {Turatto}}]{patat2010}
{Patat}, F., {Maund}, J.~R., {Benetti}, S., {et~al.} 2010, \aap, 510, A108

\bibitem[{{Pattle} {et~al.}(2023){Pattle}, {Fissel}, {Tahani}, {Liu}, \&
  {Ntormousi}}]{Pattle2023}
{Pattle}, K., {Fissel}, L., {Tahani}, M., {Liu}, T., \& {Ntormousi}, E. 2023,
  in Astronomical Society of the Pacific Conference Series, Vol. 534,
  Protostars and Planets VII, ed. S.~{Inutsuka}, Y.~{Aikawa}, T.~{Muto},
  K.~{Tomida}, \& M.~{Tamura}, 193

\bibitem[{{Pavel}(2014)}]{Pavel2014}
{Pavel}, M.~D. 2014, \aj, 148, 49

\bibitem[{{Pelgrims} {et~al.}(2020){Pelgrims}, {Ferri{\`e}re}, {Boulanger},
  {Lallement}, \& {Montier}}]{Pelgrims2020}
{Pelgrims}, V., {Ferri{\`e}re}, K., {Boulanger}, F., {Lallement}, R., \&
  {Montier}, L. 2020, \aap, 636, A17

\bibitem[{{Pelgrims} {et~al.}(2024){Pelgrims}, {Mandarakas}, {Skalidis},
  {Tassis}, {Panopoulou}, {Pavlidou}, {Blinov}, {Kiehlmann}, {Clark},
  {Hensley}, {Romanopoulos}, {Basyrov}, {Eriksen}, {Falalaki}, {Ghosh},
  {Gjerl{\o}w}, {Kypriotakis}, {Maharana}, {Papadaki}, {Pearson}, {Potter},
  {Ramaprakash}, {Readhead}, \& {Wehus}}]{Pelgrims2024}
{Pelgrims}, V., {Mandarakas}, N., {Skalidis}, R., {et~al.} 2024, \aap, 684,
  A162

\bibitem[{{Pelgrims} {et~al.}(2023){Pelgrims}, {Panopoulou}, {Tassis},
  {Pavlidou}, {Basyrov}, {Blinov}, {Gjerl{\ensuremath{\varnothing}}w},
  {Kiehlmann}, {Mandarakas}, {Papadaki}, {Skalidis}, {Tsouros}, {Anche},
  {Eriksen}, {Ghosh}, {Kypriotakis}, {Maharana}, {Ntormousi}, {Pearson},
  {Potter}, {Ramaprakash}, {Readhead}, \& {Wehus}}]{Pelgrims2023}
{Pelgrims}, V., {Panopoulou}, G.~V., {Tassis}, K., {et~al.} 2023, \aap, 670,
  A164

\bibitem[{{Pelgrims} {et~al.}(2025){Pelgrims}, {Unger}, \&
  {Mari{\c{s}}}}]{pelgrims2025}
{Pelgrims}, V., {Unger}, M., \& {Mari{\c{s}}}, I.~C. 2025, \aap, 695, A148

\bibitem[{{Planck Collaboration} {et~al.}(2016){Planck Collaboration},
  {Aghanim}, {Ashdown}, {Aumont}, {Baccigalupi}, {Ballardini}, {Banday},
  {Barreiro}, {Bartolo}, {Basak}, {Benabed}, {Bernard}, {Bersanelli},
  {Bielewicz}, {Bonavera}, {Bond}, {Borrill}, {Bouchet}, {Boulanger},
  {Burigana}, {Calabrese}, {Cardoso}, {Carron}, {Chiang}, {Colombo}, {Comis},
  {Couchot}, {Coulais}, {Crill}, {Curto}, {Cuttaia}, {de Bernardis}, {de
  Zotti}, {Delabrouille}, {Di Valentino}, {Dickinson}, {Diego}, {Dor{\'e}},
  {Douspis}, {Ducout}, {Dupac}, {Dusini}, {Elsner}, {En{\ss}lin}, {Eriksen},
  {Falgarone}, {Fantaye}, {Finelli}, {Forastieri}, {Frailis}, {Fraisse},
  {Franceschi}, {Frolov}, {Galeotta}, {Galli}, {Ganga}, {G{\'e}nova-Santos},
  {Gerbino}, {Ghosh}, {Giraud-H{\'e}raud}, {Gonz{\'a}lez-Nuevo}, {G{\'o}rski},
  {Gruppuso}, {Gudmundsson}, {Hansen}, {Helou}, {Henrot-Versill{\'e}},
  {Herranz}, {Hivon}, {Huang}, {Jaffe}, {Jones}, {Keih{\"a}nen}, {Keskitalo},
  {Kiiveri}, {Kisner}, {Krachmalnicoff}, {Kunz}, {Kurki-Suonio}, {Lamarre},
  {Langer}, {Lasenby}, {Lattanzi}, {Lawrence}, {Le Jeune}, {Levrier}, {Lilje},
  {Lilley}, {Lindholm}, {L{\'o}pez-Caniego}, {Ma}, {Mac{\'\i}as-P{\'e}rez},
  {Maggio}, {Maino}, {Mandolesi}, {Mangilli}, {Maris}, {Martin},
  {Mart{\'\i}nez-Gonz{\'a}lez}, {Matarrese}, {Mauri}, {McEwen}, {Melchiorri},
  {Mennella}, {Migliaccio}, {Miville-Desch{\^e}nes}, {Molinari}, {Moneti},
  {Montier}, {Morgante}, {Moss}, {Natoli}, {Oxborrow}, {Pagano}, {Paoletti},
  {Patanchon}, {Perdereau}, {Perotto}, {Pettorino}, {Piacentini},
  {Plaszczynski}, {Polastri}, {Polenta}, {Puget}, {Rachen}, {Racine},
  {Reinecke}, {Remazeilles}, {Renzi}, {Rocha}, {Rosset}, {Rossetti}, {Roudier},
  {Rubi{\~n}o-Mart{\'\i}n}, {Ruiz-Granados}, {Salvati}, {Sandri}, {Savelainen},
  {Scott}, {Sirignano}, {Sirri}, {Soler}, {Spencer}, {Suur-Uski}, {Tauber},
  {Tavagnacco}, {Tenti}, {Toffolatti}, {Tomasi}, {Tristram}, {Trombetti},
  {Valiviita}, {Van Tent}, {Vielva}, {Villa}, {Vittorio}, {Wandelt}, {Wehus},
  {Zacchei}, \& {Zonca}}]{planck2016}
{Planck Collaboration}, {Aghanim}, N., {Ashdown}, M., {et~al.} 2016, \aap, 596,
  A109

\bibitem[{{Rezaei Kh.} {et~al.}(2020){Rezaei Kh.}, {Bailer-Jones}, {Soler}, \&
  {Zari}}]{RezaeiOrion}
{Rezaei Kh.}, S., {Bailer-Jones}, C. A.~L., {Soler}, J.~D., \& {Zari}, E. 2020,
  \aap, 643, A151

\bibitem[{{Schlegel} {et~al.}(1998){Schlegel}, {Finkbeiner}, \&
  {Davis}}]{Schlegel1998}
{Schlegel}, D.~J., {Finkbeiner}, D.~P., \& {Davis}, M. 1998, \apj, 500, 525

\bibitem[{Schwarz(1978)}]{schwarz1978estimating}
Schwarz, G. 1978, The annals of statistics, 461

\bibitem[{{Skalidis} \& {Pelgrims}(2019)}]{Skalidis2019}
{Skalidis}, R. \& {Pelgrims}, V. 2019, \aap, 631, L11

\bibitem[{{Speagle}(2020)}]{Speagle2020}
{Speagle}, J.~S. 2020, \mnras, 493, 3132

\bibitem[{{Sun} {et~al.}(2008){Sun}, {Reich}, {Waelkens}, \&
  {En{\ss}lin}}]{Sun2008}
{Sun}, X.~H., {Reich}, W., {Waelkens}, A., \& {En{\ss}lin}, T.~A. 2008, \aap,
  477, 573

\bibitem[{{Tahani} {et~al.}(2022){Tahani}, {Glover}, {Lupypciw}, {West},
  {Kothes}, {Plume}, {Inutsuka}, {Lee}, {Grenier}, {Knee}, {Brown}, {Doi},
  {Robishaw}, \& {Haverkorn}}]{Tahani2022Orion}
{Tahani}, M., {Glover}, J., {Lupypciw}, W., {et~al.} 2022, \aap, 660, L7

\bibitem[{{Tahani} {et~al.}(2018){Tahani}, {Plume}, {Brown}, \&
  {Kainulainen}}]{Tahani2018}
{Tahani}, M., {Plume}, R., {Brown}, J.~C., \& {Kainulainen}, J. 2018, \aap,
  614, A100

\bibitem[{{Tassis} {et~al.}(2018){Tassis}, {Ramaprakash}, {Readhead}, {Potter},
  {Wehus}, {Panopoulou}, {Blinov}, {Eriksen}, {Hensley}, {Karakci},
  {Kypriotakis}, {Maharana}, {Ntormousi}, {Pavlidou}, {Pearson}, \&
  {Skalidis}}]{Tassis2018}
{Tassis}, K., {Ramaprakash}, A.~N., {Readhead}, A. C.~S., {et~al.} 2018, arXiv
  e-prints, arXiv:1810.05652

\bibitem[{{Tritsis} \& {Tassis}(2018)}]{TritsisMusca}
{Tritsis}, A. \& {Tassis}, K. 2018, Science, 360, 635

\bibitem[{{Unger} \& {Farrar}(2024)}]{Unger2024}
{Unger}, M. \& {Farrar}, G.~R. 2024, \apj, 970, 95

\bibitem[{{Uppal} {et~al.}(2024){Uppal}, {Ganesh}, {Pelgrims}, {Joshi}, \&
  {Sarkar}}]{Uppal2024}
{Uppal}, N., {Ganesh}, S., {Pelgrims}, V., {Joshi}, S., \& {Sarkar}, M. 2024,
  \aap, 690, A49

\bibitem[{Vallat(2018)}]{pingouin2018}
Vallat, R. 2018, Journal of Open Source Software, 3, 1026

\bibitem[{{Vergely} {et~al.}(2022){Vergely}, {Lallement}, \&
  {Cox}}]{vergely2022}
{Vergely}, J.~L., {Lallement}, R., \& {Cox}, N.~L.~J. 2022, \aap, 664, A174

\bibitem[{{Versteeg} {et~al.}(2023){Versteeg}, {Magalh{\~a}es}, {Haverkorn},
  {Angarita}, {Rodrigues}, {Santos-Lima}, \& {Kawabata}}]{Versteeg2023}
{Versteeg}, M.~J.~F., {Magalh{\~a}es}, A.~M., {Haverkorn}, M., {et~al.} 2023,
  \aj, 165, 87

\bibitem[{{Ward} {et~al.}(2022){Ward}, {Dilillo}, {Eckley}, \&
  {Fearnhead}}]{cumsum2022}
{Ward}, K., {Dilillo}, G., {Eckley}, I., \& {Fearnhead}, P. 2022, arXiv
  e-prints, arXiv:2208.01494

\bibitem[{{Wolleben} {et~al.}(2010){Wolleben}, {Fletcher}, {Landecker},
  {Carretti}, {Dickey}, {Gaensler}, {Haverkorn}, {McClure-Griffiths}, {Reich},
  \& {Taylor}}]{Wolleben2010}
{Wolleben}, M., {Fletcher}, A., {Landecker}, T.~L., {et~al.} 2010, \apjl, 724,
  L48

\bibitem[{Zeileis {et~al.}(2003)Zeileis, Kleiber, Kr\"{a}mer, \&
  Hornik}]{Zeileis2003}
Zeileis, A., Kleiber, C., Kr\"{a}mer, W., \& Hornik, K. 2003, Computational
  Statistics \& Data Analysis, 44, 109–123

\bibitem[{Zeileis {et~al.}(2002)Zeileis, Leisch, Hornik, \&
  Kleiber}]{Zeileis2002}
Zeileis, A., Leisch, F., Hornik, K., \& Kleiber, C. 2002, Journal of
  Statistical Software, 7

\bibitem[{{Zhang} {et~al.}(2025){Zhang}, {Su}, {Chen}, {Fang}, {Du}, {Zhang},
  {Yan}, {Liu}, {Zhang}, {Sun}, \& {Yang}}]{Zhang2025}
{Zhang}, S., {Su}, Y., {Chen}, X., {et~al.} 2025, arXiv e-prints,
  arXiv:2507.18002

\end{thebibliography}

\appendix
\section{Nature of flagged cases}\label{A1}
As described in Sect.~\ref{sec:2.3}, we flag sightlines as flag = `1', where a break passes the statistical significance test (i.e., the null hypothesis is rejected) in only one direction, either forward or backward, but not both. This situation typically arises when the breakpoint detection algorithm, prior to any rejection step, misses the true cloud location and instead identifies a nearby spurious break caused by local fluctuations in the cumulative Mahalanobis distance. Since the segmentation boundaries are updated dynamically,  rejecting a break early in the sequence reshapes subsequent segment boundaries and can affect the outcome of the test for later breaks. Hence, a break that appears significant in one direction might not be confirmed when tested from the opposite direction.

\begin{figure}
   \centering
\subfigure{\includegraphics[width=0.4\textwidth]{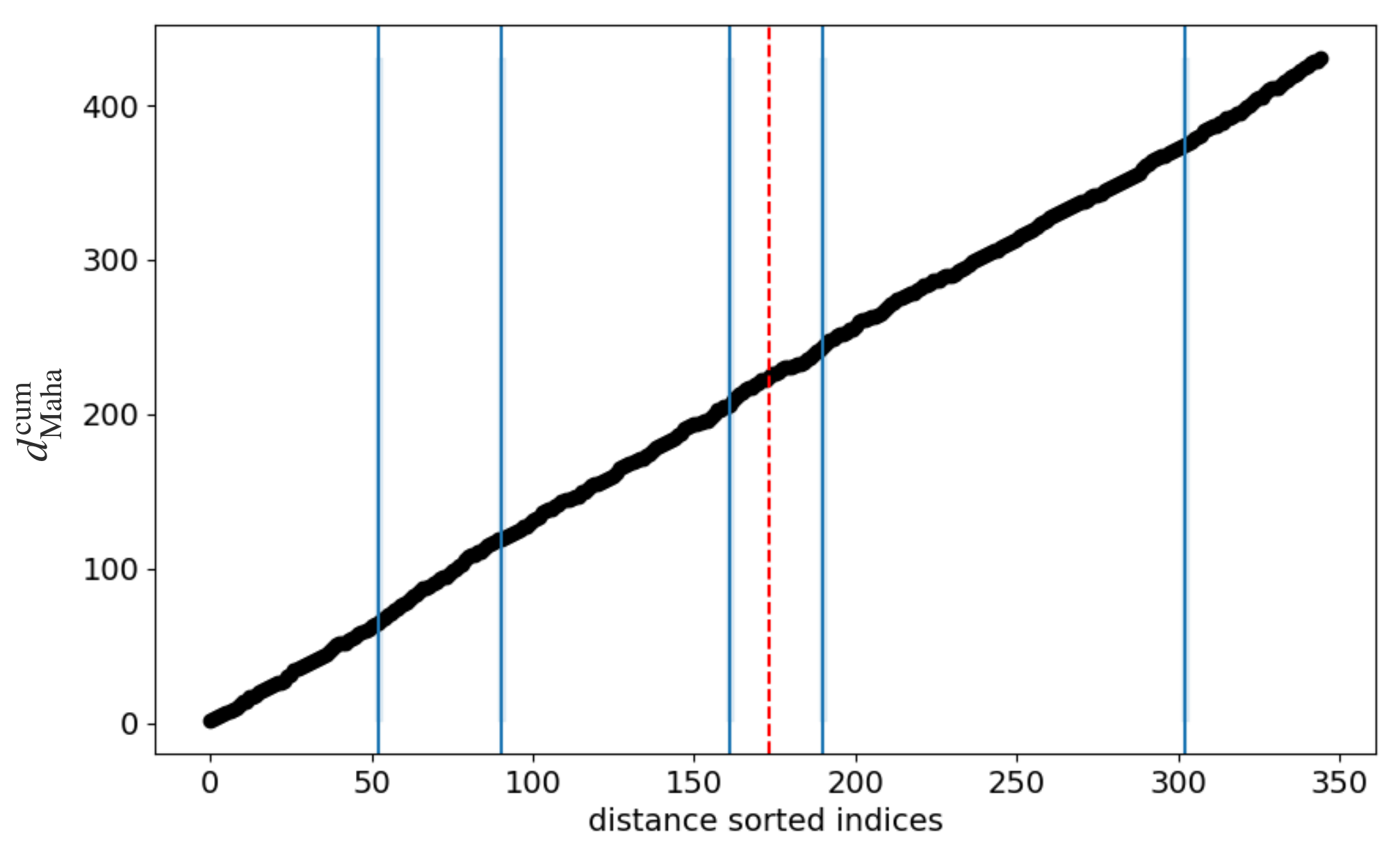}} 

\subfigure{\includegraphics[width=0.4\textwidth]{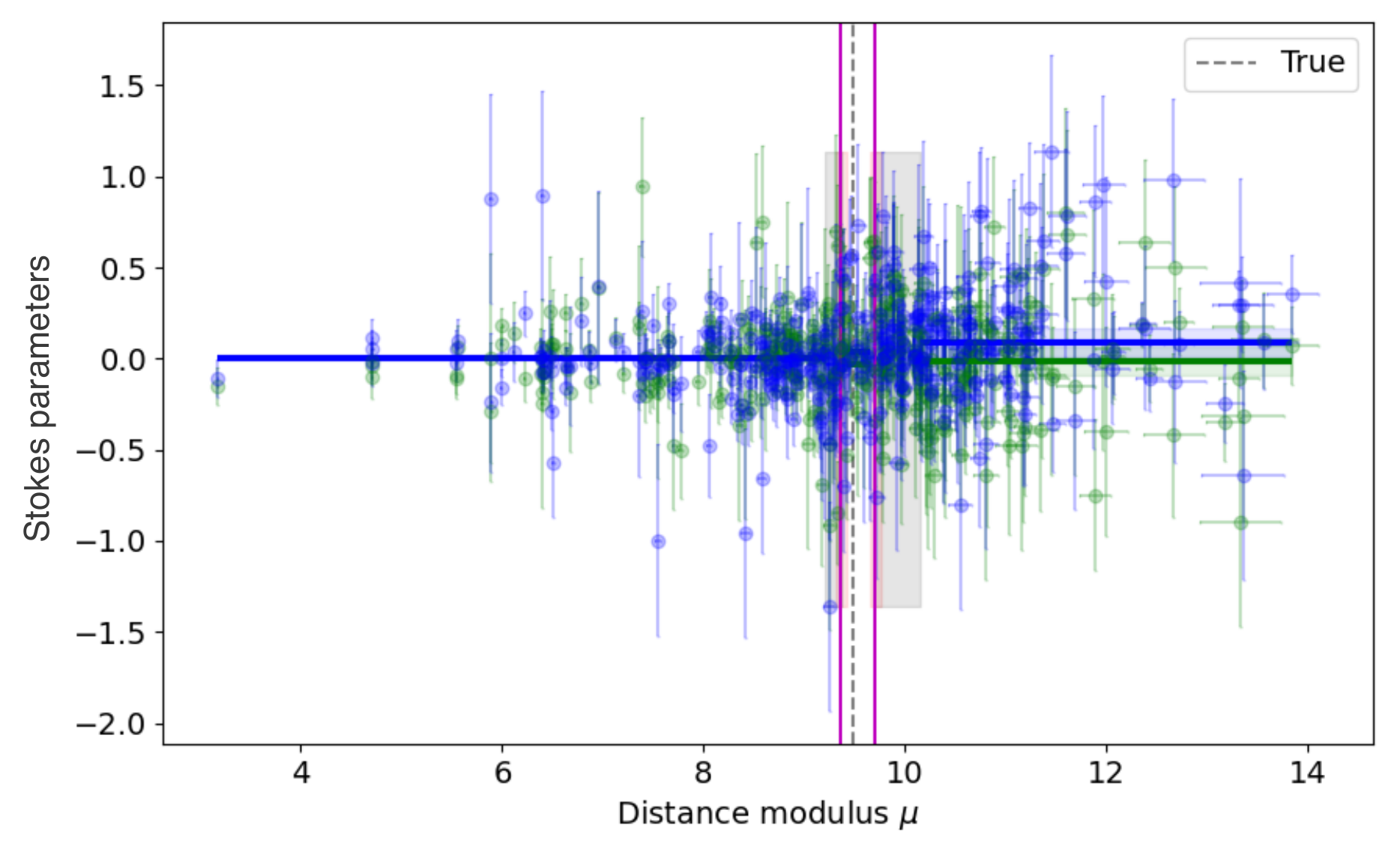}} 
      \caption{Example of a flagged reconstruction case. The top panel shows the breakpoints (blue vertical lines) identified by the \textit{strucchange} package in the distance-sorted cumulative Mahalanobis distance profile. The bottom panel displays the line-of-sight decomposition in the $qu$–$\mu$ plane ($q$ in green, while $u$ in blue). The red dashed line in the top panel and the gray dashed line in the bottom panel indicate the true location of the input dust layer.} \label{fig:A1}
   \end{figure}

In most single-cloud mock samples, a common type of flagged case is observed when a break identified in the forwards pass is rejected in the backward (or vice-versa) while a different break is selected in the reverse direction by rejecting null-hypothesis at the same confidence level.
In such cases, the algorithm outputs two different breaks, both of which are flagged as flag = `1'. Although the identified breakpoints differ, they consistently indicate the presence of only one cloud layer in both analysis direction. Therefore, such cases should not be considered as false positives. Rather, they reflect an underlying detection, where the presence of one layer is real but its precise location is uncertain due to inaccuracies in the initial identification of the breaks by the \textit{strucchange} breakpoint algorithm. An example of this behavior is shown in Fig.~\ref{fig:A1}, where the mock sample contains a single cloud at approximately $790$~pc, inducing a polarization of $0.1\%$. In this case, $50\%$ of the stars lie behind the cloud.  

We applied \texttt{TRIShUL} on this data set, and the \textit{strucchange} breakpoint detection algorithm identifies five breaks in the distance-sorted cumulative Mahalanobis distance profile, marked by blue vertical lines in the top panel.
However, none of these breaks align with the true cloud location, possibly due to the dominance of fluctuation in the cumulative profile over the actual polarization signal.  The rejection step eliminates all but the fourth break in the forward pass, while in the backward pass, the fourth break is rejected in favor of the third one. As a result, the algorithm outputs both the third and fourth breaks, but flags them as uncertain. The final solution, displayed in the $qu$–$\mu$ plane, is shown in the bottom panel of Fig.~\ref{fig:A1}.  In cases involving two or more clouds, the number of flagged cases is expected to increase, as similar ambiguity can be shown by one or more breaks along the LOS. 

In many such cases, one of the identified breaks can be discarded by examining the estimated polarization, if it is consistent with zero within the uncertainties, and recompute the cloud parameters based only on the remaining break. A more recommended approach is to rerun \texttt{TRIShUL} with modified breakpoint detection parameters, such as increasing the maximum number of allowed breaks to better capture the true transition or adjusting the minimum number of stars per segment. For the illustrated example, increasing the maximum number of breakpoints to ten successfully resolves the flagged case.

\section{Effect of inclination angle of magnetic field on the performance of our method}\label{A2}
In this section, we examine the effect of the inclination angle of the magnetic field with POS on the performance of our method. Similar simulated data is generated as described in Sect.~\ref{sec:3.1} but with inclination angle $\gamma_{B_{\rm{reg}}}$ sampled from $0^\circ$ to $75^\circ$ in $15^\circ$ steps, and $p_{\rm{max}}$ adjusted to keep the observed polarization $p_C \approx 0.2\%$. For each configuration, 10 realizations are generated by varying the POS orientation of the magnetic field $\psi_{B_{\rm{reg}}}$, resulting in a total of 300 samples.

\begin{figure}
   \centering
\subfigure{\includegraphics[width=0.4\textwidth]{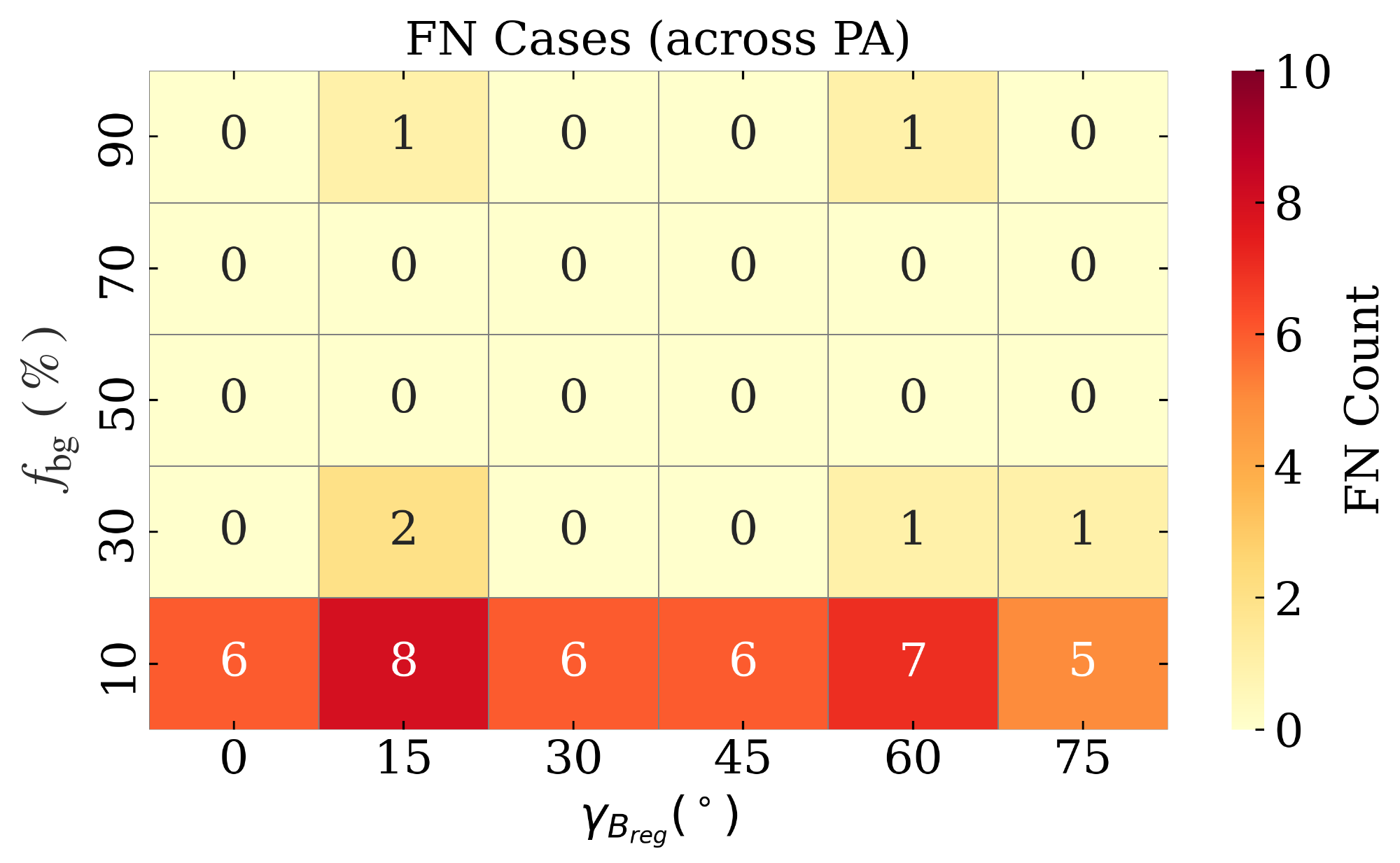}}
\subfigure{\includegraphics[width=0.4\textwidth]{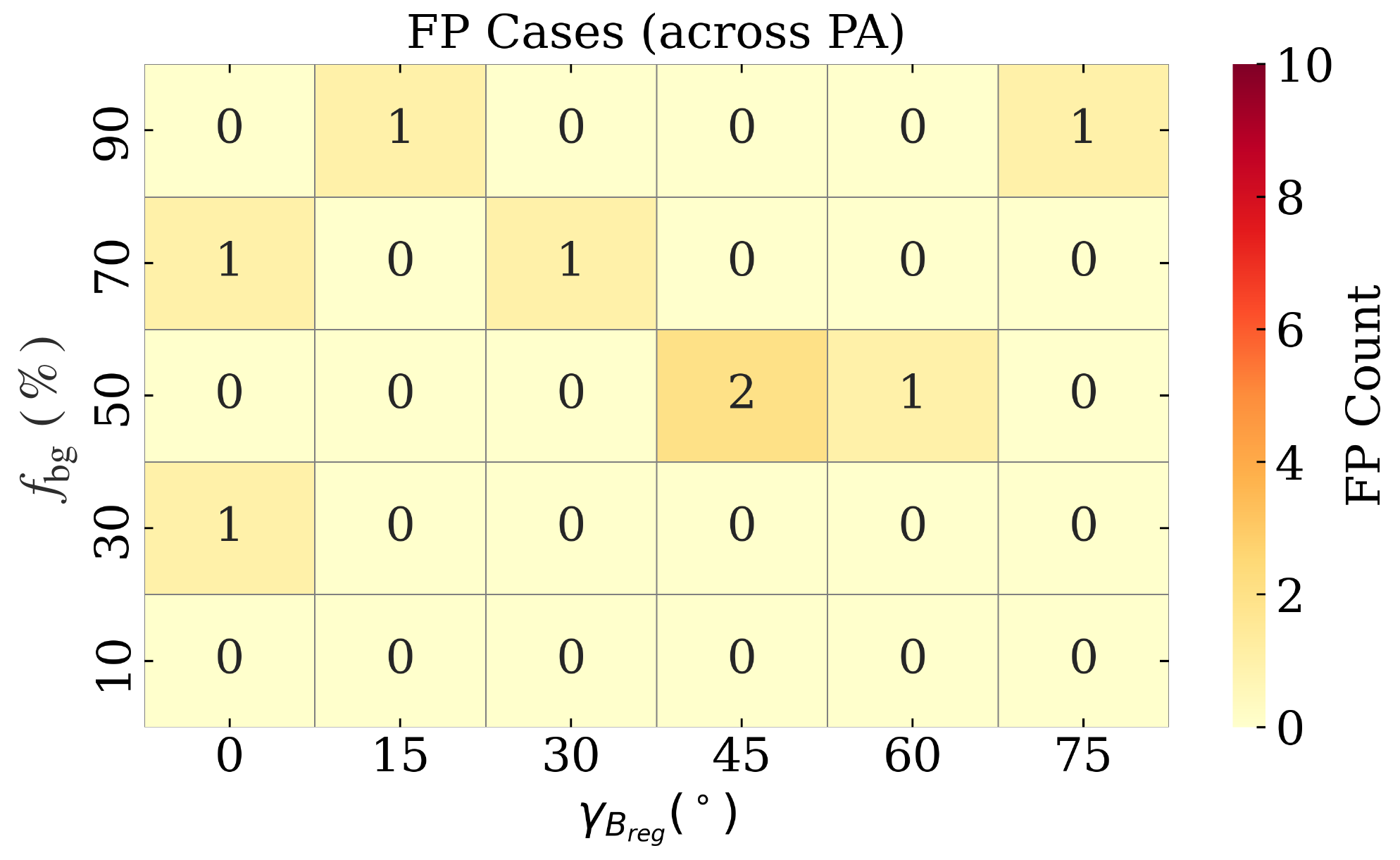}}
 \subfigure{\includegraphics[width=0.4\textwidth]{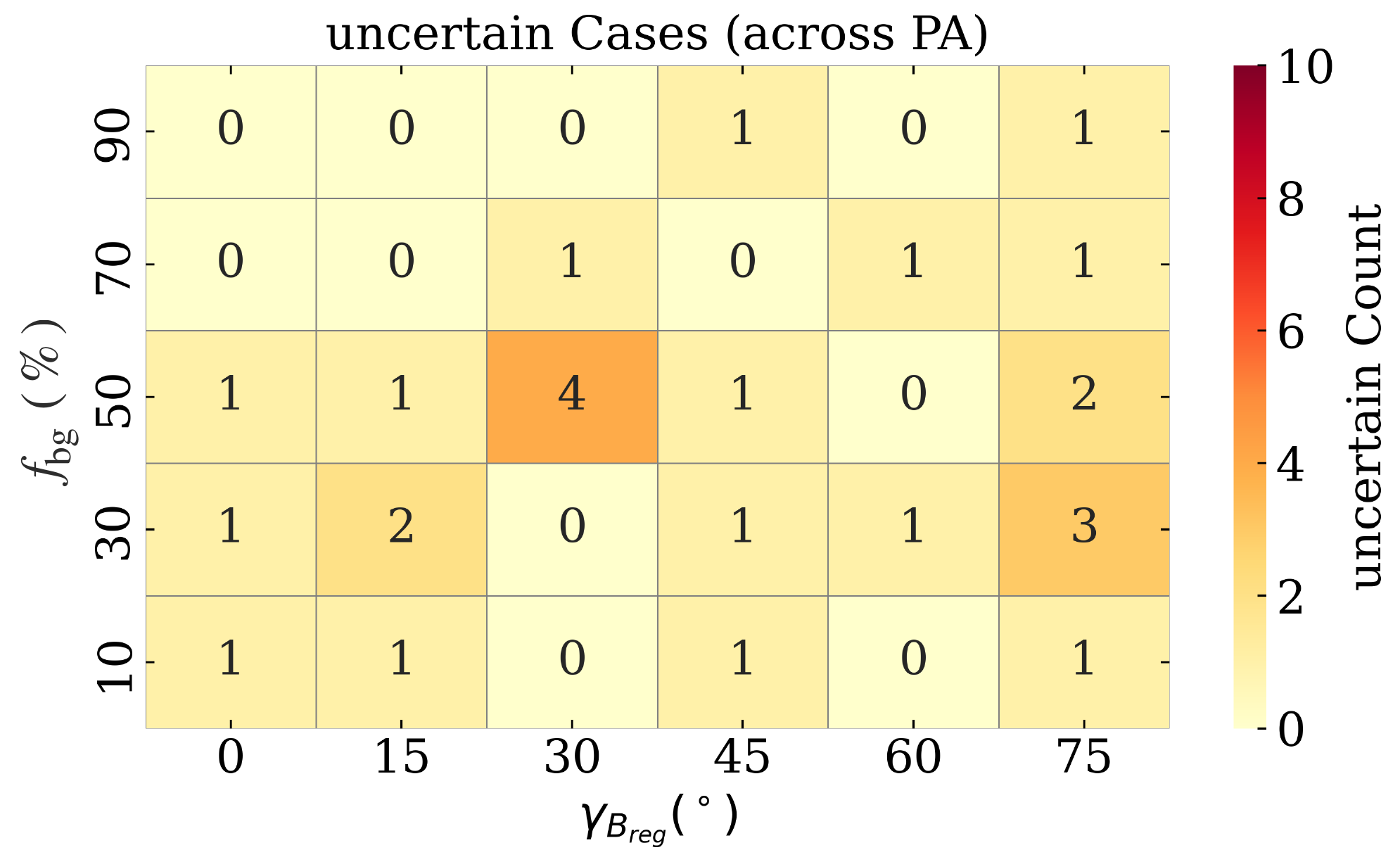}}
\subfigure{\includegraphics[width=0.4\textwidth]{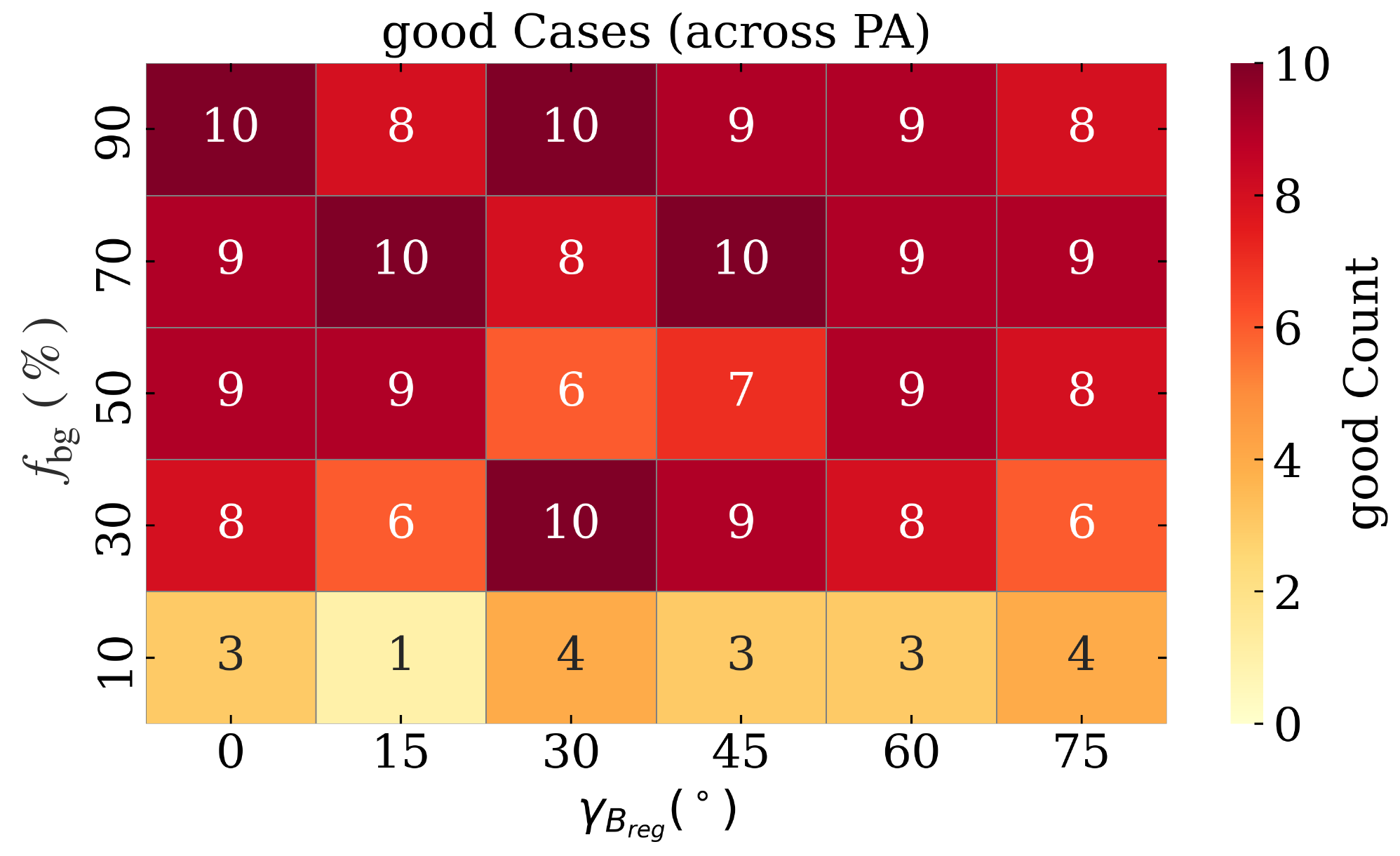}}
\caption{Distribution of decomposition outcomes (FP, FN, flagged, and reliable) for 300 single-cloud mock sightlines, plotted as a function of inclination angle of magnetic field ($\gamma_{B_{\rm{ref}}}$) and background star fraction ($f_{\rm{bg}}$) across ten different position angle of POS magnetic field.} \label{fig:A21}
   \end{figure}

We performed LOS tomography decomposition of starlight polarization with our method, testing a maximum of five breaks, with a minimum of five stars per segment. Among the 300 tested samples, approximately 15\% resulted in false negatives. As shown in the top-most panel of Fig.~\ref{fig:A21}, these failures are predominantly associated with cases where $f_{\rm{bg}}$ = 10\%,  similar to the results shown in Sect.~\ref{sec:3.1}. Only $3\%$ of the cases led to false positives, illustrated in the second panel. About $8.6\%$ of the cases are flagged as uncertain and require further analysis, while the remaining 74\% successfully identified a single cloud with high statistical confidence (see the bottom panel of Fig.~\ref{fig:A21}).

\begin{figure}
   \centering
\subfigure{\includegraphics[width=0.4\textwidth]{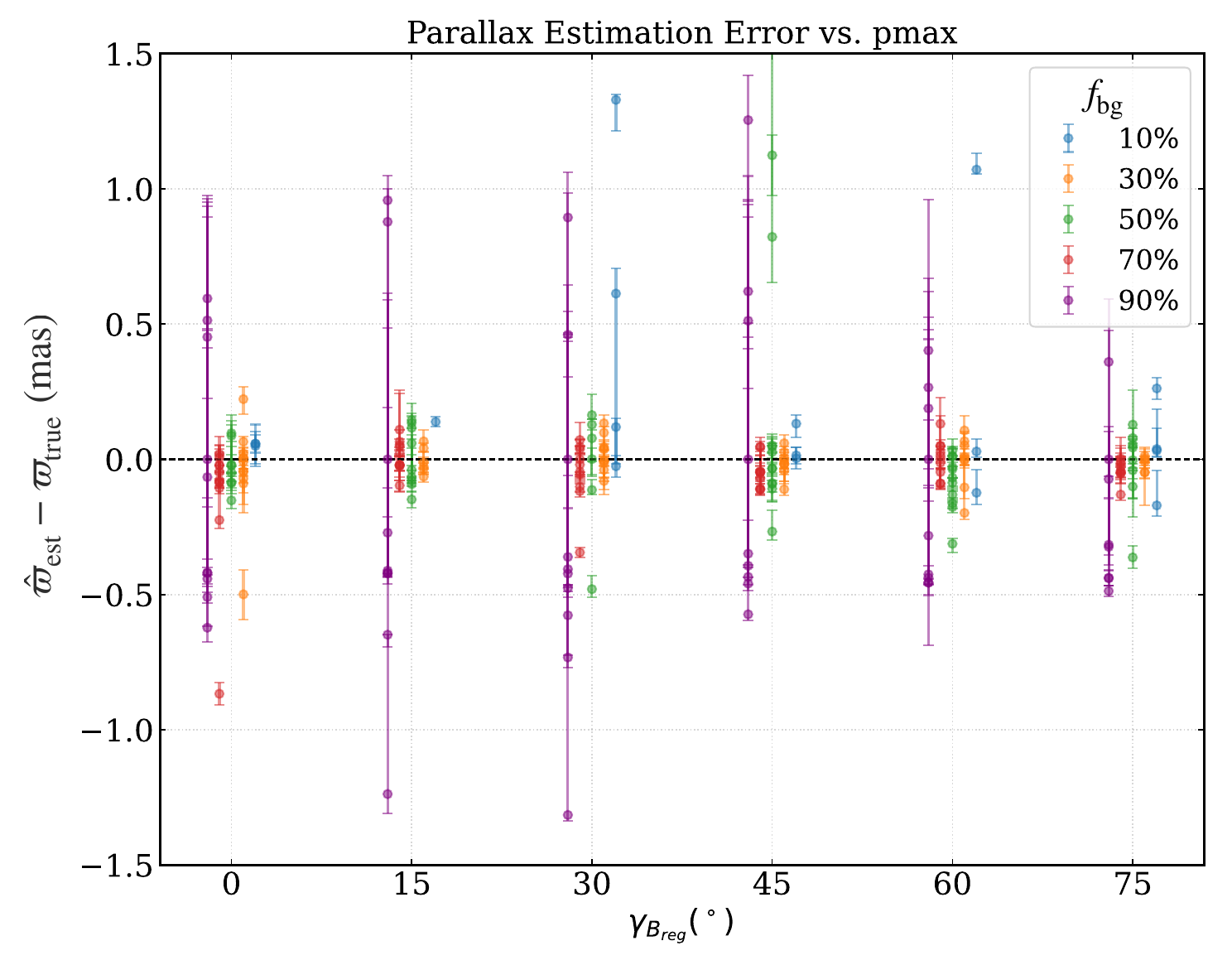}}

\subfigure{\includegraphics[width=0.4\textwidth]{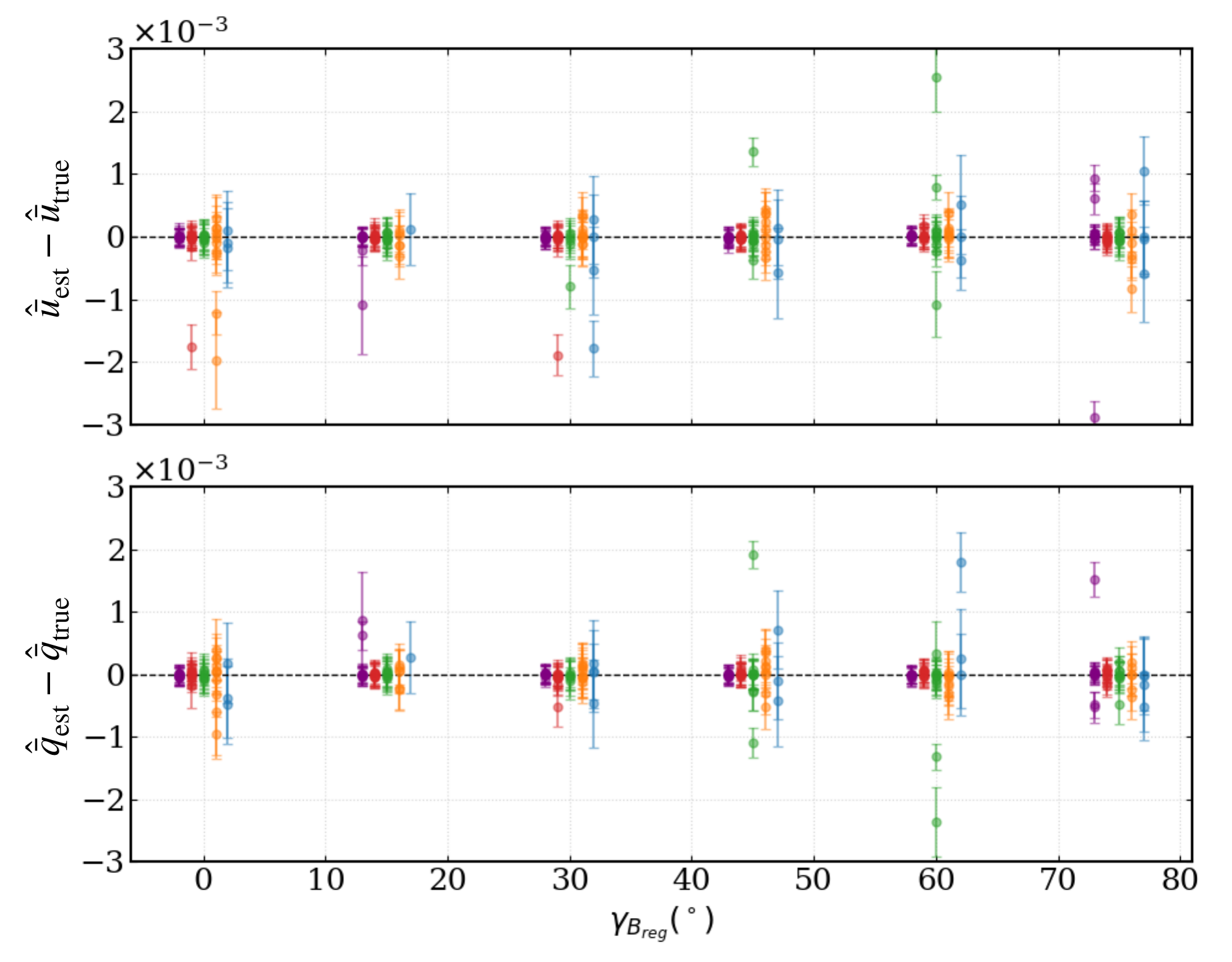}}
   \caption{Difference in the estimated parameters, parallax (top panel) and Stokes parameter (bottom panel), of the identified clouds with corresponding true values as a function of inclination angle of magnetic field ($\gamma_{B_{\rm{reg}}}$). The black dashed line is the reference for a perfect match between estimated and true values.} \label{fig: FigA22}
   \end{figure}

To assess the influence of magnetic field inclination on the accuracy of estimated parameters from our method, we computed the deviations of the estimated values (parallax and Stokes parameters) from corresponding true values, following Sect.~\ref{sec:3.1} and plotted in Fig~\ref{fig: FigA22}. The results are consistent with those shown in Fig.~\ref{fig:Fig4} for single-cloud scenarios. No clear dependency of estimated parameters on the inclination angle for a fixed $f_{\rm{bg}}$ is observed in the figure. The dominant factor affecting the accuracy of the estimated parameters remains the background star fraction ($f_{\rm{bg}}$), i.e., the method often fails to detect the layer when only 10\% of stars are located behind the cloud. 
The figure shows increased scatter and larger uncertainties in the estimated cloud parallax for $f_{\rm{bg}} = 90\%$ and the same effect in polarization properties ($q$ and $u$), but for $f_{\rm{bg}} = 10\%$; these trends are discussed in detail in Sect.~\ref{sec:3.1}.
  
\section{Classification statistics for single-cloud mock samples}
\label{A3}
\begin{figure}
   \centering
\subfigure{\includegraphics[width=0.4\textwidth]{Fig3.pdf}}
\subfigure{\includegraphics[width=0.4\textwidth]{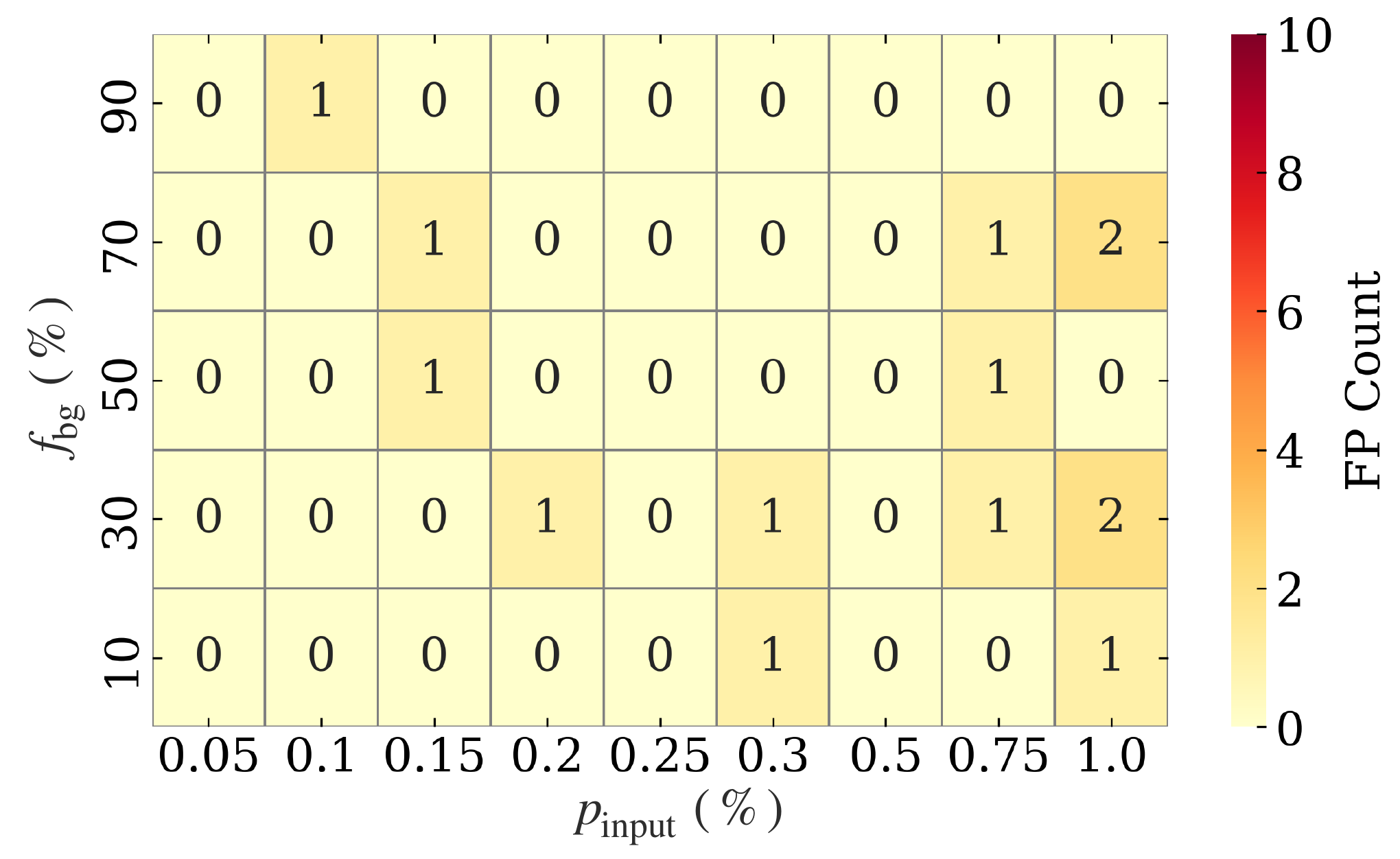}}
 \subfigure{\includegraphics[width=0.4\textwidth]{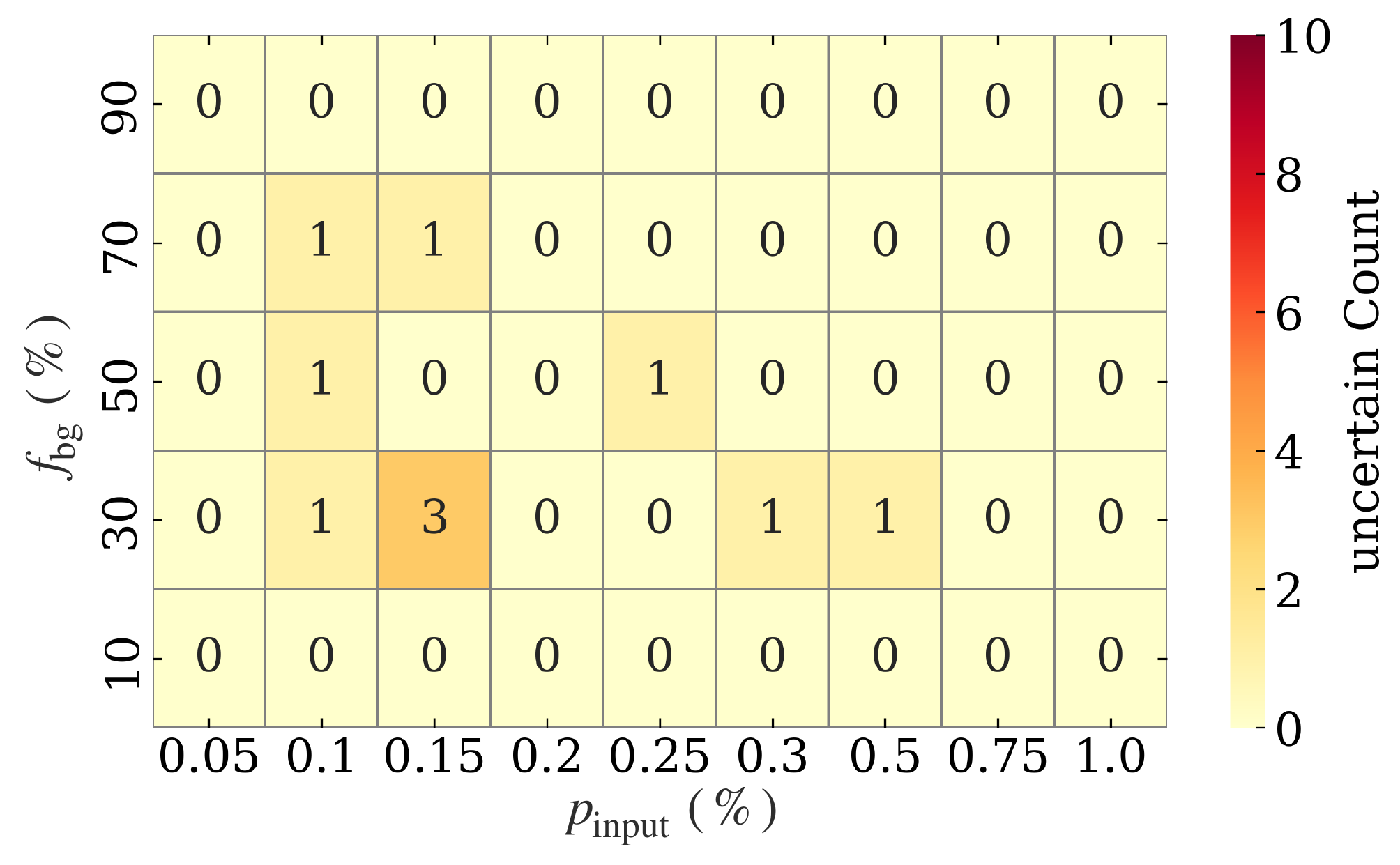}}
\subfigure{\includegraphics[width=0.4\textwidth]{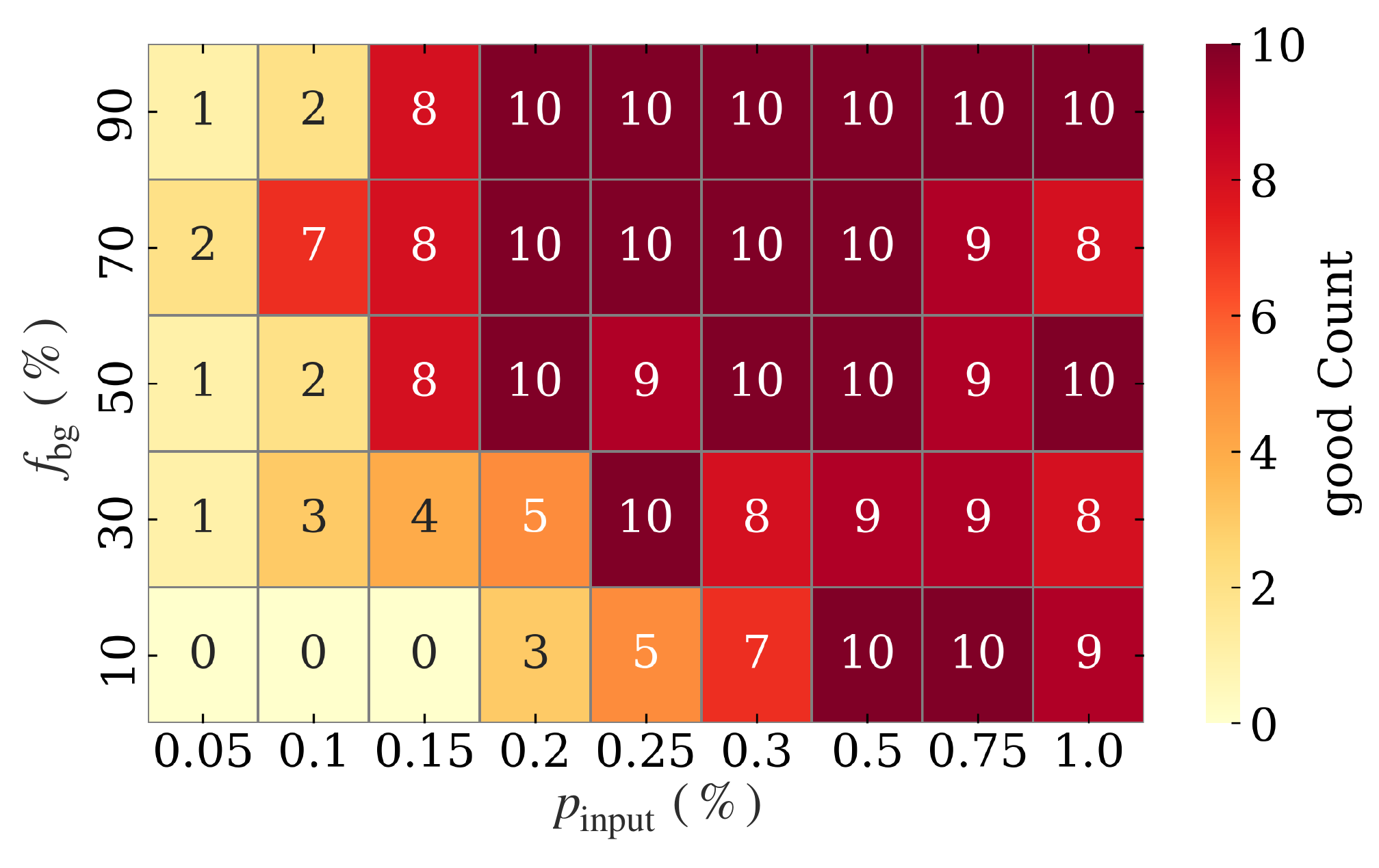}}
\caption{Distribution of decomposition outcomes (FN, FP, flagged, and reliable) for 450 single-cloud mock sightlines. Plotted as a function of input polarization strength ($p_{\rm{input}}$) and background star fraction ($f_{\rm{bg}}$).} \label{fig:A3}
   \end{figure}

The distribution of the number of FN, FP, flagged, and reliable cases based on the number of clouds recovered in the LOS tomography of starlight polarization in 450 single cloud mock samples is presented in Fig.~\ref{fig:A3}. The results are presented as a function of the input polarization signal ($p_{\rm{input}}$), i.e., the maximum polarization induced by the cloud on background stars, and the fraction of stars located behind the cloud $(f_{\rm{bg}})$. A detailed discussion and interpretation of these results is provided in Sect.~\ref{sec:3.1}.
\section{Classification statistics for two-cloud mock samples}\label{A4} 
\begin{figure} 
   \resizebox{\hsize}{!}{\includegraphics[angle=270]{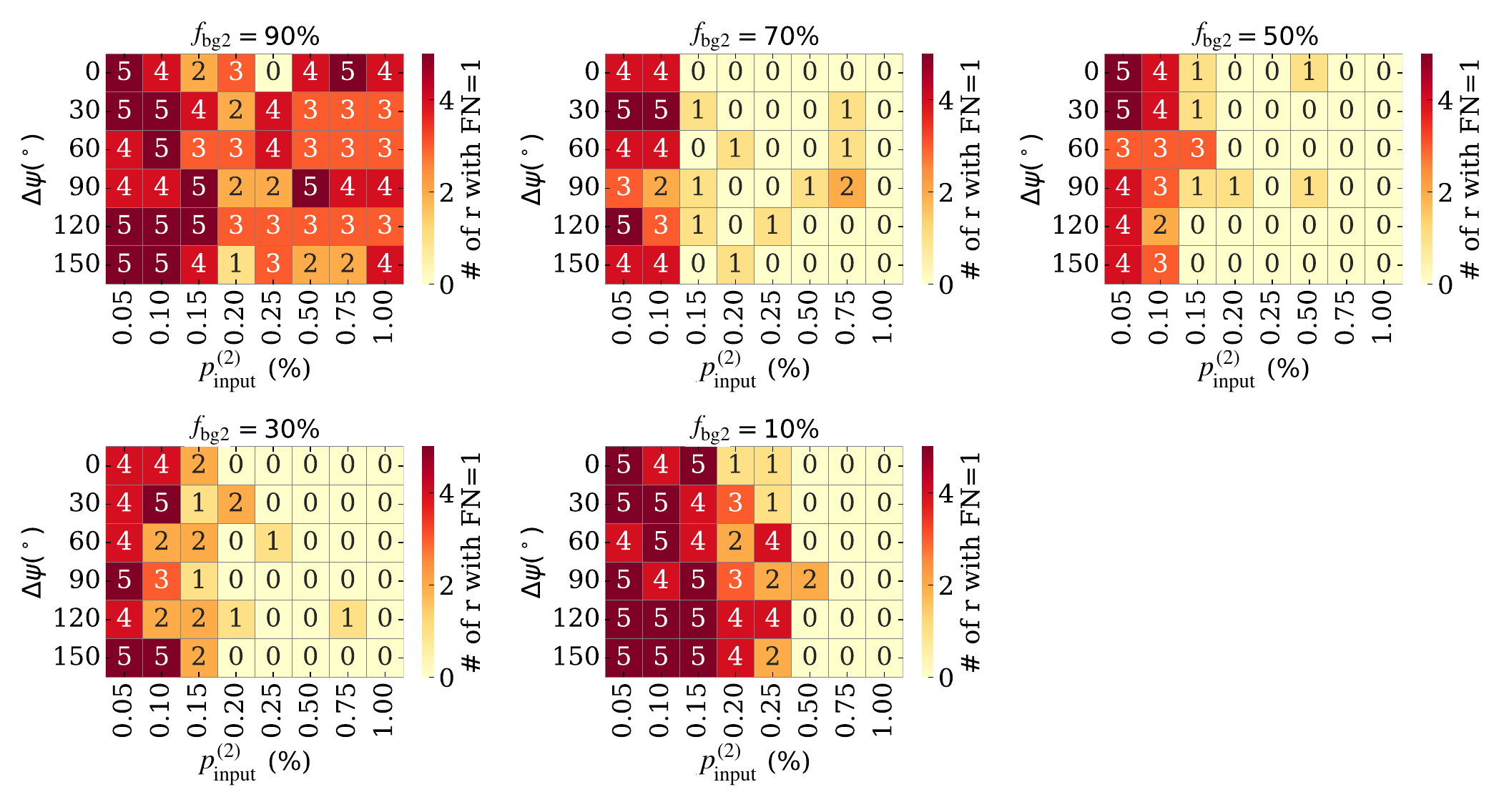}}
     \caption{Distribution of false negatives decomposition results for 1200 two-cloud mock samples using \texttt{TRIShUL}. The statistics represents the resulted FN for five random samples of input combination of background star fraction ($f_{\rm{bg2}}$, separate panels), maximum polarization of second cloud ($p_{\rm{input}}^{(2)}$, along $X$-axis), and differential polarization angle between two clouds ($\psi$, $Y$-axis).}
     \label{fig:FigA4}
\end{figure}

For the two cloud scenario, we generate 1200 sample, each corresponds to a unique combination of three input parameters: the maximum polarization induced by the second (farther) cloud ($p_{\rm{input}}^{(2)}$), the relative position angle of magnetic field in second cloud with respect to the nearby ($\Delta \psi$), and the fraction of stars that are background of nearby cloud that are also background of second cloud, ($f_{\rm{bg2}}$). For each parameter combination, five random realizations were generated. The number of false negative cases resulted from the LOS tomography decomposition of starlight polarization for using \texttt{TRIShUL} are shown in Fig.~\ref{fig:FigA4}. To visualize the joint impact of these variables, the results are displayed in panels grouped by $f_{\rm{bg2}}$, with $p_{\rm{input}}^{(2)}$ on the $X$-axis and $\Delta \psi$ on the $Y$-axis. The implications and interpretation of these results are addressed in Sect.~\ref{sec:3.2}

\end{document}